\documentclass[fleqn,usenatbib]{mnras}
\DeclareUnicodeCharacter{2212}{\textendash}
\usepackage{newtxtext,newtxmath}

\usepackage[T1]{fontenc}

\DeclareRobustCommand{\VAN}[3]{#2}
\let\VANthebibliography\thebibliography
\def\thebibliography{\DeclareRobustCommand{\VAN}[3]{##3}\VANthebibliography}

\usepackage{graphicx}	
\usepackage{amsmath}	
\usepackage{hyphenat}
\usepackage{natbib}
\usepackage{booktabs,caption}
\usepackage{pdflscape}
\usepackage{hyperref}
\usepackage[export]{adjustbox}

\title[Globular Cluster Candidates in the M81 group]{New Globular Cluster Candidates in the M81 group}
\author[Pan et al.]{Jiaming Pan,$^{1}$\thanks{E-mail: jiamingp@umich.edu}
    Eric F.\ Bell,$^{1}$
    Adam Smercina,$^{1,2}$
    Paul Price,$^{3}$
    Colin T.\ Slater,$^{2}$
    Jeremy Bailin,$^{4}$
    \newauthor
    Roelof S.\ de Jong,$^{5}$
    Richard D'Souza,$^{6}$
    In Sung Jang,$^{7}$ and 
    Antonela Monachesi$^{8,9}$
\\ 
$^{1}$Department of Astronomy, University of Michigan, 1085 S.\ University Ave., Ann Arbor, 48109 MI, USA\\
$^{2}$Department of Astronomy, University of Washington, Box 351580, Seattle, 98195-1580 WA, USA\\
$^{3}$Department of Astrophysical Sciences, Princeton University, Princeton, 08544 NJ, USA\\
$^{4}$Department of Physics and Astronomy, University of Alabama, Box 870324, Tuscaloosa, AL 35487-0324, USA\\
$^{5}$Leibniz-Institut für Astrophysik Potsdam (AIP), An der Sternwarte 16, 14482 Potsdam, Germany\\
$^{6}$Vatican Observatory, Specola Vaticana, V-00120, Vatican City\\
$^{7}$Dept.\ of Astronomy \& Astrophysics, Univ.\ Chicago, 5640 S.\ Ellis Avenue, Chicago, IL 60637, USA\\
$^{8}$Instituto de Investigación Multidisciplinar en Ciencia y Tecnología, Universidad de La Serena, Raúl Bitrán 1305, La Serena, Chile\\
$^{9}$Departamento de Astronomía, Universidad de La Serena, Av. Juan Cisternas 1200 N, La Serena, Chile\\
}

\date{Accepted XXX. Received YYY; in original form ZZZ}

\pubyear{2022}

\begin{document}
\label{firstpage}
\pagerange{\pageref{firstpage}--\pageref{lastpage}}
\maketitle

\begin{abstract}
The study of outer halo globular cluster (GC) populations can give insight into galaxy merging, globular cluster accretion and the origin of GCs.  We use archival Subaru Hyper Suprime-Cam (HSC) data in concert with space-based GALEX, IRAC and Gaia EDR3 data to select candidate Globular clusters (GCs) in the outer halo of the M81 group for confirmation and future study.  We use a small sample of previously-discovered GCs to tune our selection criteria, finding that bright already-known GCs in the M81 group have sizes that are typically slightly larger than the Subaru PSF in our fields.  In the optical bands, GCs appear to have colours that are only slightly different from stars.  The inclusion of archival IRAC data yields dramatic improvements in colour separation, as the long wavelength baseline aids somewhat in the separation from stars and clearly separates GCs from many compact background galaxies.  We show that some previously spectroscopically-identified GCs in the M81 group are instead foreground stars or background galaxies.  GCs close to M82 have radial velocities suggesting that they fell into the M81 group along with M82.  The overall M81 GC luminosity function is similar to the Milky Way and M31.  M81's outer halo GCs are similar to the Milky Way in their metallicities and numbers, and much less numerous than M31's more metal-rich outer halo GC population.  These properties reflect differences in the three galaxies' merger histories, highlighting the possibility of using outer halo GCs to trace merger history in larger samples of galaxies.  

\end{abstract}

\begin{keywords} 
globular clusters: general
 -- galaxies: individual: M81 -- galaxies: star clusters: general  
\end{keywords}


\section{Introduction}

In what kind of galaxies do globular clusters (GCs) form, and what can they tell us about the growth of galaxies? A great deal of progress has been made towards understanding the answer to this question. While many GCs are formed {\it in-situ} in the main body of galaxies during periods of intense star formation \citep[e.g.,][]{Li2014,Kruijssen2015}, it is now also clear that many clusters are accreted from satellite galaxies as those satellites tidally disrupt, forming an extended distribution of {\it ex situ} (or accreted) GCs. While such a picture emerged from the joint consideration of the spatial, chemical and kinematic properties of metal-poor and metal-rich GCs \citep[e.g.,][]{Forbes2018}, perhaps the most direct evidence for GC accretion is in the Local Group, where significant numbers of outer GCs are still associated with recognizable stellar streams in both the Milky Way (MW; e.g., Sagittarius' GC population; \citealt{Law2010,Massari2017}) and the Andromeda Galaxy (M31; e.g., \citealt{Veljanoski2014,Mackey2019}). Accordingly, one might expect outer GCs to be particularly sensitive to the accretion history of a galaxy; indeed, Andromeda's outer GC population is around six times more numerous and more metal-rich than the MW's \citep{Huxor2014,Wang2019}, reflecting Andromeda's much more active merger history (M31: \citealt{D'SouzaBell2018,Hammer2018,McConnachie2018} c.f., MW: \citealt{Helmi2018,Belokurov2018}). Consequently, while outer halo GCs are dramatically outnumbered by those at smaller radii, it is primarily by studying these outermost GCs that one can obtain particular insight into a galaxy's accreted GC population.  

While GCs are particularly numerous in massive, typically elliptical or lenticular galaxies, the GC systems of star-forming spiral galaxies are more difficult to study \citep{Harris2013} --- not only are GCs typically much less numerous, but clusters are projected onto a much more structured disk, making their recovery much more challenging. 
Because of their likely accretion origin, the study of spiral galaxies' outer halo GCs may aid in developing tools to characterize their merger histories \citep[e.g.,][]{Kruijssen2019} and dark matter properties \citep[e.g.][]{Hudson2014,Harris2017BH}. 
The most promising galaxies for such studies are nearby galaxies that have {\it independent} constraints on merger histories from their stellar halos \citep[e.g.,][]{Harmsen2017,Bell2017}, allowing direct correlation of merger histories with outer halo GC numbers and properties. These studies can be done only on galaxies that are so nearby that their outer GC populations are spread out over square degrees or more, dramatically worsening the impact of foreground star and background galaxy contamination of GC candidate catalogues \citep[e.g.,][]{Zinn2014,Hughes2021}.  In this work, we advance toward this broad goal by surveying for GCs in the outer halo of the M81 group.

The M81 group presents a particularly promising opportunity to gain insight into the relationship between outer halo GCs and mergers and accretions. Analyses of ground-based \citep{Okamoto2015,Smercina20} and HST data \citep{Barker2009,Monachesi2013,Harmsen2017} suggest that the M81 galaxy itself --- a Sab galaxy with stellar mass similar to the MW --- has had a quiet merger history, not dissimilar to the MW's. However, M81 is currently interacting with M82 and NGC 3077 and will merge with them in the next few Gyr, likely resulting in a system with a large, metal-rich stellar halo similar to M31's \citep{Smercina20}. In addition, the M81 group is close enough to survey for GCs with imaging \citep[e.g.,][]{Chandar04,Nantais11} and spectroscopy \citep{JeanP.Brodie1991,Nantaisa,Lee2020}.   


Searches for GCs in the M81 group, individual members of the M81 group, and in the main body of M81 itself have together identified hundreds of candidate GCs. On the group scale, joint photometric (for discovery) and spectroscopic (for follow-up) investigations are the primary mode of operation. 
\citet{JeanP.Brodie1991} found eight GCs in the M81 galaxy using the MMT. \citet{Perelmuter95} determined three of these to be GCs according to their study, and discovered another 22 GCs. \citet{schroder02} added 16 more GCs. \citet[hereafter NH10]{Nantaisa} spectroscopically confirmed 62 new GCs that used HST imaging for initial selection, and combined them with the samples drawn from previous studies. They provide a total list of 108 GCs. Later, \citet[hereafter DZ15]{Zinn2015} found five M81 bright group GC candidates using multi-wavelength photometric data. Furthermore, \citet{Jang12} found GC-1 and GC-2 in the remote halo of the M81 group using HST images.  \citet[hereafter L18]{Lee2020} presented a wide-field spectroscopic survey of the M81 group GC candidates using MMT/Hectospec. They catalogue 113 new candidate GCs in the M81 group drawn from a Canada-French-Hawaii Telescope MegaCam photometric catalogue. We will show later that this catalogue in particular is extremely contaminated with foreground stars, in great part because the metallicity ([Fe/H]\,$\sim -1$) and low velocities of faint MW halo stars is similar to the metallicity and velocity $v_{r,M81} = -34$\,km/s signatures expected of M81 GC candidates. Finally, a recent study used medium-band data to select 642 candidate GCs in the M81 group \citep{JPLUS}; again, with the scale of this catalogue contamination is highly likely and spectroscopic, multi-wavelength or high-resolution imaging follow-up investigation will be necessary. 

Work in the main bodies of M81 group satellites often relies on HST imaging to cope with the difficult challenges of dust and star formation characteristic of the M81 group's larger satellites. Four old GCs in M82 galaxy have been explored using spectroscopy; two from \citet{Saito05} and two from \citet{Konstantopoulos2009}. 
\cite{Lim2013} used HST images to identify star clusters in M82; 35 of them had colours red enough to be identified as candidate GCs.  \citet{Davidge2004} identified some candidate old GCs and young star clusters in the central parts of NGC 3077 based on near-infrared colours and brightness. Other M81 group members host a few GCs; for example, \cite{Georgiev2009b} found 5 GCs in the IKN dwarf spheroidal with HST imaging data, and KDG61 has a single known GC \citep{Sharina05}\footnote{\citet{Sharina05} also report cluster candidates in HoIX, also in our survey footprint; all of these clusters are substantially bluer than typical GCs and are young or intermediate age star clusters.}. 

In the main body of M81, great progress has been made using high-resolution HST imaging to directly resolve star clusters. \citet{Chandar01a} identified 114 compact star clusters using HST Wide Field Planetary Camera 2 (WFPC2) imaging. 59 of these are old GCs based on an analysis of their colours \citep{Chandar2001b}. \citet{Chandar04} extended this work, deriving a GC luminosity function, finding that it peaks at $M_{V}$ $\sim$ $-$7.5, similar to the early findings of \citet{P&R95} and the GC luminosity functions of the MW and M31. Most recently, \citet{Nantais11} (hereafter NH11) analyzed 419 GC candidates in M81 identified by Hubble Space Telescope's Advanced Camera for Surveys with BVI imaging. They found 136 good cluster candidates; combined with previous samples, this yields a set of 221 highly probable new GCs. The luminosity function of these GCs peaks at $M_{V}$ = $-$7.53 $\pm$ 0.15 mag, adjusting from their assumed distance modulus of 27.67 \citep{Freedman2001} to our adopted value of 27.79 \citep{Radburn2011}. Yet, these works are unable to survey large enough areas to strengthen the census of outer halo GCs in the M81 group --- precisely those GCs which promise to offer the most insight into GC accretion and its relationship with merger history. 


Therefore, in this paper we augment and vet the census of outer halo GCs in the M81 group using archival data, laying the important observational ground work for a more detailed investigation of GC accretion. We combine Subaru Hyper Suprime-Cam (HSC) survey data with imaging catalogues from longer wavelength Spitzer Infrared Array Camera (IRAC) data to identify candidate clusters (see \citealt{Cantiello2018} for a similar method for discovering new globular cluster candidates in NGC 253). For some candidates, supporting data are available from The Galaxy Evolution Explorer (GALEX), Gaia's early data release 3 (EDR3), and existing spectroscopy. We show that the combination of HSC's PSF quality and optical colour-colour constraints provides a strong starting point, to which longer wavelength information from IRAC increases the purity of the sample. Candidates' Subaru images are visually inspected to reject obvious galaxies from consideration. A total of 24 GC candidates are presented. Ten of these have confirming Gaia EDR3 information and spectroscopy and are our strongest candidates. We also identify a number of contaminants in previously published catalogues of candidate GCs in the M81 group. With these candidates in hand, we explore the kinematic, metallicity, and other physical properties of the GC system in the outer halo of the M81 group, finding evidence for coherent velocity structure in the M81 group GC system, with a number of clusters clearly being associated with the accretion of M82 into the M81 group. 

This paper is structured as follows. In section~\ref{sec:method} we describe the key datasets, and in section~\ref{sec:selection} we describe how we combine insights from these datasets to identify GC candidates. Section~\ref{sec:result} presents some analysis of GC candidate luminosities, metallicities and radial velocities, and presents the catalogue of GC candidates. In section~\ref{sec:discussion} we connect our results to considerations of the growth of the M81 group, and highlight the importance of multi-wavelength data in this kind of study. We summarize in section~\ref{sec:conclusion}. In what follows, we use AB magnitudes and adopt a distance to M81 of 3.6\,Mpc ($m-M = 27.79$; \citealt{Radburn2011}).  

\section{Datasets}
\label{sec:method}

Globular clusters distinguish themselves from stars in that they are more extended than a point source (requiring data with high angular resolution), and that their spectral energy distributions are broader than those of stars, as they are a mix of different star types (requiring multi-wavelength data). On the other hand, while GCs and high redshift galaxies are both compact but measurably extended, GCs distinguish themselves from high redshift galaxies in that they have relatively red optical--near-IR colours compared to low redshift stars or GCs. 
Consequently, in order to detect globular cluster candidates and distinguish them from foreground stars and background galaxies, we combine deep, high-resolution wide-field optical imaging with space-based near-infrared data to provide the necessary combination of resolution and wavelength coverage. In addition, to build confidence in the GC candidates, we also used less sensitive but high spatial resolution space-based data from Gaia and (again, less sensitive) near-ultraviolet imaging from GALEX. We describe these datasets here, before discussing how they are combined to isolate a set of globular cluster candidates in \S \ref{sec:selection}.

\subsection{Subaru HSC survey}
\label{sec:hscmethod}In order to distinguish stars from globular cluster candidates morphologically and provide accurate optical spectral energy distributions, we rely on deep wide-field imaging from the Subaru Telescope's Hyper Suprime Cam 
 \citep[HSC,][]{HSC12,HSC18}, taken through the Gemini-Subaru exchange program (PI: Bell, 2015A-0281). The images were taken in the “classical” observing mode over two nights 2015 March 26-27 in three passbands (g, r, and i). The survey covers two pointings exploring outer regions of the M81 group, centred approximately on NGC 3077 and M82, for a survey footprint area of 3.5\,deg$^2$ in three filters: g, r, and i, with exposure times ranging between 3300$-$5400\,s per filter per pointing. The average FWHM is around 0\farcs7--0\farcs8 across the entire survey, giving limiting 5$\sigma$ point source magnitudes of g $\sim$ 27, r $\sim$ 26.5, and i $\sim$ 26.2. The dataset is described in more depth in \citet{Smercina20}. 

These data were reduced with the HSC optical imaging pipeline \citep{Bosch2018}. The pipeline performs photometric and astrometric calibration using the Pan-STARRS1 catalogue \citep{Magnier2013} but reports the final magnitudes in the HSC natural system. Because we are interested in compact sources and wish to avoid contamination from diffuse light (from the outskirts of galaxies, foreground Milky Way cirrus, or dwarf galaxies), the photometric pipeline subtracts from the source flux a very local sky measurement, using the clipped mean of (deblended) pixels within an annulus that extends from 7 to 15 times the PSF sigma. Sources are detected in all three bands. We use the i-band detection to determine the reference position and structural parameters for forced photometry in the g, r, and i-band stacks. As we are interested in slightly extended objects, we adopt composite model (CModel) magnitudes for this work, which takes the best fit PSF-convolved exponential and de Vaucouleurs profiles and returns the linear combination of the two that best fit the image. Foreground galactic extinction is corrected for following \citet{SFD98}, adopting adjusted coefficients for translating $E(B-V)$ into $A_{g,r,i}$ following \citet{Schlafly2011}. 

\subsection{Space-based survey data}
The Subaru data, by itself, has an angular resolution that only marginally resolves globular clusters and has insufficient wavelength coverage to distinguish stars from clusters decisively. Data that allow a wider wavelength baseline with sufficient sensitivity are required. The best available combination of sensitivity with wavelength coverage is given by
InfraRed Array Camera (IRAC) coverage \citep{Fazio2004} of around 1.75 square degrees in the region of M81, M82, and NGC 3077 in the Spitzer Heritage Archive (SHA). Most of the observations that we use here were taken on 6 November 2003, and each of 23 positions was observed with six 12\,s frames \citep{Willner2004}. Each pointing provided simultaneous 5\farcm2 $\times$\ 5\farcm2 images in 4 bands (3.6 $\mu$m, 4.5 $\mu$m, 5.8 $\mu$m, 8.0 $\mu$m). We used the Spitzer Enhanced Imaging Products (SEIP) Source List for this work. This source list --- made with the final version of the IRAC pipeline --- weighs reliability more than completeness or areal coverage,  requiring $>10\sigma$ detection in at least one IRAC band, and a minimum of three images (in high dynamic range or mapping mode) contributing to the coverage of the pixels in an object to report a detection. Excessively highly-peaked objects (identified by comparing the peak magnitude and the 3.8 arcsec diameter aperture magnitude), highly asymmetric objects, and moving objects were rejected. This dataset does not cover the full field of view of the Subaru observations, but instead is confined to the central parts of the M81 group; the overlap between the IRAC and Subaru data is around 1 square degree. In this paper, we used the 3.6 $\mu$m band in concert with other surveys to select globular cluster candidates. See the explanatory supplement of IRAC for more details about the photometry\footnote{Spitzer Enhanced Imaging Products Explanatory Supplement (Suppl.), available from \url{https://irsa.ipac.caltech.edu/data/SPITZER/Enhanced/SEIP/docs/seip_explanatory_supplement_v3.pdf}}. We note that because of IRAC's larger PSF and lower S/N, GC candidates are sometimes be confused with nearby sources, contaminating the IRAC flux or leading to a sufficient positional offset that the Subaru GC candidate has no IRAC counterpart. Further discussion of this source of incompleteness and error will be in \S \ref{sec:incompelteness}.   

While we use these Subaru and IRAC data in concert to select globular cluster candidates, we also have made use of two less sensitive datasets to confirm and characterize the selections (for brighter cluster candidates) that we performed with the deeper Subaru and IRAC datasets.

We used the Gaia Early data release three, Gaia EDR3 \citep{Gaia2016b, Gaia2020a} --- an all-sky survey covering more than a billion point-like or extremely compact sources --- to confirm the extended nature of cluster candidates. 
Photometric information is reported in three passbands: G, BP, and RP, with photometry reported in EDR3 for objects to an approximate magnitude limit of G $\approx 21$ magnitude. 
Magnitudes in G-band are determined by fitting a stellar point spread function to the imaging data, whereas BP and RP are low-resolution spectrophotometric measurements determined from an aperture, which in the case of a slightly extended source (like a globular cluster) contains more light. The comparison of the G-band photometry with the BP and RP photometry,  therefore, offers a way to distinguish compact objects from stars. The point-spread function fit has been improved in Gaia EDR3, so the difference between extended and compact objects is amplified \citep{Riello2021}. In particular, following \citet{Voggel2020} and \citet{Hughes2021}, we use  BP/RP Excess Factor ($ \textrm{BP}_{\textrm{excess}}$) ---  where extended objects have a higher $ \textrm{BP}_{\textrm{excess}}$ due to the higher sum of flux in BP and RP compared with G-band flux --- and Astrometric Excess Noise (AEN) --- where high values, indicating a poor fit between the object and the expectations of a point source, can indicate that an object is partially resolved.

We have also used data from the Galaxy Evolution Explorer (GALEX; \citealt{Martin2005}). It used microchannel plate detectors to take images in the near-UV (NUV, $\lambda_{eff} \sim$232 nm) and far-UV (FUV, $\lambda_{eff} \sim$ 154 nm) bands
\citep{Morrissey2007}. M81 was observed by GALEX on 2003 December 8 for 3089 seconds across two orbits. The GALEX instrument has a circular 1$\degr$.2 diameter field of view. This allowed it to observe M81, M82 and NGC 3077 in a single pointing \citep{Perez2006}. In this paper, we made use of data in the NUV band, downloaded from the GALEX archive using MAST, to help characterize brighter globular cluster candidates identified using Subaru and IRAC. See technical documentation of GALEX for more details\footnote{“GALEX Technical Documentation”, available from \url{http://www.galex.caltech.edu/researcher/techdocs.html}}. 

\section{Selecting Globular cluster candidates}
\label{sec:selection}

Outer halo globular clusters are outnumbered dramatically by foreground stars and background galaxies with similar sizes and optical colours. Furthermore, the selections required to prune a sample down to very likely globular clusters depend on factors that vary from case to case --- e.g., the PSF of the imaging dataset determines if clusters will appear extended vs.\ point-like, or the most appropriate colour separations in a dataset's particular passbands. 

Consequently, it is useful to have a set of certain (or very likely) globular clusters in the dataset with which one can empirically derive the best set of selections in size, colour and magnitude spaces. The challenge is that many existing catalogues of globular cluster candidates contain either galaxies or foreground stars as contaminants. This introduces an inevitable level of circularity in cluster selection --- catalogues of cluster candidate are examined in multi-dimensional size, magnitude and colour space, and the clumping of slightly extended objects with cluster-like colours and magnitudes, and optical--near-IR colours only a little redder than stars (following e.g., Figs.\ 14 and 15 of \citealt{Munoz2014}) are identified as the strongest candidates. These strong candidates then define the multi-dimensional selection that we can use to then select a wider set of candidates (without any confirming spectroscopic or high-resolution data). We describe our cuts determined in this way in this section. 

We have a two-stage selection for M81 group GC candidates. Our first step is to select an initial (extremely contaminated) sample from Subaru HSC data. We then dramatically refine the selection using space-based IRAC data to cull objects with colours that are star-like or similar to galaxies at a higher redshift. A subset of these candidates are confirmed to be extended sources using Gaia EDR3 data and have spectroscopy from L18; we call this subset {\it strong} candidates (will be discussed further in section~\ref{sec:space}). While the HSC data does not resolve especially fainter GC candidates into stars, we have also visually inspected the {\it strong} candidates in HSC imaging to ensure that they do not look like obvious galaxies or stars (section~\ref{sec:image}).

\subsection{Initial selection}
\begin{figure*}
	\includegraphics[width=15cm]{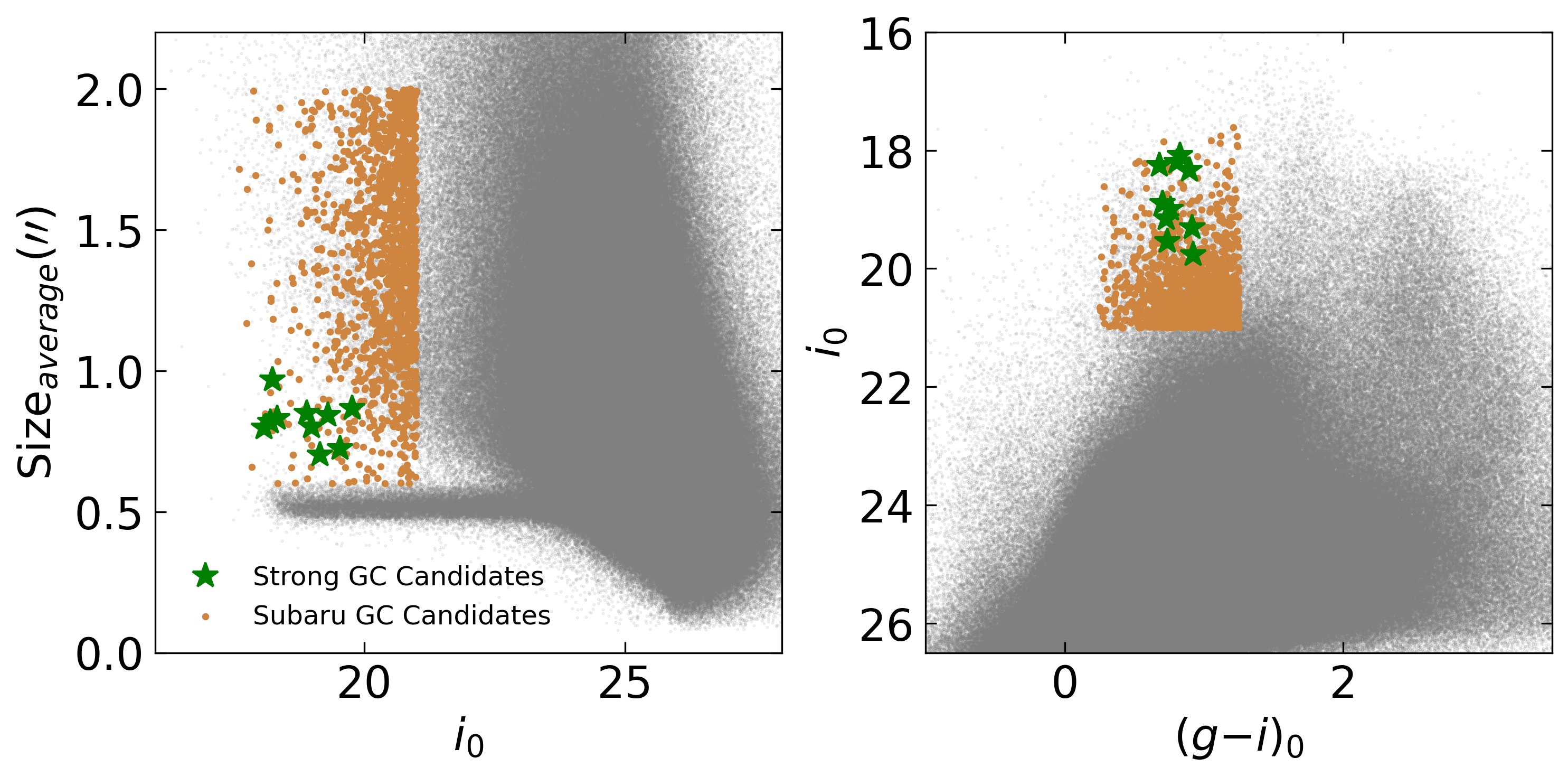}
    \caption{1458 Subaru GC candidates (orange dots) and 10 {\it strong} GC candidates (green stars) that act as guidance for the selection of the candidates. The grey points are all the sources from Subaru HSC data. The GC candidates have a size bigger than stars, bright magnitude, and lie in a constrained colour region of $0.25 < g-i < 1.25$. This is the starting sample for us to choose a small number of candidate GCs that are relatively clean. 
    }
    \label{fig:HSC GC candidates}
\end{figure*}

First, we select M81 halo GC candidates based on their sizes, magnitudes, and colours from Subaru HSC survey.
The GC candidates are chosen to have an average size between 0.6$\arcsec$ and 2.0$\arcsec$, $0.25 < g-i < 1.25$, and $17.5 < i < 21$, where sizes are the second-order adaptive moments calculated by the “shapeHSM” measurement HSC pipeline plugin \citep{Bosch2018,Hirata2003}. The PSF is just over 0.5$\arcsec$ in size (this measure is a Gaussian sigma rather than a FWHM, and so is smaller than the seeing FWHM of $\sim 0.7\arcsec$). Assuming that PSF size combines with intrinsic size in quadrature, then one expects intrinsic sizes between $\sim0.2 \arcsec$ and 2.0$\arcsec$, corresponding to $\sim 4-26$ pc in size. This selection comfortably selects all {\it strong} GC candidates, and selects a large number of other objects (Fig.~\ref{fig:HSC GC candidates}). We extend the size cut to 2$\arcsec$ to include possible extended GCs, as some HST-confirmed GCs in IKN have sizes between 1$\arcsec$ and 2$\arcsec$ in our Subaru data. The size cut removes stars which mostly have average sizes between 0.4$\arcsec$ and 0.6$\arcsec$. The colour cut in $g-i$ is in a range similar to the colour range of GC candidates in \citet{Nantaisbcan}. The magnitude limit is chosen to be $i = 21$, corresponding approximately to the limits fainter than which supporting information (e.g., IRAC fluxes, image morphologies) becomes too poor quality to differentiate between GC candidates and background galaxies.

Through this method, 1468 objects are selected, out of which 10 are {\it strong} GC candidates. We call the remaining 1458 objects Subaru GC candidates. This sample thus contains extended, bright objects lying within the typical colour range of GCs. There are still many contaminants and space-based surveys are utilized to further clean the sample, as discussed in the following sections.

\subsection{Removing galaxies and stars}
\label{sec:space} 
\begin{figure*}
	\includegraphics[width=15cm]{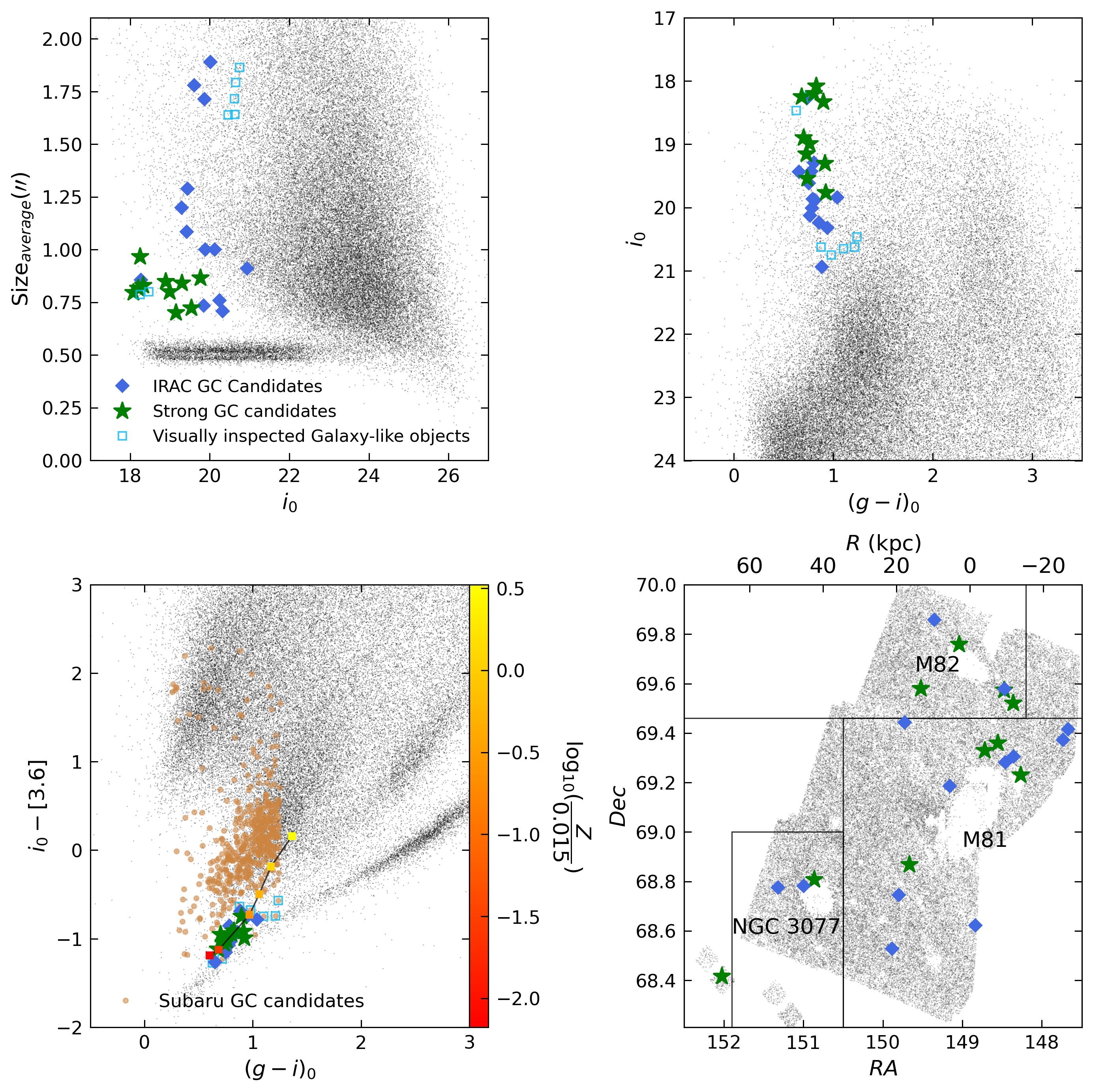}
    \caption{IRAC 3.6 $\mu$m data combined with Subaru to provide decisive insight into GC candidates. The first two panels show the size/magnitude, and colour magnitude trends of the candidates. The lower left-hand panel shows the $g-i/i$--[3.6 $\mu$m] colour-colour diagram, clearly illustrating the impact of IRAC 3.6 $\mu$m data --- it clearly distinguishes between the much redder background galaxies that dominate the `Subaru GC candidate' pool (orange points) and the much smaller set of objects with GC-like colours (green and shades of blue). Thirteen IRAC GCs candidates in M81 group were selected using the combination of Subaru HSC data, IRAC data, and visual inspection (blue diamonds), along with 10 {\it strong} GC candidates (green stars). The 7 blue squares were selected using the same colour criterion, but were visually classified as galaxies based on HSC images. The fourth panel shows the distribution on the sky of the GC candidates and {\it strong} GC candidates; the spatial locations of GC candidates tend to clump around each primary interacting group member (M81, M82, and NGC 3077). Rectangles denote the boundaries for the assignment of GCs to their host galaxies described in section \ref{sec:finalcl}.}
    \label{fig:IRAC GC candidates}
\end{figure*}

We use the IRAC 3.6 $\mu$m band to clean the Subaru GC candidate sample. First, we cross-match IRAC and Subaru HSC data to find sources with a separation smaller than 0.75$\arcsec$. We found 473 Subaru GC candidates that have an IRAC 3.6 $\mu$m band measurement; if instead maximum separations for IRAC matching were chosen to be any value between 0.5$\arcsec$ and 2$\arcsec$,  the set of Subaru GC candidates with IRAC matches would differ by 3\% or less. Then, we select the GCs based on the $i$ - [3.6 $\mu$m]/$g - i$ diagram, as shown in Fig.~\ref{fig:IRAC GC candidates}. This selection resembles the colour selection for GC candidates in \citet{Munoz2014} and \citet{Powalka2016}. We choose optical--IRAC colours in the following colour regions:
\begin{equation}
\begin{split}
(i-[3.6 \mu m]<-0.55) \& 
(i-[3.6 \mu m]>-1.4)\& \\
(i-[3.6 \mu m]>(-1.92+0.95\times(g-i)))\&\\
(i-[3.6 \mu m]<(-2.06+1.75\times(g-i))).\\
\label{iracselection}
\end{split}
\end{equation} 
Through these selections, we obtain 20 GC candidates out of the 473 Subaru GC candidates that have IRAC photometric measurements with colours similar to {\it strong} GC candidates. We classified 7 of these GC candidates as galaxies based on visual inspection which will be discussed in detail in section~\ref{sec:image}. The 13 GC candidates left are called IRAC GC candidates hereafter. IRAC GC candidates lie in a narrow area between galaxies and stars at the top and bottom in the colour-colour diagram, as shown in the lower three panels of Fig.~\ref{fig:IRAC GC candidates}, in analogy with the colour selections of \citet{Munoz2014} and \citet{Powalka2016}. In particular, stars lie in a tight stellar locus with relatively blue $i$-[3.6 $\mu$m] colours for a given $g-i$, while galaxies (with composite stellar populations and higher redshift) have much redder $i$-[3.6 $\mu$m] colour. 

For illustrative purposes, we add the sequence of 10\,Gyr old simple stellar population (SSP) models with metallicity ($Z$) of 0.0001, 0.0004, 0.004, 0.008, 0.02, and 0.05 in the colour-colour diagram panel of Fig.~\ref{fig:IRAC GC candidates}. We use the PEGASE \citep{Fioc1997,Fioc1999} stellar population model, converting SDSS magnitudes into HSC magnitudes following \citet{Akiyama2018}. SSPs with $Z$ lower than 0.004 have colours that agree very well with the optical--IRAC colours of the {\it strong} GC candidates. At higher metallicity, the colours of SSPs start to overlap strongly with galaxies.



While the joint Subaru--IRAC selection is sufficient to isolate a sample of candidate globular clusters, in order to isolate a set of GC candidates with the highest chances of being GCs, we make use of available Gaia, GALEX, and spectroscopic data. We discuss the further confirmation of these GC candidates using Gaia and GALEX data in Appendix \ref{appendix:GG}.

\subsection{Examination of GCs in the literature}
\label{sec:spectro}

It is useful and illuminating to cross-check the objects with HSC $g$, $i$, and IRAC 3.6$\mu$m data with objects with existing literature spectroscopy and/or are confirmed GCs on the basis of much higher angular resolution HST imaging \citep{Georgiev2009b,Jang12} --- we will call this the literature GC catalogue hereafter. There are 56 objects in the literature GC catalogue, some of them without $r$ band measurements in HSC data, owing to saturation of their central parts. In the literature GC catalogue, 51 objects have entries in L18's catalogue, 6 objects are from the NH10 catalogue (three overlap with L18), and 2 objects are HST-confirmed GCs from IKN. Many objects in L18 and NH10 samples were not identified by our work because they lie close to the centre of M81, M82 and NGC 3077, where the Subaru and IRAC data are too crowded to identify secure GC candidates. In IKN, IRAC catalogues have entries for only 2 out 5 HST-confirmed IKN GCs in 3.6 $\mu$m band; this is likely due to crowding and the influence of the very bright foreground star superimposed on IKN. 



\begin{figure*}
	\includegraphics[width=17cm]{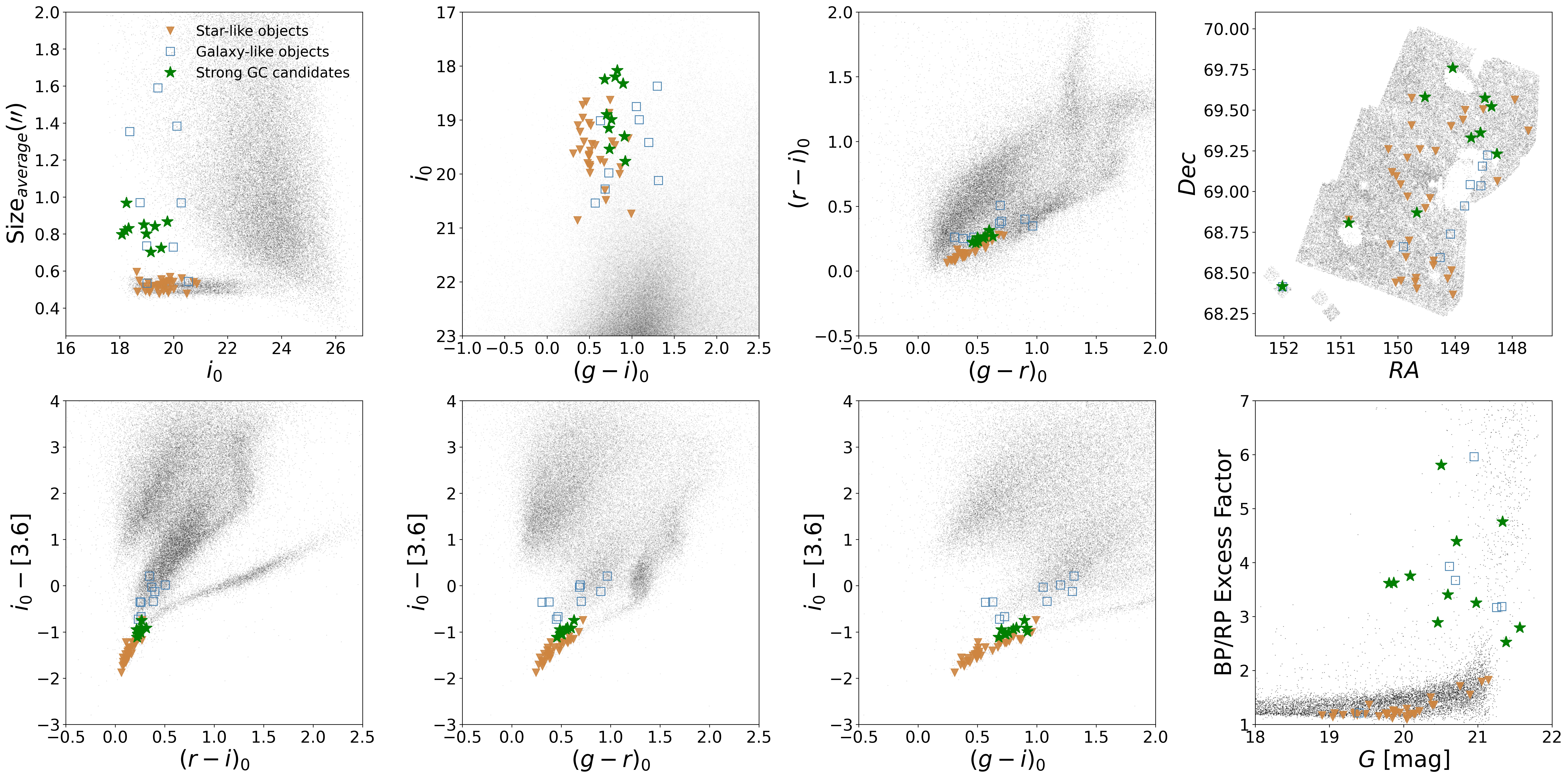} 
    \caption{Subaru, IRAC and Gaia allow a consistent and clear classification of literature GC candidates as likely stars, clusters or background galaxies.  The black datapoints show all objects from Subaru matched with IRAC. Brown triangles are literature objects which have $g-i$/$i-$[3.6$\mu$m] colours similar to stars; they also appear to be unresolved by both HSC and Gaia. Green stars show our strong GC candidates with cluster-like colours and extended sizes. Blue squares are literature GCs that have Subaru--IRAC colours similar to galaxies; in practice, most of these objects have measurably extended sizes. The first three panels present size, CMD, and colour--colour diagrams using just Subaru HSC data. The fourth panel is the map of the identified literature GCs. The lower panels include three colour--colour diagrams involving the IRAC 3.6 $\mu$m band and the last one shows Gaia EDR3 constraints for the literature GCs. The L18 sample makes up most of the literature GC sample, and is clearly highly contaminated by stars. The six objects from the NH10 catalogue have no stars; instead, the main contaminant is background galaxies. }
    \label{fig:IRAC confirmed GC candidates}
\end{figure*}

The IRAC data combined with Subaru HSC data of the aforementioned sample for strong GC candidates is shown in Fig.~\ref{fig:IRAC confirmed GC candidates}. We classified candidates to be cluster-like, galaxy-like, and star-like based on colour and size. Cluster-like literature objects are chosen to satisfy the same colour cuts in $i -$ [3.6 $\mu$m]$/g - i$ as before, and to have an average size > 0.6$\arcsec$ in Subaru HSC data. Star-like literature objects have colours bluer in $i-$[3.6 $\mu$m] in the $i -$ [3.6 $\mu$m]$/g - i$ diagram or average sizes $< 0.6\arcsec$ in Subaru HSC data and $i-[3.6 \mu m]<-0.55$. Galaxy-like literature objects have $i-$[3.6 $\mu$m] colours redder than strong and IRAC-selected GC candidates. While most clear GCs (based on HST imaging) are correctly classified as cluster-like candidates, one IKN HST-confirmed GC and three NH10 GCs that are clearly GCs based on HST imaging have galaxy-like colours owing to a relatively red $i -$ [3.6 $\mu$m] colour. This underlines the impact of source confusion on GC recognition --- while many GCs are well-measured, in crowded regions both Subaru and especially IRAC suffer source confusion, resulting in no measurements being reported or erroneously bright fluxes being recovered.  


Gaia EDR3 measurements of the literature GC catalogue are presented in the lower right-most panel of Fig.~\ref{fig:IRAC confirmed GC candidates}. Every object with optical-IRAC colours and HSC sizes like stars has a Gaia EDR3 $ \textrm{BP}_{\textrm{excess}}$ indicating that they are likely to be unresolved stars. Five objects identified with optical--IRAC colours indicating that they are likely to be galaxies have Gaia EDR3 $ \textrm{BP}_{\textrm{excess}}$ values indicating that they are likely to be extended sources; these five objects also appear to be extended ($> 0.6\arcsec$) in Subaru HSC data. One object with galaxy-like colours has a small $ \textrm{BP}_{\textrm{excess}}$ value ($< 2$) and small average size ($< 0.6\arcsec$) in Subaru HSC data --- it is unresolved, despite its unusual colours (perhaps owing to a red companion in an unresolved binary system). 

Taken together, it is clear that many literature `confirmed' GC candidates --- particularly those from L18 --- lie in either the star or the galaxy region of the colour--colour diagram, with sizes commensurate with their colour--colour classification (stars are compact in HSC and Gaia EDR3; galaxies being more extended). While overly-red candidates in $i-[3.6\mu m]$ are sometimes GCs with confused IRAC photometry (see Sections \ref{sec:incompelteness} and \ref{sec:besteffort}), some objects are indeed galaxies. In practice, stars dominate L18’s sample. 
This indicates that spectroscopic classification of GC candidates into stars, galaxies and background galaxies is disappointingly challenging; in particular, M81's low radial velocity and the GC's low metallicities appear similar enough to stars that L18 mistakenly classified many stars as GC candidates.  

This contamination motivated our iterative approach to sample selection --- we needed as large a sample as possible of real GCs to tune selection criteria, but samples were (often dramatically) contaminated. By combining the claims from the literature with insights from Subaru, IRAC, Gaia and GALEX that were not used by these authors, we were able to isolate a very secure sample of ten strong GC candidates. These ten objects resemble star clusters in terms of size, colour, and magnitude using Subaru HSC, IRAC, and Gaia EDR3 data. In concert, we were able to use these datasets to shed light on contamination in previous samples of GC candidates.  

This contamination is particularly devastating at a large galactocentric radius. However, we note that IRAC photometry sometimes confused GCs with galaxies as mentioned above. We incorporated HST images of some identified contamination for further confirmation. If IRAC colour classification disagrees with it, we use HST images classification instead (further discussion on section \ref{sec:incompelteness}). If we consider only the ten {\it strong} candidates to be true GCs, then identified objects in literature GC catalogue ($N = 56$) is 18\% true GCs, and 61\% foreground stars (further visual inspection of these foreground stars is discussed in section \S\ref{sec:image}). If instead we consider cluster-like objects (with IRAC colours or HST images suggesting that they are GCs, but without Gaia EDR3 confirmation) to be true GCs, then the literature sample has 34\% true GCs. Either way, the contamination of the literature GC sample is between 66\% and 82\%, requiring an iterative and information-rich approach to GC candidate selection.   

A few strong GC candidates have parallaxes and proper motions that look, at first blush, to be inconsistent with zero to within their reported error bars in Gaia's EDR3. They have magnitudes close to the limit of Gaia EDR3 with 19.5 to 21 in G band magnitude. As a check of whether these measurements should be taken particularly seriously, we checked Gaia EDR3 parallax and proper motion of galaxy-like objects with large $i-$[3.6 $\mu$m] colours (where these objects are so distant as to actually have zero parallax and proper motions). We found that most of these clear galaxies also have reported Gaia EDR3 parallaxes that appear to be inconsistent with zero to within the reported error bars, and are comparable to the measures of the {\it strong} GC candidates. Therefore, the positive parallax and proper motion estimates for {\it strong} GC candidates in Gaia EDR3 are likely to be incorrect; it may be that the slightly extended sizes of these objects cause erroneous measurements to be reported.  


\subsection{Visual inspection of HSC imaging}
\label{sec:image}

While the resolution of HSC imaging does not permit unambiguous identification of globular clusters from imaging, the imaging is of sufficient quality to permit many candidates to be classified on their degree of resolution into individual stars in their outskirts. We do not use this as a necessary requirement for inclusion in the set of GC candidates or strong candidates; rather, this is a supplementary piece of information that helps to characterize the candidates. 

We use both the raw image data and a residual image for classification. We used the \texttt{imcascade} multi-gaussian expansion code from \citet{imcascade} to generate a circularly-symmetric model that fits the inner parts of cluster candidates reasonably well, with some small residuals reflecting the transitions between different Gaussians in the expansion. Despite these small residuals, subtraction of this fit substantially improves our sensitivity to faint partially-resolved stars in clusters, improving the fidelity of the classifications. 

In practice, objects that appear from their size and multi-wavelength data to be stars (e.g., from the literature GC catalogues; section~\ref{sec:spectro}) indeed appear unresolved and free of structure. Objects with sizes larger than stars in the galaxy part of the colour-colour diagram indeed are extended and are either smooth or have structure (such as tidal tails, spiral arms), suggesting a galaxy rather than a collection of partially-resolved point sources. Clusters have clear contributions from partially-resolved point sources, particularly so for the brighter, larger clusters. A few objects are resolved but are unclear in their classification and could be either galaxies or clusters; we classify these as ambiguous. We present our classifications as an aid to those using this catalogue in Table \ref{tab:GCcatalog} and the HSC images and residual images are in the appendix \ref{appendix:B}, where Fig.\ \ref{fig:GC image} are strong GC candidates (one strong GC has saturated imaging in one passband and so is not classified), and Fig.\ \ref{fig:GC image IRAC} are IRAC GC candidates. Seven objects that satisfied Subaru--IRAC size and colour-colour cuts that nonetheless clearly appear to be galaxies are dropped from further consideration; we present these in Table \ref{tab:GCgalaxy} and Fig.\ \ref{fig:Galaxyimage IRAC}.

\section{Results}
\label{sec:result}

In the following sections, we present and explore our catalogue of 24 M81 group outer halo GCs: 10 strong candidates, 13 IRAC GC candidates, and GC-2 (included despite its saturated HSC data as it is a HST-confirmed outer halo GC). We start with a discussion of M81 group GC radial velocities --- M82 globular clusters clearly stand out from the rest and motivate a tentative assignment of candidate GCs to M81, M82 or NGC 3077. Given the maturity of the published measurements of M81's GC population, we compile what we consider to be a trustworthy catalogue of M81 candidates and a supplementary list of literature GCs that are likely to be contaminants; M82's and NGC 3077's cluster populations are less well-studied, and we do not attempt to compile a catalogue of their GCs beyond our own. We then use these catalogues to explore radial trends in luminosity function and metallicity in the M81 group.

\subsection{Radial velocity substructure in the M81 outer halo GC population}
\label{sec:radial}
\begin{figure*}
	\includegraphics[width=15cm]{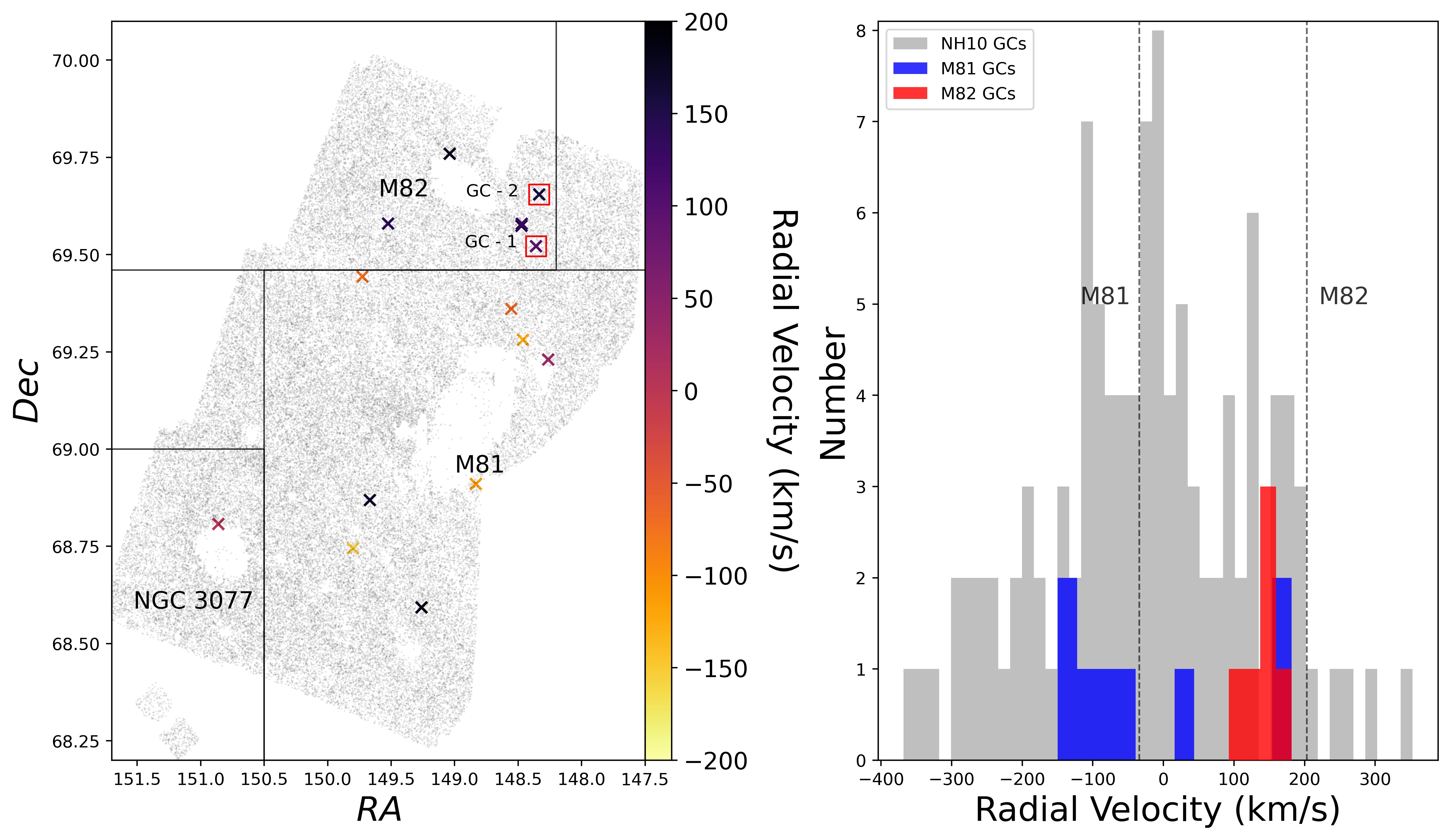}
    \caption{The radial velocities of candidate GCs with spectroscopy from L18. The left panel shows the map of the GC candidates, and they are colour coded to show the radial velocity in km/s. Rectangles in the left-hand panel are the same as those in Fig.~\ref{fig:IRAC GC candidates}. The right panel shows the radial velocity distribution of M81 (blue histogram) and M82 (red histogram) GCs. M81 GCs from NH10 are also plotted for comparison (grey histogram). Radial velocities of M81 and M82 galaxy are indicated by the two dotted lines. While M81's outer halo GCs have similar velocities to the rest of M81's GCs, M82 GCs show a significantly higher mean radial velocity than that of the M81 GCs, indicating that they were accreted along with M82 and have not yet kinematically mixed with the M81 outer halo GC population. GC-1 and GC-2 are highlighted with red squares.}
    \label{fig:radial velocity}
\end{figure*}

To motivate the analysis that follows, it is important to explore the radial velocities of the subset of our GCs that have spectroscopic observations from L18 (the overlap between our GC catalogue and the literature GC catalogue) in Fig.~\ref{fig:radial velocity}. 

The most obvious feature of Fig.~\ref{fig:radial velocity} is that the six GCs closest to M82 have a higher mean radial velocity (141.22 $\pm$ 12.10 km/s) than that of M81, or the mean velocity of M81's GCs from NH10 (mean $v_r = -23 \pm 4$\,km/s). The chance of drawing 6 GCs with a higher mean velocity than those near M82 from the NH10 velocity distribution is about 0.27\%. This implies the M82 halo GCs are likely to have belonged to M82 and have not yet kinematically mixed with the other M81 group GCs. 

In contrast, the remaining GCs of M81 show no evidence of kinematic structure. Their mean radial velocity is $v_r = -18.39 \pm11.37$\,km/s, similar to the mean radial velocity of NH10 GCs. Furthermore, there is no evidence for rotation in these GCs, consistent with the weak rotation in metal-poor GCs reported in NH10's M81 GC population.


\subsection{Our catalogue of Globular cluster candidates}
\label{sec:finalcl}
We present our catalogue of GC candidates in Table \ref{tab:GCcatalog}, along with tentative associations with members of the M81 group. Table \ref{tab:GCcatalog} contains 24 GC candidates, combining the 10 {\it strong} GC candidates (ID 1 - 10) with our 13 IRAC GC candidates (ID 11 - 23) that have cluster or ambiguous/unclear morphology and GC-2 (ID 24). Nine of these candidates appear in no other existing catalogue of M81 group GCs. We consider these GC candidates --- particularly the strong candidates GCs --- to be likely to be GCs, but acknowledge that further confirmation is very desirable, ideally very high-resolution imaging to clearly resolve the outskirts of clusters into stars. We also include a list of objects selected by our Subaru--IRAC size and colour--colour cuts that appear to be clear background galaxies in Table \ref{tab:GCgalaxy}, and a list of objects in literature GC catalogues that Subaru--IRAC size and colour--colour analyses show are stars or galaxies in Table \ref{tab:GCcanta}. Table \ref{tab:GCgalaxy} and Table \ref{tab:GCcanta} columns have the same meanings with those in Table \ref{tab:GCcatalog}.

Based on the radial velocity substructure discussed in section \ref{sec:radial}, we have devised spatial selections for assigning M81 GCs to a likely primary galaxy. We assign GCs with RA$<150.5^{\circ}$ and Dec$<$69.46$^{\circ}$ to have M81 as their likely host, and GC candidates with RA$>148.2^{\circ}$ and Dec$>$69.46$^{\circ}$ to have M82 as their likely host. 
On this basis, we assign 13 M81 group GCs to M81 as their primary (9 IRAC GCs + 4 Strong GCs), all of which have projected radii greater than 8\,kpc; 7 of them are outside of 20\,kpc. We assign 7 GC candidates to M82 (2 IRAC GCs + 4 Strong GCs + GC-2), all of which are at projected radii greater than 8\,kpc from M82. We assign the three GC candidates with  $150.5^{\circ}<$RA$<151.9^{\circ}$ and Dec$<$69$^{\circ}$ to NGC 3077 as their primary (2 IRAC GCs + 1 Strong GCs); two are within 5\,kpc projected radius from NGC 3077, and one of them is outside of 10kpc. These boundaries are highlighted in Figs. \ref{fig:IRAC GC candidates} and Fig. \ref{fig:radial velocity}. Additionally, there is one IKN confirmed GC (ID 10). 


\subsection{Incompleteness caused by confusion in the IRAC data}
\label{sec:incompelteness}

Owing to IRAC's larger PSF, the IRAC data suffers particularly from crowding and confusion, which can lead to incompleteness, both through object positions being offset too far from the optical detection to register as a match to the Subaru GC candidate, and because IRAC fluxes are recovered too bright and a GC is classified as a galaxy owing to its red optical--near-IR colour. We explore this issue in this section. 

A first test is to check for high-resolution imaging for colour-selected stars and galaxies. We focus on HST, as its 0.09" image quality can resolve GCs into stars at this distance, offering a clear and decisive test of colour-based classifications. Eight objects out of 41 potential galaxies and stars in the literature GC catalogue (section \ref{sec:spectro}) were found to have archival HST imaging in the Hubble Legacy Archive\footnote{Based on observations made with the NASA/ESA Hubble Space Telescope, and obtained from the Hubble Legacy Archive, which is a collaboration between the Space Telescope Science Institute (STScI/NASA), the Space Telescope European Coordinating Facility (ST-ECF/ESA) and the Canadian Astronomy Data Centre (CADC/NRC/CSA).}. Two observed star-like objects are clear stars in HST images; two are clear galaxies (classified as such based on colours). Four out of these eight sources --- all classified on the basis of optical--IRAC colours as galaxies --- appear to be clusters based on visual inspection. Three of these four wrongly classified objects lie close to the disk of the M81 galaxy, and the IRAC catalogue flagged them as being in a bright region where the IRAC photometry may be suspect. The fourth source was in a crowded region of the halo, but did not have a flag indicating any problem with the IRAC photometry.  

While HST imaging is informative, the fractions of colour-misclassified galaxies should not be taken to be representative of the whole catalogue. Note that HST imaging is concentrated towards the main bodies of galaxies, probing areas that are more likely to suffer from contamination. To better estimate the fraction of misclassified galaxies, we used the \texttt{fluxflag} provided as part of the IRAC photometry, which accounts for contaminating flux from nearby objects. Considering the entire catalogue of matches between Subaru and IRAC, only  4.4\% were flagged as near bright or extended sources. This suggests that at least 4.4\% of true clusters may be misclassified as galaxies, but likely more as GCs should be clustered more towards the bright parts of galaxies than foreground stars or background galaxies.


It is also important to consider sources whose IRAC positions have been affected by confusion. Recalling that the IRAC survey coverage is smaller than Subaru's coverage, there are 77 sources out of 583 Subaru candidates within the IRAC field of view missed by IRAC data. We visually inspected images of those 77 Subaru candidates; all of them appear to be close pairs of objects (typically galaxies) that are recovered as a single IRAC object, and none of them appears to be a clear GC candidate, but we cannot rule out the loss of fainter ambiguous GC candidates. Combining this with the incompleteness estimated using fluxflag, the incompleteness caused by IRAC crowding is at least 18\%. 

We note that the classification of stars should be correct. Not only do all stars appear to be point sources in the Subaru data based on sizes and visual inspection of images, but also all Subaru and colour-classified stars with Gaia EDR3 or HST data appear to be unresolved point sources. This underlines the utility of both the excellent 0.7'' seeing in the parent Subaru dataset being good enough to partially resolve halo GCs at the distance of 3.6 Mpc, but also reflects that IRAC source confusion does not cause sources to have fainter than expected IRAC fluxes. Consequently, the combination of Subaru and IRAC are sufficient to reliably classify objects as stars. 

To conclude --- while it is possible to reliably classify objects as stars, confusion of the IRAC images will likely cause the loss of some bone fide GCs either through affecting the IRAC positions or fluxes. This loss is likely to be at least 18\%, but should not be dramatically higher for the outer halo GCs that are the focus of this paper, as these are in the uncrowded regions of the Subaru and IRAC imaging. 



\subsection{A best-effort compilation of M81 GC candidates} \label{sec:besteffort}

In addition to the 13 GC $R_p>8$\,kpc candidates that we assign to M81 as their primary galaxy (Table \ref{tab:GCcatalog}), our work also allows us to update existing catalogues of M81 GC candidates. 

A first obvious conclusion is that the catalogue of L18 must be used with particular caution, as most of their objects turn out to be foreground stars. NH10, in contrast, is a reasonable but not perfect basis for a list of reliable GC candidates in M81. Of the six NH10 candidates that we have in our catalogue, we choose to exclude two objects (one-third of those for which we have information) as likely galaxies (their 70319 and 50357), yielding a total of 106 objects. Note that only NH10 ID 50357 appears as a clear galaxy based on HST images. The classification of NHID 70319 is mainly based on colours and sizes in the Subaru HSC and IRAC data, which could be affected by crowding/confusion in the IRAC photometry. Its Subaru HSC data is marginally resolved, but NHID 70319 visually appears to be more likely to be a galaxy than a star cluster. We will assume that NHID 70319 is a galaxy for the purposes of this study; however, further confirmation is necessary.
We also add two M81 GC candidates M81-C1 and M81-C2 from DZ15 catalogue (these two candidates were outside of our imaging footprint).

We also have been careful to explore as much as is possible {\it our} misclassification of clusters as galaxies (on the basis of optical--IRAC colours) owing to IRAC crowding. NHID 464 and 34 were classified as galaxies based on photometric data from Subaru and IRAC, but are clear clusters in HST images. They were, along with two other wrongly classified clusters L18 ID 102 and 124, counted as reliable M81 GCs. 



These 121 candidates are, we consider, one of the most reliable samples of GC candidates in the M81 galaxy. We include this list of candidates in Appendix \ref{appendix:A}. Note that we have not included in this the extra GC candidates from \citet{Nantais11}. Their sample extends to significantly fainter limits, increasing completeness but also risking increased contamination (suggested by \citealt{Nantais11} to be $\sim 8 \%$ for bright objects and much higher at fainter limits, largely owing to background galaxies). 

\subsection{GC candidates in M82 and NGC 3077}
\label{sec:M82NGC3077}
The GC populations of M82 and NGC 3077 have received considerably less attention than that of M81 and are much more difficult to reliably study owing to the active star formation and dust in both galaxies. Consequently, we do not attempt the exercise of recommending a `best effort' GC catalogue for these two galaxies. 

In M82, \cite{Saito05} reported two globular cluster candidates within a few kpc of M82's centre with spectroscopic data;  \cite{Konstantopoulos2009} reported two more potential old globular clusters projected onto the inner parts of M82 and their ages were estimated using spectroscopic data. 
\cite{Lim2013} found 1105 star clusters in M82 based on HST images, and most of them are estimated to be intermediate-age star clusters with ages from 100 Myr to 1 Gyr and mostly located in the disk region. In their sample, 35 star clusters are in the inner parts of M82's halo and had colours red enough to be old globular clusters. All but one of their clusters are too close to M82 to be covered by our study; our ID 6 overlaps with one of the clusters from \citet{Lim2013}. In our catalogue, ID 5, 6, 7, 8, 20, 21, and 24 belong to M82. Strong candidate ID 5 was listed as GC candidate M81-C3 in the DZ15 catalogue. All of our candidates have $r_{proj,M82} > 5$\,kpc. Given the red colours and large projected distances of our GC candidate sample, it is very likely that our GC candidates are globular clusters in M82's halo, adding 6 very likely GCs to M82's GC population. We note that ID 21 has never been in any previous catalogue of M82 GCs.

We found three GC candidates around NGC 3077, and to date, there are few studies of GC systems in NGC 3077. \citet{Davidge2004} identified some candidate old GCs and young star clusters in the central parts of NGC 3077 based on near-infrared colours and brightness (listed in their Table 1). In our sample, NGC 3077 GC candidates include IDs 9, 22, and 23. GC candidate ID 23 is identified here for the first time. GC candidate ID 22 has also been identified in the DZ15 catalogue, as their M81-C4 in their Table 4\footnote{We note that DZ15's M81-C5 lies $\sim300$ projected kpc from M81, and we do not include it in any of the catalogues in this study.}.

\subsection{Globular cluster luminosity function}
\label{sec:GCLF}
\begin{figure}
	\includegraphics[width=\columnwidth]{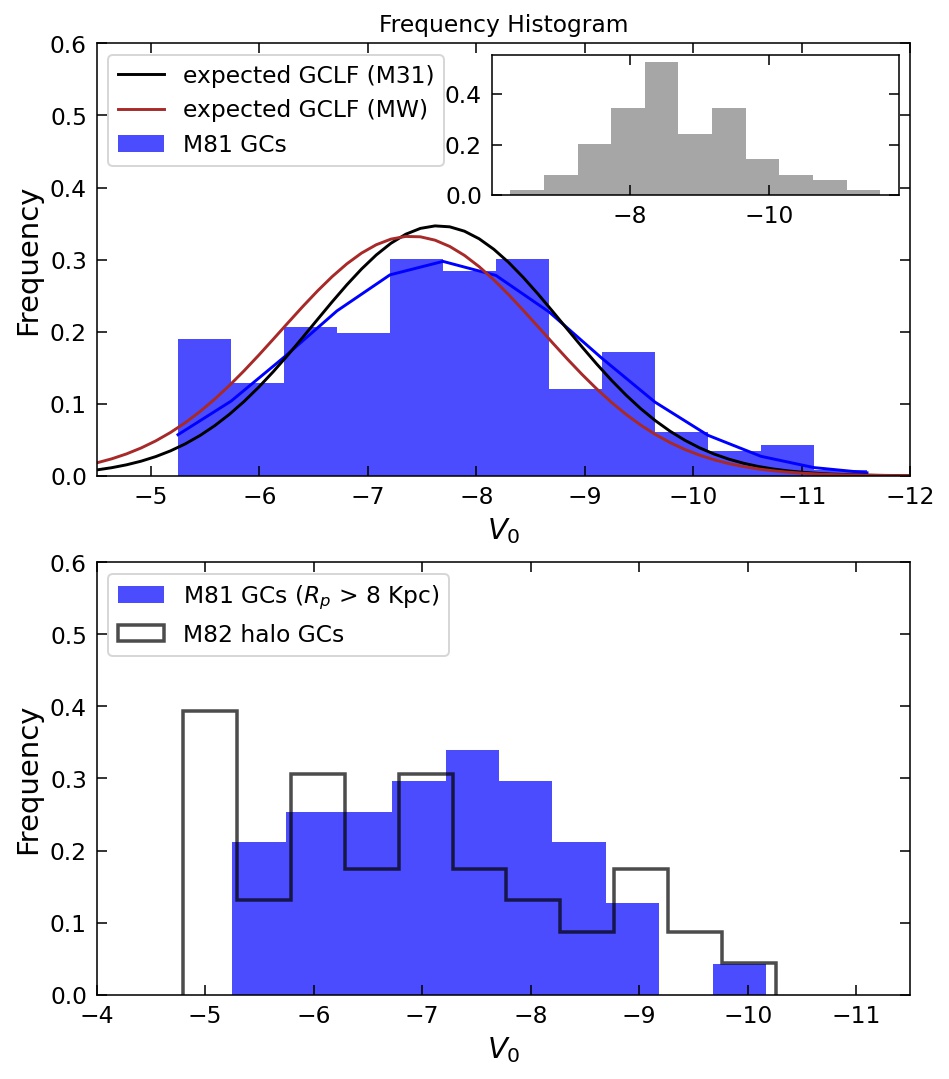}
    \caption{The $V$-band globular cluster luminosity function of M81 GCs including our catalogue and probable GCs in the NH11 catalogue (upper panel). The luminosity function of sample in Appendix A is shown in the upper right corner of the upper panel, which missed many faint objects in the NH11 catalogue. The blue line is the best-fitted Gaussian distribution for the M81 GC luminosity function. The brown line and black line show Gaussian fits to the Milky Way GCLF and M31 GCLF respectively. 71\% of probable NH11 probable GCs are at small projected radii $R_p<8$\,kpc . The lower panel compares the M81 GCLF for GCs with $R_p$ greater than 8 Kpc with the M82  GC luminosity function.    
}
    \label{fig:GCLF}
\end{figure}

We present the luminosity function of all M81 GCs in Fig.~\ref{fig:GCLF}, comparing it with the subset of M81 GCs with $R_p$ > 8 Kpc and M82 GCs; bin sizes of 0.5\,mag were adopted. For this purpose, we expand the combined sample (Appendix A) to include probable GCs from \citet{Nantais11}; NH10's sample has few fainter GCs, biasing the luminosity function significantly towards brighter magnitudes. This sample thus contains 238 GCs of M81 galaxy in total. The luminosity function of the sample in Appendix A is shown in the upper right corner of the upper panel of the Fig.~\ref{fig:GCLF}. The V magnitude of our M81 GC candidates is estimated from Subaru $g$- and $i$-band magnitudes: 
\begin{equation}
  V = g + 0.067 - 0.439\times(g-i), 
  \label{eq:coloreq}
\end{equation}
and we correct our $m_V$ estimates for foreground extinction following \citet{SFD98} using updated extinction coefficients from \citet{Schlafly2011}, as our candidates are so isolated as to make extinction from M81 itself unimportant. NH11's tabulated magnitudes $m_V$ do not correct for extinction (galactic foreground + that from M81 itself); we adopt a uniform $A_{V}$ of 0.73 for their sample --- their mean V-band extinction  \citep{Nantais11}. For the purpose of estimating absolute magnitude, we adopt a distance modulus of M81 group of 27.79  \citep{Radburn2011}.  
Note that we are unable to correct the luminosity function of our sample for completeness, as our selection is sufficiently dependent on multiple data sources (Subaru, IRAC archival data, Gaia) to make simulating globular cluster recovery impractical. 

Fig.~\ref{fig:GCLF} shows also a Gaussian fit to the M81 GC luminosity function; M81 has a GCLF peak of $M_{V}= -7.69 \pm 0.09$ and a standard deviation of $1.34$. \footnote{If instead we only consider the sample in Appendix A, then the GCLF peaks at $M_{V}= -8.68 \pm 0.10$ with a standard deviation of $0.98$ mag.} The median of M81 GCLF is $M_{V} = -7.69$. The GCLF peak and standard deviations do not depend on binning, and the Gaussian model is an appropriate fit (a Kolmogorov-Smirnov test comparing the M81 GCLF with its Gaussian model fit gave a $p=0.54$ of the observed LF being drawn from the fitted distribution). This is similar but not identical to NH11's own analysis --- \citep{Nantais11} found a turnover magnitude of $M_{V} = -7.53 \pm 0.15$, translating their result to our adopted distance modulus (they had assumed a $m-M=27.67$; \citealt{Freedman2001}) --- or a re-analysis of that dataset by \citet{Lomel2022}, who found $M_{V} = -7.52 \pm 0.16$. 
Restricting ourselves to just NH11's clusters gives a peak $M_{V} = -7.68$; we speculate that the small difference (within our and their quoted joint uncertainties) stems from our adoption of a uniform $A_V$ estimate and the difference between their `turnover' magnitude and the peak of our Gaussian fit. M81's GCLF peak magnitude is similar to the LF peak of metal-poor clusters in many galaxies $M_{V} = -7.66$  \citep{Rejkuba2012}, and the MW's GCLF peak $M_{V} = -7.4 (\sigma = 1.15)$ \citep{harris2001} and that of M31, at $M_{V} = -7.65 \pm 0.65 (\sigma = 1.2)$ \citep{Barmby2001}, assumes a M31 distance modulus of 24.49 \citep{Joshi2003}. Both the MW and M31 GCLF are indicated in Fig.~\ref{fig:GCLF} with peak and standard deviation from the literature. 

In the lower panel of Fig.~\ref{fig:GCLF}, we compare the luminosity function of M81 GCs with $R_p > 8$\,kpc with M82 GCs. The M82 GCs consists of seven GCs in our catalogue, 2 GCs from \cite{Saito05}, 2 GCs from \cite{Konstantopoulos2009}, and 35 halo GCs from \cite{Lim2013}. The $V$-band absolute magnitude of GC-2 was taken from \cite{Jang12}. The mean of the M81 at $R_p > 8$\,kpc is $M_{V} = -7.20 \pm 0.16$ and the median is $M_{V} = -7.30$. The M82 halo GCs have a mean $M_{V} = -6.84 \pm 0.17$ and a median of $M_{V} = -6.76$. We did a two-sample KS test of these two luminosity functions, and the obtained p value is 0.24. This means that M81 GCLF ($R_p > 8$\,kpc) and M82 halo GCLF are consistent with being drawn from the same distribution. 


\subsection{Metallicity}
    \label{sec:metal}
\begin{figure*}
	\includegraphics[width=17.5 cm]{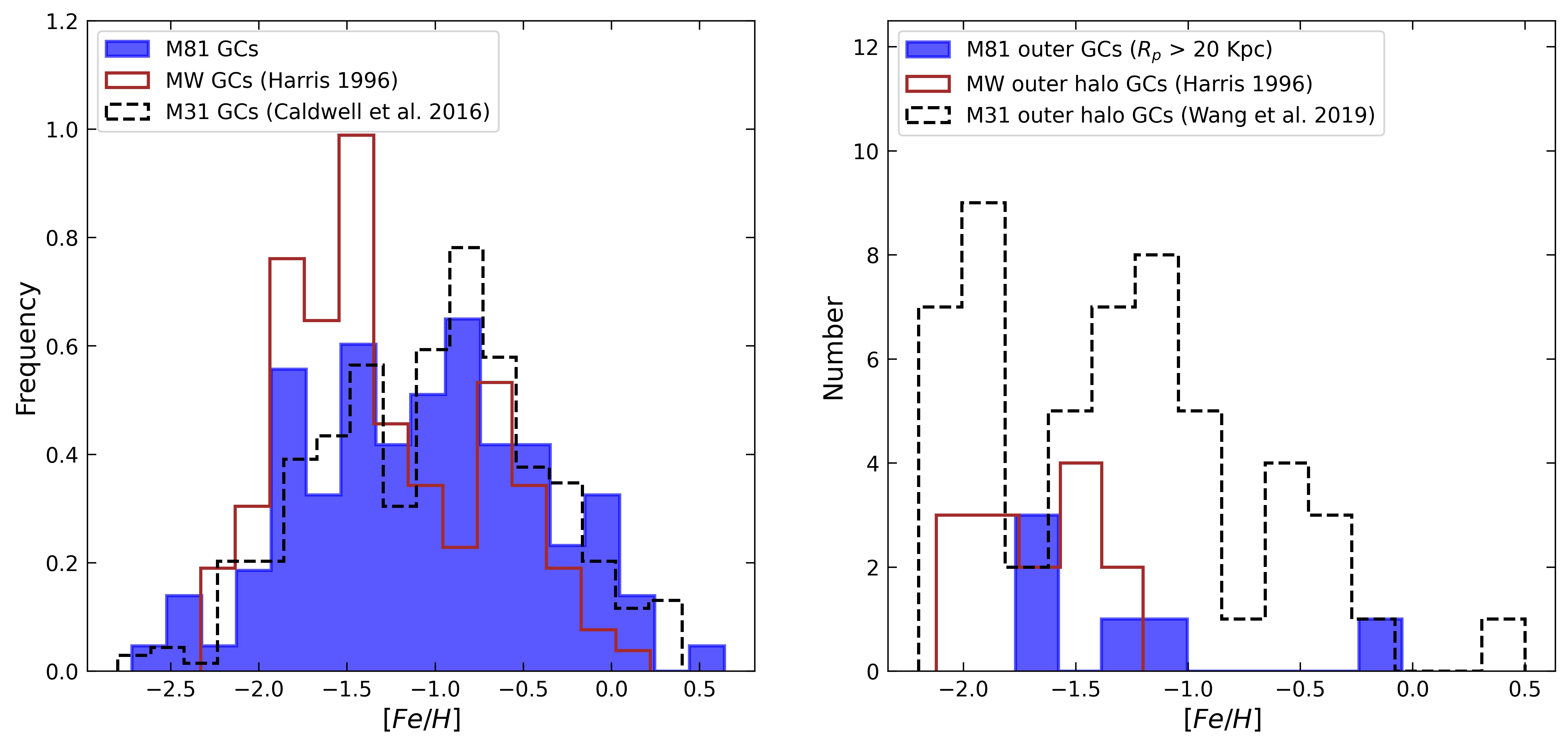}
    \caption{The left panel shows the histogram of metallicity of M81 GCs with spectroscopy from L18 and NH10. MW and M31 GC metallicities were plotted for comparison. Blue, brown, and black distributions show the metallicity of M81 GCs, MW GCs, and M31 GCs respectively. The right panel shows the histogram of M81 outer GC at $R_p$ > 20 kpc, along with M31 and galactic outer halo GCs at $R_p$ > 20 kpc for comparison.}
    \label{fig:meta}
\end{figure*}

Fig.~\ref{fig:meta}'s left panel shows the distribution of the metallicities [Fe/H] of the subset of M81 GCs with metallicity estimates from L18 and NH10.
Five NH10 GCs were excluded due to large metallicity uncertainties; potential galaxies in NH10 were also excluded. In total, there are 109 M81 GCs with useful metallicity estimates, nine of which are from our catalogue. MW and M31 GC metallicities are plotted for comparison \citep{Harris1996,caldwell2016}; we adopt the compilation of \citet{caldwell2016} for M31 for this comparison as it contains the largest existing set of spectroscopic M31 GC metallicities. The mean metallicities are $-1.08 \pm 0.06$, $-1.30 \pm 0.05$, and $-1.02 \pm 0.03 $ respectively for M81, MW, and M31. The difference between M81's and the MW's rather metal-poorer GC metallicity distribution is significant: the chance of drawing both distributions from a common parent distribution via a two-sample KS test is only 0.7\%. There is little difference between M81's and M31's distributions, with a 24.1\% chance of drawing them both from a common parent. We also note that there is no obvious bimodality in the metallicity distribution of the M81 galaxy. A similar conclusion has also been drawn in the NH10 catalogue \citep{Nantaisa}. However, both MW and M31 GC systems are widely thought to show bimodal metallicity distributions \citep{Barmby2000A}.

The mean metallicity of the 85 M81 GCs with $R_{p}<8$\,kpc is <[Fe/H]> = $-1.02 \pm 0.06$. The mean metallicity of 24 outer GCs in M81 at $R_{p}>8$\, is $-1.30 \pm 0.19$. The mean metallicity of the outer GCs in M81 galaxy is thus lower than that of M81 GCs with $R_{p,M81} < 8$\,kpc; M81's GCs show a metallicity gradient, in the sense that the typical GC metallicity is lower at a larger radius. This confirms the claim of NH10 of a metallicity gradient in M81's GC system (e.g., their Fig.\ 10). A metallicity gradient was not unexpected; both the MW and M31 also show metallicity gradients \citep{harris2001,Lee2008,caldwell2016}.

We dig deeper into this issue in the right panel of Fig.~\ref{fig:meta}. In the MW and M31, 20\,kpc is often used as a divider between inner halo GCs and outer halo GCs. Fig.~\ref{fig:meta} compares the metallicity distributions of just the very outermost GCs in M81 ($R_{p,M81}>20$\,kpc) with 14 MW \citep{Harris1996} and 49 M31 \citep{Wang2019} outer halo GCs with projected radius $R_{p}>20$\,kpc. Projected radii for MW GCs are calculated by projecting along the vector between the Sun and the Galactic Center; in Galactic Cartesian coordinates $r_p = \sqrt{Y^{2}+Z^{2}}$ using the positions from the \citet{Harris1996} catalogue. MW globular cluster metallicities come from \citet{Harris1996}; we adopt colour-derived metallicities from \citet{Wang2019} for M31 outer halo GCs as they have many more outer halo GC metallicity estimates than \citet{caldwell2016}. 
We find that the six M81 outer GCs with $R_p > 20$\,kpc have a mean metallicity of [Fe/H] $= -1.27 \pm 0.27$. The mean metallicity of MW and M31 outer halo GCs are $-1.68 \pm 0.09$ and $-1.34 \pm 0.09$ respectively. While M31's outer halo GCs appear to have mean metallicity not that different from M81's, it is important to note that M31's outer GC population is quite different from M81's --- nearly 1/3 of M31's outer halo GCs have metallicities higher than the most metal-rich of M81's outer halo GCs. We will return to this issue in \S \ref{sec:m81merger}.

\subsection{The radial distribution of M81 GCs}
\label{sec:radiusdis}
In Fig. \ref{fig:radi}, we present the number of M81 GCs (from our combined catalogue in Appendix \ref{appendix:A}), MW GCs \citep{Harris1996}, and M31 GCs \citep{caldwell2016} as a function of the projected radius from their respective galactic centers. There are 143 MW GCs and 441 M31 GCs. 

In M81, 18\% of its GCs lie between 8 and 20\,kpc, and 9\% GCs lie beyond 20\,kpc. The MW has a similar distribution --- 21\% of the MW's GCs lie between 8 to 20\,kpc and 10\% of its GCs lie beyond 20 kpc. Lastly, in M31, 20\% of its GCs have projected distances of 8 -- 20\,kpc, and 19\% lie more than 20\,kpc from M31's centre. Unlike the MW or M31, no M81 GC lies more than 41 kpc in projected radius. Number statistics and limited survey areas with sufficient depth are clearly relevant factors, but this hints at a possible paucity of very distant M81 GCs. One other obvious feature of the radial distribution is the large number of M31 GCs compared with that of MW and M81 GCs. The plot shows that M31 GCs clearly outnumber MW and M81 GCs at all projected distances, but particularly outside of 20\,kpc. Further discussion of the number of outer halo GCs will be made in section \ref{sec:m81merger}.

\begin{figure}
	\includegraphics[width=\linewidth]{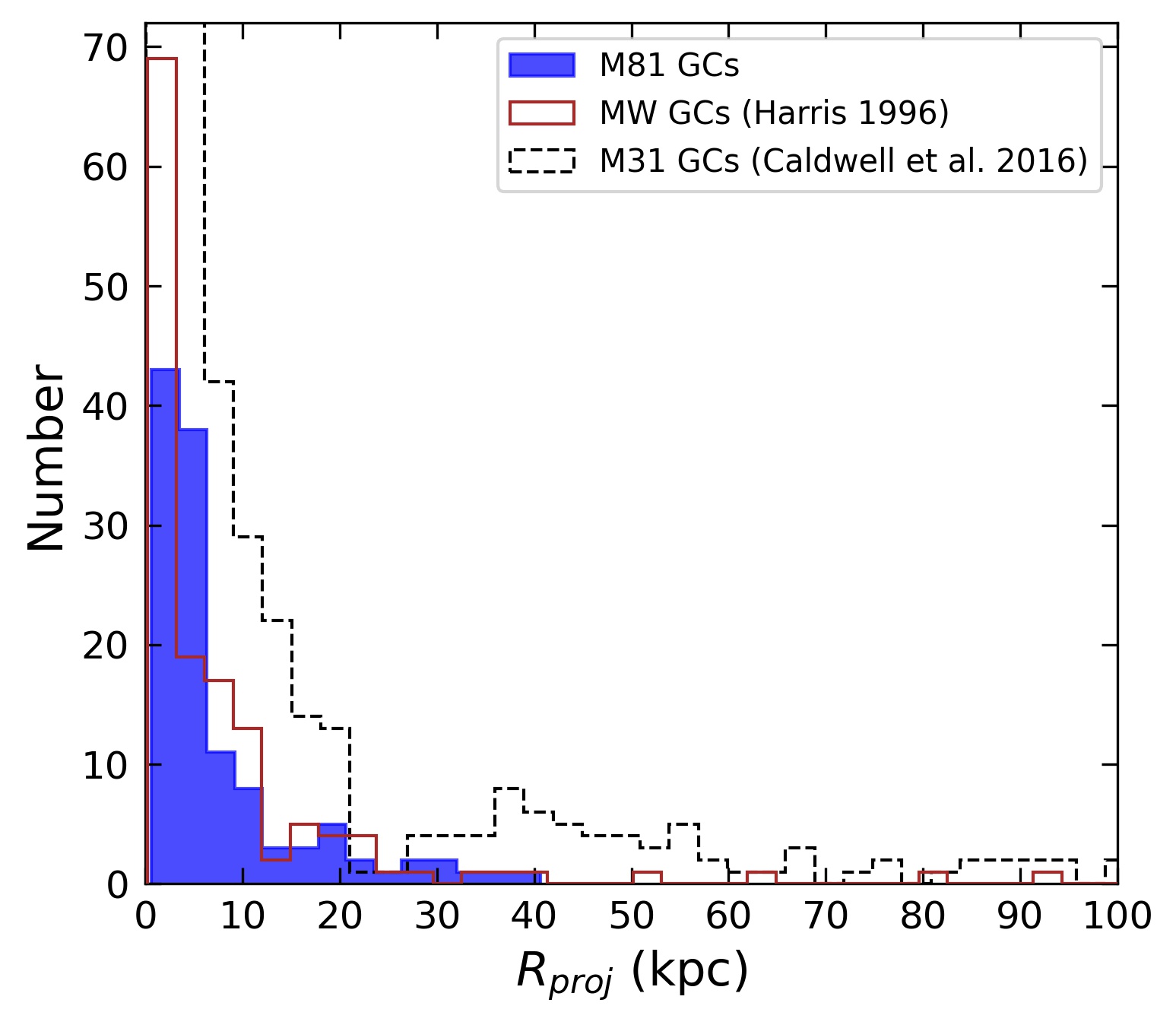}
    \caption{Histograms of projected radius (kpc) of 121 M81 GCs, along with that of MW and M31 for comparison. GCs with projected radii larger than 100kpc are not shown.}
    \label{fig:radi}
\end{figure}

\section{Discussion}
\label{sec:discussion}
In the following section, we discuss the implications of our results. We focus on the implications for the origin of massive GCs, galaxy evolution history, and future observations.
\subsection{The origin of two remote GCs in the halo}
Two globular clusters in the remote halo of M81 group have been reported based on Hubble Space Telescope archive images \citep{Jang12}: GC-1 and GC-2 in Table \ref{tab:GCcatalog}. These clusters were identified through visual inspection of the ACS/WFC and WFC3 UVIS fields respectively. Radial surface brightness of the two globular clusters shows no clear sign of tidal truncation in the outer part, indicating that they are located in an isolated environment. Their distances were derived using the tip of the red giant branch (TRGB) method: GC-1 is at the same distance as M81 (distance $\sim$ 3.6 Mpc), and GC-2 is likely behind the M81 along the line of sight (distance $\sim$ 4 Mpc). \citet{Jang12} suggested that both clusters were likely to belong to M81, as M81 is substantially more massive than M82. 

The positions of the above mentioned two GCs are marked in the upper right panel of Fig.~\ref{fig:radial velocity} in red squares. We found that GC-1, GC-2, and the four other GCs close to M82 have consistent radial velocities (mean radial velocity = 141.22 $\pm$ 72.59 km/s) that are much more similar to M82's radial velocity  (203 $\pm$ 4 km/s) than M81's ($v_r = -34$\,km/s) \citep{deVaucouleurs91}. This tentatively suggests that GC-1 and GC-2 are more likely to be associated with M82, perhaps in the process of being tidally stripped as M82 and its clusters tidally interact with the rest of the M81 group. GC-2 may still be an extremely isolated GC; M82's distance appears to be similar to M81's ($D_{M82} = 3.53$\,Mpc; \citealt{Tully2013}), and so GC-2 may lay considerably in the background, despite its similar velocity to the rest of M82's GCs. 

\subsection{M81 galaxy merger history} 
\label{sec:m81merger}

In sections \ref{sec:metal} and \ref{sec:radiusdis}, we showed that M81 has relatively few ($N=10$ with projected radius >20\,kpc), relatively metal-poor outer halo globular clusters. In this respect, M81 is rather similar to the Milky Way ($N=14$ with projected radius >20\,kpc) and very different from M31, which has numerous outer halo GCs ($N=113$ with projected radius >20\,kpc; \citealt{Wang2019}), where a third of them are more metal rich than the most metal-rich M81 halo GC. 

Some of this deficiency likely reflects an incomplete census of M81 GCs. Our survey does not cover part of M81's halo (we are missing 40\% of its area) --- this geometric factor alone would suggest that we might be missing several GCs, although doubling the sample seems to be an upper limit to the possible magnitude of this effect. We note that the M81 GC census extends well below the GCLF peak, to fainter than $M_V \sim -6$. In the MW and M31, clusters fainter than $M_V \sim -6$ comprise $\sim20$\% of the overall cluster population, so we do not consider it likely that the numbers of M81 GCs will be increased significantly if one were to push the census to deeper limits. We conclude that while M81's outer halo may contain as many GCs as the MW, neither galaxy hosts as many GCs, or as metal-rich outer halo GCs, as M31 does.  

There are differences in GC halo kinematics between these three systems also. Beyond the GCs that are clearly associated with the Sagittarius stream \citep[e.g.,][]{Law2010,Massari2017}, the kinematic signatures of possible groups of MW GCs that might have been accreted together are modest and require full orbital information \citep[e.g.,][]{Kruijssen2019,Massari2019}. M81 group GCs show the clear signature of the accretion origin of the subset of GCs that came in with M82 (section \ref{sec:radial}); otherwise, no clear kinematic substructure is seen --- in many respects similar to the situation in the MW. This contrasts with M31's outer halo GC population, where subsets of GCs are clearly associated with stellar streams in its halo using both kinematic and spatial evidence (e.g., \citealt{Veljanoski2014,Mackey2019}).

It is tempting to interpret these differences in the outer halo GC population in the light of the knowledge of these galaxies' accretion histories. Both the MW and M81's stellar halos are relatively low mass ($\sim 10^9 M_{\odot}$; \citealt{Deason2019,Smercina20}), have moderate metallicity ($\sim 1/10$ solar; \citealt{Durrell2010,Monachesi2013,Conroy2019,Smercina20}), and are dominated by ancient stars ($\sim 9$\,Gyr; \citealt{Gallart2019,Durrell2010}). The interpretation of both halos is similar --- both galaxies merged early with metal-poor moderately-massed galaxies \citep{Bell2017,Helmi2018,Smercina20}, and only now are interacting with much larger satellites (the LMC and M82 respectively; \citealt{Besla2007,Okamoto2015,Smercina20}). The MW's and M81's outer halo GCs, in their sparseness, lack of coherent kinematics, and relatively low metallicities would reflect these quiet merger histories. 

M31, on the other hand, has had a much more active late-time  merger history. M31 has a metal-rich and massive stellar halo \citep[e.g.,][]{Ibata2014}, with a giant stream that wraps at least once around M31 \citep[e.g.,][]{Ibata2001,Gilbert2009,Fardal2013} and multiple smaller streams \citep{McConnachie2018}. The stellar halo and giant stream both contain intermediate-age stars (up to a few Gyr old; \citealt{Brown2006,Brown2007}), indicating ongoing late merging and accretion. While there is discussion in the community about the best interpretation of these features, a large late merger has been suggested \citep{Hammer2018,D'SouzaBell2018}, and it is clear that there have been other smaller accretions \citep{McConnachie2018}. M31's outer halo globular cluster population, in its richness, kinematic substructure, and the substantial number of high metallicity GCs, mirrors M31's active merger history. Taken together, these three galaxies paint a picture in which outer halo GC populations are sensitive to galactic merger histories, possibly paving a path towards using them quantitatively to learn about a galaxy's largest mergers. 

It is interesting to consider the future of the M81 group's GC population. M81 is currently merging with M82, NGC 3077, and presumably some other dwarf galaxies in the M81 group \citep{Okamoto2015,Smercina20}. As these galaxies are tidally disrupted, their GCs (which are much denser and more tightly bound than the galaxies themselves) will likely remain intact. This will significantly enrich M81's GC population, particularly at large radius, making it more similar to M31's rather richer GC system; this may give insight into the drivers of GC system growth in a hierarchical context. 


\subsection{Reflections on M81's dark matter halo mass} 
\label{sec:m81halo}

It is interesting to reflect on the implications of the proposed empirical correlation between the total mass of GCs and the galaxy halo's mass (including both dark matter and baryonic matter; e.g., \citealt{Hudson2014}, \citealt{Harris2015}, \citealt{Harris2017BH}), where the ratio of GCs total mass and galaxy halo's mass is roughly 3-4$\times 10^{-5}$. The sheer number of GCs in M31 suggests that it has a substantially larger dark halo mass than either the MW or M81. As a rough estimate of the total number of GCs in M81 to compare to the MW, we combine the NH11 catalogue within a projected radius of 7 kpc (where it is the most complete) with GCs from the best-effort M81 GC catalogue in Appendix \ref{appendix:A} with $r_p>7$\,kpc. We compare this with the number of MW GCs \citep{Harris1996} within the projected radius of 41 kpc, within which 97\% of the MW's GC population resides and beyond which there are no M81 GCs. With these choices, the ratio of the number of M81 GCs and MW GCs is $\sim$1.5. While the number of GCs is possibly overestimated in the central parts of M81 owing to contamination, this will be offset by likely incompleteness in our best-effort catalogue in M81's outer parts; we estimate an uncertainty in GC number of $\sim 20$\%. This implies that M81's halo mass is not considerably larger than that of MW, in contrast to previous findings using kinematic data where M81 mass is more than three times more massive than MW mass \citep[e.g.][]{Karachentsev2014}.

\subsection{Towards a fuller census of GCs in the Local Volume}

This work clarifies some of the challenges that face those wishing to complete a census of GCs in the Local Volume. 
Fig.\ \ref{fig:IRAC confirmed GC candidates} emphasizes a key point --- high spatial resolution (either good ground-based seeing or higher resolution from space) coupled with wide wavelength baselines (either near-IR or UV) are required to dramatically narrow the pool of objects to be considered for further study. Candidates from previous works with poorer imaging and no wide-wavelength baseline photometry end up being strongly contaminated with foreground stars and background galaxies, and spectroscopy alone is insufficient to purify the sample of candidates, especially in systems like M81 where the systemic velocity is close to the velocity of Galactic stars. The necessity of a deep wide-field, high resolution and multi-wavelength approach has been emphasized before, although generally in systems with much richer GC systems, e.g., the joint deep optical/Gaia study of NGC 5128's GC system \citep{Hughes2021} or studies of the GC systems of Virgo cluster GCs \citep[e.g.,][]{Munoz2014,Powalka2016}. Wide-field multi-wavelength information has proven decisive even in works where candidates are nearby enough that their imaging is well-resolved, e.g., the identification of M31 GC candidates at large projected distances by \citet{Zinn2014}. Gaia has important limitations as a foundational dataset for GC searches --- its magnitude limit is rather bright, failing to detect many M81 GC candidates, and many nearby GCs (e.g., those in M31) are so large as to not appear in the Gaia catalogue \citep{Voggel2020}. 
In contrast, the Vera Rubin Observatory's combination of good angular resolution, depth and wide area will be an important foundation on which to build. We note that it is as yet unclear how much longer wavelength near-infrared imaging will be a limiting factor for reducing contamination in samples of GC candidates to manageable dimensions. 

The need for high angular resolution is the greatest in the main body of galaxies. Owing to the rich structure in the main body of even early-type spiral galaxies like M81, and the effects of dust and crowding on multi-wavelength photometry, HST-resolution imaging is essential for identifying GCs and differentiating them from background galaxies and younger clusters and star forming regions (e.g., \citealt{Harris2009},
\citealt{Nantais11}, \citealt{Lim2013}, \citealt{M101_GCs}). HST-like imaging is important for even well-characterized outer halo clusters; our candidates with ambiguous Subaru imaging will require HST images to decisively differentiate them from background galaxies. Gaia has proved already helpful for brighter clusters in this regard \citep{Hughes2021}, and the Nancy Grace Roman Space Telescope, with its wide field, high resolution and longer wavelengths, will likely prove very valuable. 

This and similar works (e.g., \citealt{Zinn2014}, \citealt{Hughes2021}) sketch a roadmap towards assembling a more substantial sample of galaxies with well-measured outer halo GC populations. A census of outer halo GCs for samples of Local Volume galaxies with independent constraints on merger histories will help to illuminate the origin of outer halo GCs. If these works reveal a similar correspondence between outer halo GC population as seen in the MW, M81 and M31, this will open an important possibility. It is currently only feasible to resolve stars in stellar halos out to distances of $\sim 10$\,Mpc. In contrast, because GCs are bright and can be resolved with space-based imaging quality out to distances of 20\,Mpc or more, outer halo GCs may prove to be a powerful and relatively observationally inexpensive measure of galactic merger histories for large samples of galaxies, allowing investigation with statistical samples of the relationships between merger histories and galaxy properties (e.g., \citealt{Bell2017}, \citealt{D'SouzaBell2018}, \citealt{Gallart2019}) or satellite populations (e.g., \citealt{Smercina22}).

\section{Conclusions}
\label{sec:conclusion}

Because outer halo GCs are particularly sensitive to galactic merger and accretion histories, their study in nearby galaxies helps us to both learn about the origin and growth of GC populations and develop a tool that may illuminate galactic merger histories. Outer halo GC inventories of galaxies with constraints on their merger histories from other sources (e.g., their stellar halos) are therefore particularly helpful. In this work, we have combined wide-field and relatively high resolution Subaru HSC imaging of the M81 group with IRAC 3.6$\mu$m data to select candidate GCs; Gaia EDR3, GALEX, spectroscopic information and visual classification helped to confirm and characterize these candidates. We came to the following conclusions.

\begin{enumerate}
  \item The combination of extended size and a narrow and characteristic range in optical--near-IR colours are powerful discriminants that allow the identification of GC candidates with high confidence. Despite modest incomplateness caused by crowding in IRAC images, a total of 23 M81 GC candidates were identified using this technique, supplemented by one known GC in the outer halo from the literature. Ten candidates have confirming spectroscopic and Gaia EDR3 information, and are our strongest candidates; the other 13 are visually consistent with being GCs, and are a set of good candidates. 
  \item Our method also allows us to cull galaxies and stars from studies of outer halo GC candidates in the M81 group. Many candidates in literature GC catalogues (mostly from L18) are actually foreground stars; galaxies are also an important contaminant. This highlights the difficulty of GC identification in nearby galaxies --- even spectroscopy is by itself insufficient to differentiate between GCs and contaminants. 
  \item The subsample of 13 GCs with spectroscopic data shows a kinematic substructure. The GCs closest to M81 have a mean radial velocity ($-36.0$\,km/s) consistent with GCs in the main body of M81 (mean $v_r = -23$\,km/s). GCs closer to M82 have a distinctly higher mean radial velocity (141\,km/s), much more similar to M82's radial velocity of 203\,km/s. On this basis, we assign our sample to M81, M82, or NGC 3077 on the basis of their position within the group.  
   \item M81 GCs have a luminosity function peaking at $M_V \sim -7.7$ and consistent with a Gaussian distribution, similar to the MW and M31. M81's outer halo GCs and M82's GC population both have a slightly fainter peak magnitude. Only 9\% of M81's GCs lie beyond 20\,kpc in projected radius, similar to the MW's fraction (13\%) but significantly less than the fraction of M31 GCs with radii larger than 20 projected kpc (19\%).
  \item The mean metallicity [Fe/H] of M81 GCs is $-1.11 \pm 0.06$, intermediate between the metallicity of the MW and M31 GC systems; at larger radii $R_p > 8$\,kpc the metallicity is lower at [Fe/H]$\sim -1.4$; M81's GC system has a metallicity gradient.  
  \item In its combination of relative sparseness, low metallicity and lack of clear kinematic substructure M81's GCs resemble the MW's outer halo GCs. This contrasts with M31, which has a much richer outer halo GC population, including a substantial fraction of metal-rich GCs, and showing signs of the kinematic substructure. These differences mirror what has been inferred about these three galaxies' merger histories from their stellar halos: the MW and M81's largest accretions happened long ago and had relatively low masses and metallicities ($<$ $10^{9}$\,$M_{\odot}$ and [Fe/H]$\sim -1$), while M31's most important merger event was more recent (few Gyr ago) and much higher mass and metallicity ($>$ $10^{10}$\,$M_{\odot}$ and [Fe/H] $>$ -0.5). This correspondence between outer halo GC properties and existing {\it independent} constraints on merger histories suggests that outer halo GCs may be able to provide interesting constraints on galactic merger histories.
\end{enumerate}

\begin{landscape}

  \begin{table}

  \caption{Final catalogue of M81 group GC candidates. ID from 1 - 10 are strong GC candidates, and ID from 11 -23 are IRAC GC candidates. ID 24 is GC-2 which is a confirmed GC, but it does not have r and i band measurement due to saturation. The classification is based on our visual inspection of the residual of GC candidates after the subtraction of MGE fitting in the HSC imaging}
 \label{tab:GCcatalog}
  \begin{minipage}{0.7\textwidth}
 \begin{tabular}{cccccccccccccc}
  
  \hline
ID  & NH10 ID \footnote{NH10 catalogue ID number} & L18 ID\footnote{L18 catalogueue ID number} & T15 ID \footnote{ID of IKN GCs in \citet{Tudorica2015}}& RA (J2000) \footnote{Right ascension in degrees of the Subaru HSC survey data (J2000)} & DEC (J2000) \footnote{Declination in degrees of the Subaru HSC survey data (J2000)}  &g  \footnote{CModel in the g band of the Subaru HSC survey data} 
& r \footnote{CModel in the r band of the Subaru HSC survey data} 
& i  \footnote{CModel in the i band of the Subaru HSC survey data} 
& metallicity \footnote{Metallicity [Fe/H] values in the L18 catalogue for a subset of overlapping GC candidates}& radial velocity \footnote{Radial velocity values in the L18 catalogue for a subset of overlapping GC candidates}&classification \footnote{Classification of GC candidates by visually inspecting HSC images; str. stands for strong GC candidates. ID 6 does not have a classification due to saturated imaging.}&$R_p$ \footnote{Projected radius from the assigned host galaxy in column (14)} &Assignment \footnote{Tentative assignment of the galaxy (M81, M82, and NGC 3077) that is associated with the GCs}\\
&&&& (deg) & (deg) & (mag)& (mag) & (mag) & [Fe/H] & (km/s)&&(Kpc) \\
(1) &(2)& (3) & (4) & (5)& (6) & (7)&(8)&(9)&(10)&(11) &(12)&(13)&(14)  \\
  \hline

1 & 70349 & 4 & $\dots$ & 148.26297 & 69.23004 & 20.22 & 19.62 & 19.31 & -1.29 $\pm$ 0.42 & 27.5 $\pm$ 25.07 & cluster (str.)&17.4&M81\\
2 & $\dots$ & 60 & $\dots$ & 149.66775 & 68.86872 & 18.91 & 18.35 & 18.08 & -1.65 $\pm$ 0.15 & 171.77 $\pm$ 46.33& cluster (str.)&21.5&M81\\
3 & $\dots$ & 8 & $\dots$ & 148.55384 & 69.36056 & 20.28 & 19.77 & 19.54 & -2.72 $\pm$ 0.82 & -66.09 $\pm$ 24.16 &cluster (str.)&20.0&M81\\
4 & BH91 HS01 & $\dots$ & $\dots$ & 148.71575 & 69.32930 & 19.23 & 18.6 & 18.33 & $\dots$ & $\dots$ &cluster (str.)&17.0&M81\\

5 & $\dots$ & 106 & $\dots$ & 149.52396 & 69.57972 & 18.93 & 18.47 & 18.25 & -1.97 $\pm$ 0.48 & 140.96 $\pm$ 25.79 &cluster (str.)&13.7&M82 \\
6 & $\dots$ & 67 & $\dots$ & 149.03884 & 69.75925 & 20.69 & $\dots$ & 19.77 & -1.07 $\pm$ 0.48 & 181.77 $\pm$ 31.27 &$\dots$  (str.)&5.2&M82\\
7 & $\dots$ & 84 (GC-1) & $\dots$ & 148.35920 & 69.52151 & 19.0 & 18.45 & 18.2 & -1.53 $\pm$ 0.3 & 92.9 $\pm$ 30.37&cluster (str.)&16.6&M82 \\
8 & $\dots$ & 89 & $\dots$ & 148.47364 & 69.57391 & 19.89 & 19.39 & 19.16 & -1.92 $\pm$ 0.86 & 151.69 $\pm$ 20.36&cluster (str.)&12.7&M82 \\
9 & $\dots$ & 74 & $\dots$ & 150.86093 & 68.80736 & 19.75 & 19.25 & 18.99 & -1.53 $\pm$ 0.29 & 1.93 $\pm$ 24.86 &cluster (str.)&4.7&NGC 3077\\
10 & $\dots$ & $\dots$ & GC-5 & 152.02292 & 68.41594 & 19.6 & 19.12 & 18.9 & $\dots$ & $\dots$ &cluster (str.)&1.1&IKN \\
 \hline
11 & $\dots$ & $\dots$ & $\dots$ & 148.35462 & 69.30494 & 20.20 & 19.69 & 19.43 & $\dots$ & $\dots$ &cluster&19.2&M81\\
12 & $\dots$ & $\dots$ & $\dots$ & 147.73330 & 69.37192 & 21.10 & 20.51 & 20.24 & $\dots$ & $\dots$ &cluster&32.2&M81\\
13 & $\dots$ & $\dots$ & $\dots$ & 147.66701 & 69.41603 & 21.26 & 20.63 & 20.32 & $\dots$ & $\dots$&cluster &35.0&M81\\
14 & $\dots$ & $\dots$ & $\dots$ & 149.88266 & 68.52754 & 20.70 & 20.10 & 19.88 & $\dots$ & $\dots$ &cluster&40.6&M81\\
15 & $\dots$  & 114 & $\dots$ & 149.72654 & 69.44327 & 20.09 & 19.6 & 19.44 & -0.05 $\pm$ 0.57 & -74.89 $\pm$ 24.86 &cluster &30.2&M81\\
16 & $\dots$ & $\dots$ & $\dots$ & 148.83868 & 68.62236 & 20.36 & 19.94 & 19.61 & $\dots$ & $\dots$ & unclear&27.9&M81\\
17 & $\dots$ & 6 & $\dots$ & 148.46144 & 69.28084 & 20.88 & 20.21 & 19.84 & -0.93 $\pm$ 0.62 & -124.11 $\pm$ 25.07 &ambiguous&16.6&M81 \\
18 & $\dots$ & 118 & $\dots$&  149.80041 & 68.74459 & 20.8 & 20.24 & 20.01 & -1.11 $\pm$ 0.95 & -149.55 $\pm$ 46.33&ambiguous&28.8&M81\\
19 & $\dots$ & $\dots$ & $\dots$ &149.15786 & 69.18650 & 21.83 & 21.24 & 20.94 & $\dots$ & $\dots$&ambiguous&9.7&M81 \\
20 & $\dots$ & 88 & $\dots$ & 148.47111 & 69.57919 & 20.89 & 20.35 & 20.13 & -1.21 $\pm$ 0.59 & 123.27 $\pm$ 24.86&cluster&12.6&M82 \\
21 & $\dots$ & $\dots$ & $\dots$ & 149.35077 & 69.85822 & 20.66 & 20.14 & 19.87 & $\dots$ & $\dots$ &cluster&14.0&M82\\
22 & $\dots$ & $\dots$ & $\dots$ & 151.32148 & 68.77562 & 19.01 & 18.5 & 18.27 & $\dots$ & $\dots$ &cluster&11.5&NGC 3077\\
23 & $\dots$ & $\dots$ & $\dots$ & 150.99786 & 68.78257 & 20.10 & 19.56 & 19.29 & $\dots$ & $\dots$ &cluster&4.9&NGC 3077\\

 \hline
24 & $\dots$ & 83 (GC-2) & $\dots$ & 148.33404 & 69.65455 & 18.04 & $\dots$ & $\dots$ & -1.13 $\pm$ 0.11 & 156.72 $\pm$ 25.07 &cluster&13.9&M82\\

   \hline
  \end{tabular}%
   \end{minipage}

  \end{table}

\end{landscape}

\begin{table*}
 \caption{Background galaxies in the selection using IRAC 3.6 $\mu$m band based on HSC imaging}
 \label{tab:GCgalaxy}
 \begin{tabular}{lccccccccccccc}
  \hline
ID  & NH10 ID & L18 ID& T15 ID & RA (J2000) & DEC (J2000) &g  & r & i & metallicity & radial velocity &classification \\
&&&& (deg) & (deg) & (mag)& (mag) & (mag) & [Fe/H] & (km/s) \\
(1) &(2)& (3) & (4) & (5)& (6) & (7)&(8)&(9)&(10)&(11) &(12)  \\
  \hline

1& $\dots$ & $\dots$ & $\dots$ & 151.55603 & 68.68293 & 21.84 & 20.99 & 20.63 & $\dots$ & $\dots$&galaxy \\
2 & $\dots$ & $\dots$ & $\dots$ & 150.45692 & 69.07153 & 21.73 & 21.09 & 20.75 & $\dots$ & $\dots$ &galaxy\\
3 & $\dots$ & $\dots$ & $\dots$ & 148.52366 & 69.03110 & 21.70 & 20.85 & 20.46 & $\dots$ & $\dots$ &galaxy \\
4 & $\dots$ & $\dots$ & $\dots$ & 148.63562 & 69.40445 & 21.50 & 20.93 & 20.63 & $\dots$ & $\dots$ &galaxy\\
5 & $\dots$ & $\dots$ & $\dots$ & 148.87478 & 68.76366 & 21.75 & 21.01 & 20.65 & $\dots$ & $\dots$&galaxy  \\
6 & $\dots$ & $\dots$ & $\dots$ & 150.13364 & 69.44617 & 19.09 & $\dots$ & 18.47 & $\dots$ & $\dots$ & galaxy\\
7 & $\dots$ & $\dots$ & $\dots$ &150.06420 &69.24925 & 18.96 & 18.44 & 18.25 & $\dots$ & $\dots$ &galaxy \\

   \hline
  \end{tabular}%

\end{table*}

\begin{table*}
 \caption{Potential contaminants in the catalogue of known GCs. The classification was based on colours and size using Subaru and IRAC dataset. If classifications disagree with that of HST images, the HST images are used for the final classification.}
 \label{tab:GCcanta}
  \begin{minipage}{0.7\textwidth}
 \begin{tabular}{lccccccccccccc}
  \hline
ID&NH10 ID &L18 ID & RA (J2000) & DEC (J2000) &g  & r & i &3.6 $\mu$m \footnote{IRAC magnitude in the 3.6 $\mu$m band }&Classification \\
& & & &(deg) & (deg) & (mag)& (mag) & (mag)  \\
(1) &(2)& (3) & (4) & (5)& (6) & (7)&(8)&(9) \\
  \hline

1&$\dots$ & 42 & 149.07864 & 68.73776 & 20.97 & 20.51 & 20.28 & 21.01 & galaxy \\
2&$\dots$ & 124 & 149.90157 & 68.6602 & 19.64 & 19.27 & 19.02 & 19.36 & star \\
3&70319 & $\dots$ & 148.42752 & 69.22336 & 20.62 & 19.92 & 19.42 & 20.9 & galaxy \\
4&50357 & $\dots$ & 148.54564 & 69.03524 & 19.68 & 18.77 & 18.37 & 19.36 & galaxy \\
5 & $\dots$ & 110 & 149.68311 & 68.46808 & 20.75 & 20.15 & 19.88 & 21.06 & star \\
6 & $\dots$ & 109 & 149.66941 & 68.40286 & 19.15 & 18.83 & 18.72 & 20.35 & star \\
7 & $\dots$ & 112 & 149.69588 & 68.44566 & 20.17 & 19.81 & 19.67 & 21.19 & star \\
8 & $\dots$ & 103 & 149.37101 & 68.57431 & 20.38 & 20.0 & 19.87 & 21.35 & star \\
9 & $\dots$ & 101 & 149.1275 & 68.46473 & 19.94 & 19.62 & 19.55 & 21.29 & star \\
10 & $\dots$ & 99 & 149.03502 & 68.36543 & 20.99 & 20.49 & 20.3 & 21.53 & star \\
11 & $\dots$ & 100 & 149.06091 & 68.51418 & 20.49 & 20.12 & 19.98 & 21.32 & star \\
12 & $\dots$ & 104 & 149.37752 & 68.54588 & 19.37 & 18.81 & 18.63 & 19.87 & star \\
13 & $\dots$ & 120 & 149.83002 & 68.96764 & 20.14 & 19.75 & 19.65 & 21.22 & star \\
14 & $\dots$ & 132 & 150.10266 & 69.1181 & 19.97 & 19.58 & 19.45 & 20.96 & star \\
15 & $\dots$ & 73 & 150.85863 & 68.82589 & 19.55 & 19.22 & 19.06 & 20.54 & star \\
16 & $\dots$ & 121 & 149.83487 & 69.20601 & 19.84 & 19.51 & 19.4 & 21.07 & star \\
17 & $\dots$ & 126 & 149.95046 & 69.0432 & 19.61 & 19.3 & 19.22 & 20.87 & star \\
18 & $\dots$ & 3 & 148.25888 & 69.06385 & 21.18 & 20.68 & 20.49 & 21.72 & star \\
19 & $\dots$ & 134 & 150.16667 & 69.25866 & 20.86 & 20.24 & 20.01 & 21.16 & star \\
20 & $\dots$ & 127 & 150.02707 & 69.09375 & 20.08 & 19.7 & 19.57 & 21.05 & star \\
21 & $\dots$ & 122 & 149.85895 & 68.59598 & 21.22 & 20.95 & 20.86 & 22.57 & star \\
22 & $\dots$ & 128 & 150.03814 & 68.43935 & 19.6 & 19.08 & 18.87 & 20.11 & star \\
23 & $\dots$ & 125 & 149.94393 & 68.45088 & 20.3 & 19.62 & 19.34 & 20.34 & star \\
24 & $\dots$ & 133 & 150.13406 & 68.67416 & 21.73 & 21.01 & 20.74 & 21.48 & star \\
25 & $\dots$ & 119 & 149.8051 & 68.69657 & 20.34 & 19.94 & 19.83 & 21.06 & star \\
26 & $\dots$ & 58 & 149.51809 & 68.89751 & 20.17 & 19.81 & 19.68 & 21.08 & star \\
27 & $\dots$ & 107 & 149.61376 & 69.2586 & 19.46 & 19.18 & 19.1 & 20.66 & star \\
28 & $\dots$ & 54 & 149.33741 & 69.24866 & 20.17 & 19.61 & 19.4 & 20.6 & star \\
29 & $\dots$ & 55 & 149.43687 & 68.95878 & 20.27 & 19.69 & 19.47 & 20.57 & star \\
30 & $\dots$ & 79 & 147.71454 & 69.3734 & 19.12 & $\dots$ & 18.66 & 20.17 & star \\
31 & $\dots$ & 97 & 148.822 & 69.50034 & 19.38 & $\dots$ & 18.96 & 20.52 & star \\
32 & $\dots$ & 81 & 147.95216 & 69.56327 & 20.46 & 19.98 & 19.79 & 21.12 & star \\
33 & $\dots$ & 31 & 148.85781 & 69.43994 & 20.28 & 19.91 & 19.8 & 21.29 & star \\
34 & $\dots$ & 39 & 149.06514 & 69.40068 & 20.04 & 19.61 & 19.47 & 20.8 & star \\
35 & $\dots$ & 116 & 149.7556 & 69.57397 & 19.93 & 19.69 & 19.63 & 21.5 & star \\
36 & $\dots$ & 90 & 148.50513 & 69.5053 & 19.62 & 19.23 & 19.1 & 20.53 & star \\
37 & $\dots$ & 117 & 149.76082 & 69.40618 & 20.37 & 19.89 & 19.74 & 21.15 & star \\

   \hline
    
  \end{tabular}%
   \end{minipage}
\end{table*}

\section*{Acknowledgements}

We acknowledge useful conversations with Bill Harris. This work was partly supported by the National Science Foundation through grant NSF-AST 2007065 and by the WFIRST Infrared Nearby Galaxies Survey (WINGS) collaboration through NASA grant NNG16PJ28C through subcontract from the University of Washington. AM gratefully acknowledges support by FONDECYT Regular grant 1212046 and by the ANID BASAL project FB210003, as well as funding from the Max Planck Society through a “PartnerGroup” grant. The authors wish to recognize and acknowledge the very significant cultural role and reverence that the summit of Mauna Kea has always had within the indigenous Hawaiian community.  We are most fortunate to have the opportunity to conduct observations from this mountain.

This work has made use of data from the European Space Agency (ESA) mission
{\it Gaia} (\url{https://www.cosmos.esa.int/gaia}), processed by the {\it Gaia}
Data Processing and Analysis Consortium (DPAC,
\url{https://www.cosmos.esa.int/web/gaia/dpac/consortium}). Funding for the DPAC
has been provided by national institutions, in particular the institutions
participating in the {\it Gaia} Multilateral Agreement.

This work is based in part on observations made with the Spitzer Space Telescope, which was operated by the Jet Propulsion Laboratory, California Institute of Technology under a contract with NASA.

The Hyper Suprime-Cam (HSC) collaboration includes the astronomical communities of Japan and Taiwan, and Princeton University. The HSC instrumentation and software were developed by the National Astronomical Observatory of Japan (NAOJ), the Kavli Institute for the Physics and Mathematics of the Universe (Kavli IPMU), the University of Tokyo, the High Energy Accelerator Research Organization (KEK), the Academia Sinica Institute for Astronomy and Astrophysics in Taiwan (ASIAA), and Princeton University. Funding was contributed by the FIRST program from Japanese Cabinet Office, the Ministry of Education, Culture, Sports, Science and Technology (MEXT), the Japan Society for the Promotion of Science (JSPS), Japan Science and Technology Agency (JST), the Toray Science Foundation, NAOJ, Kavli IPMU, KEK, ASIAA, and Princeton University. 

This paper makes use of software developed for the Rubin Observatory. We thank the Rubin Observatory project for making their code available as free software at \url{http://pipelines.lsst.io}.

The Pan-STARRS1 Surveys (PS1) have been made possible through contributions of the Institute for Astronomy, the University of Hawaii, the Pan-STARRS Project Office, the Max-Planck Society and its participating institutes, the Max Planck Institute for Astronomy, Heidelberg and the Max Planck Institute for Extraterrestrial Physics, Garching, The Johns Hopkins University, Durham University, the University of Edinburgh, Queen’s University Belfast, the Harvard-Smithsonian Center for Astrophysics, the Las Cumbres Observatory Global Telescope Network Incorporated, the National Central University of Taiwan, the Space Telescope Science Institute, the National Aeronautics and Space Administration under Grant No. NNX08AR22G issued through the Planetary Science Division of the NASA Science Mission Directorate, the National Science Foundation under Grant No. AST-1238877, the University of Maryland, and Eotvos Lorand University (ELTE) and the Los Alamos National Laboratory.

SDSS-IV is managed by the 
Astrophysical Research Consortium 
for the Participating Institutions 
of the SDSS Collaboration including 
the Brazilian Participation Group, 
the Carnegie Institution for Science, 
Carnegie Mellon University, Center for 
Astrophysics | Harvard \& 
Smithsonian, the Chilean Participation 
Group, the French Participation Group, 
Instituto de Astrof\'isica de 
Canarias, The Johns Hopkins 
University, Kavli Institute for the 
Physics and Mathematics of the 
Universe (IPMU) / University of 
Tokyo, the Korean Participation Group, 
Lawrence Berkeley National Laboratory, 
Leibniz Institut f\"ur Astrophysik 
Potsdam (AIP),  Max-Planck-Institut 
f\"ur Astronomie (MPIA Heidelberg), 
Max-Planck-Institut f\"ur 
Astrophysik (MPA Garching), 
Max-Planck-Institut f\"ur 
Extraterrestrische Physik (MPE), 
National Astronomical Observatories of 
China, New Mexico State University, 
New York University, University of 
Notre Dame, Observat\'ario 
Nacional / MCTI, The Ohio State 
University, Pennsylvania State 
University, Shanghai 
Astronomical Observatory, United 
Kingdom Participation Group, 
Universidad Nacional Aut\'onoma 
de M\'exico, University of Arizona, 
University of Colorado Boulder, 
University of Oxford, University of 
Portsmouth, University of Utah, 
University of Virginia, University 
of Washington, University of 
Wisconsin, Vanderbilt University, 
and Yale University.

Software: Astropy \citep{Astropy2013,Astropy2018}, SciPy \citep{2020SciPy-NMeth}, Numpy \citep{numpy}, Matplotlib \citep{matplotlib}, Pandas \citep{pandas}

\section*{Data Availability}

The catalogue of globular cluster candidates of the M81 group in this study is available in this article and online supplementary material. The Subaru raw data is available in the Subaru archive, and the photometric catalogue is available upon request by contacting the authors. The IRAC, Gaia, and GALEX datasets are freely publicly available.


\bibliographystyle{mnras}
\bibliography{reference} 




\appendix

\section{Confirming evidence from Gaia and GALEX}
\label{appendix:GG}

While the joint Subaru--IRAC selection is sufficient to isolate a sample of candidate globular clusters, in order to isolate a set of GC candidates with the highest chances of being GCs, we make use of available Gaia, GALEX and spectroscopic data.

\begin{figure*}
	\includegraphics[width=15cm]{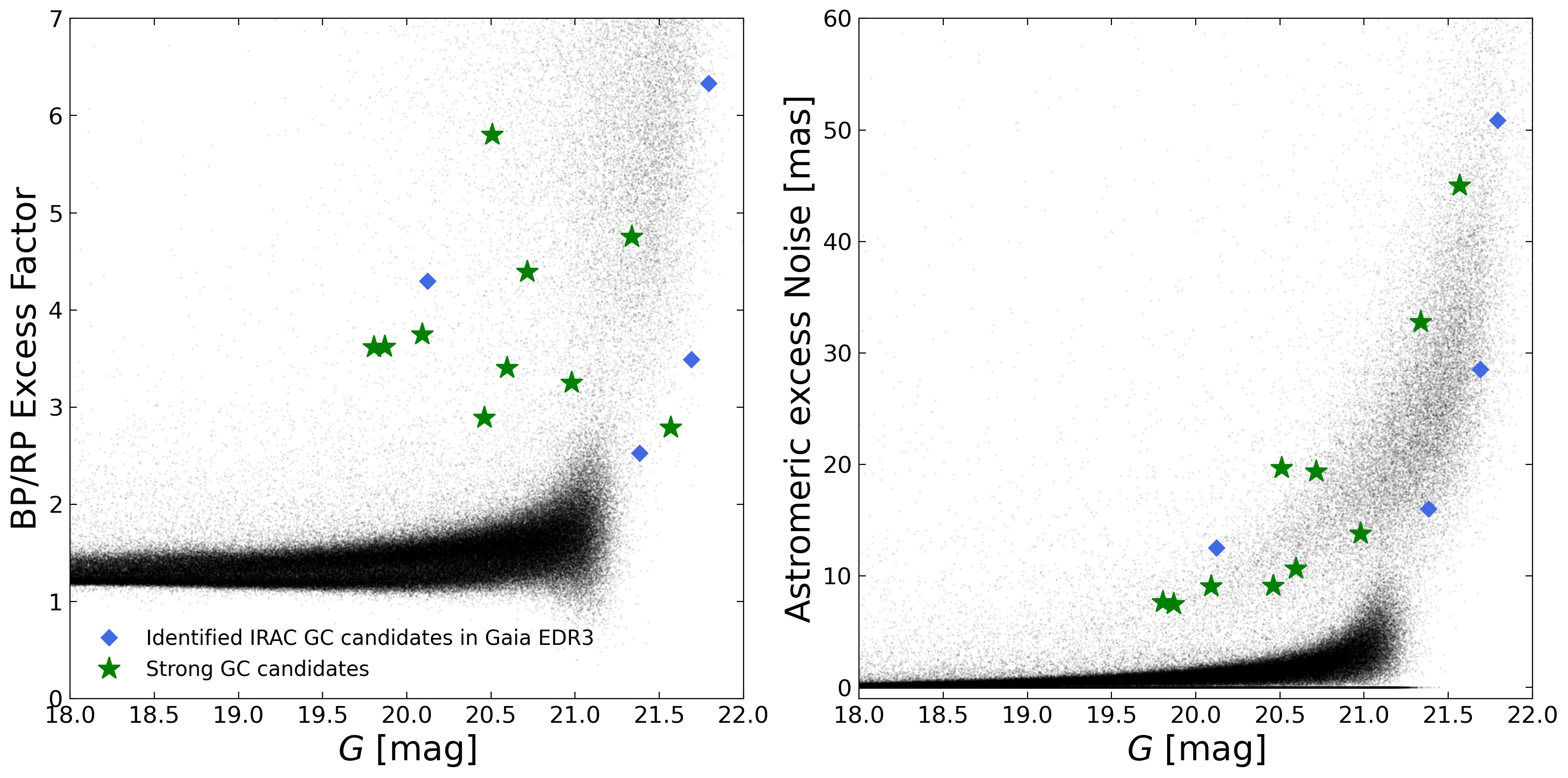}
    \caption{$ \textrm{BP}_{\textrm{excess}}$ and AEN as a function of G band magnitude of a subset of the four IRAC GC candidates bright and compact enough to have Gaia catalogue detections, and ten {\it strong} GC candidates, shown as blue diamonds and green stars respectively. Black dots are all other objects in the IRAC field with data from Gaia EDR3. The IRAC GC candidates and {\it strong} GC candidates can also be distinguished from foreground stars in Gaia EDR3 data set using $ \textrm{BP}_{\textrm{excess}}$ and AEN. 
    }
    \label{fig:Gaia Dr2 GC candidates}
\end{figure*}

Gaia EDR3 reports BP/RP Excess Factor ($ \textrm{BP}_{\textrm{excess}}$) and Astrometric Excess Noise (AEN). $ \textrm{BP}_{\textrm{excess}}$ is the ratio of the sum of BP and RP flux with G flux, and extended objects have a higher $ \textrm{BP}_{\textrm{excess}}$ due to the higher sum of flux in BP and RP compared with G flux (Fig.~\ref{fig:Gaia Dr2 GC candidates}; left hand panel). 
AEN measures the disagreements between the positions of observed sources and their best-fitted astrometric model. Larger AEN values mean poorer fits between the objects and point sources, so we can use AEN values to select marginally resolved extended objects (right hand panel of Fig.~\ref{fig:Gaia Dr2 GC candidates}). All ten strong GC candidates (by definition), and four sufficiently bright and compact IRAC GC candidates have $ \textrm{BP}_{\textrm{excess}}$ and AEN too large to be unresolved point sources, and are therefore extended, confirming the Subaru measurement of a slightly extended source for selection as a GC candidate. 
Gaia $ \textrm{BP}_{\textrm{excess}}$ and AEN values have been used to select GC candidates (without the much more stringent IRAC selections) in NGC 5128, at a similar distance of $D \sim 3.5$\,Mpc, by \citet{Voggel2020} and \citet{Hughes2021}.

\begin{figure}
	\includegraphics[width=\columnwidth]{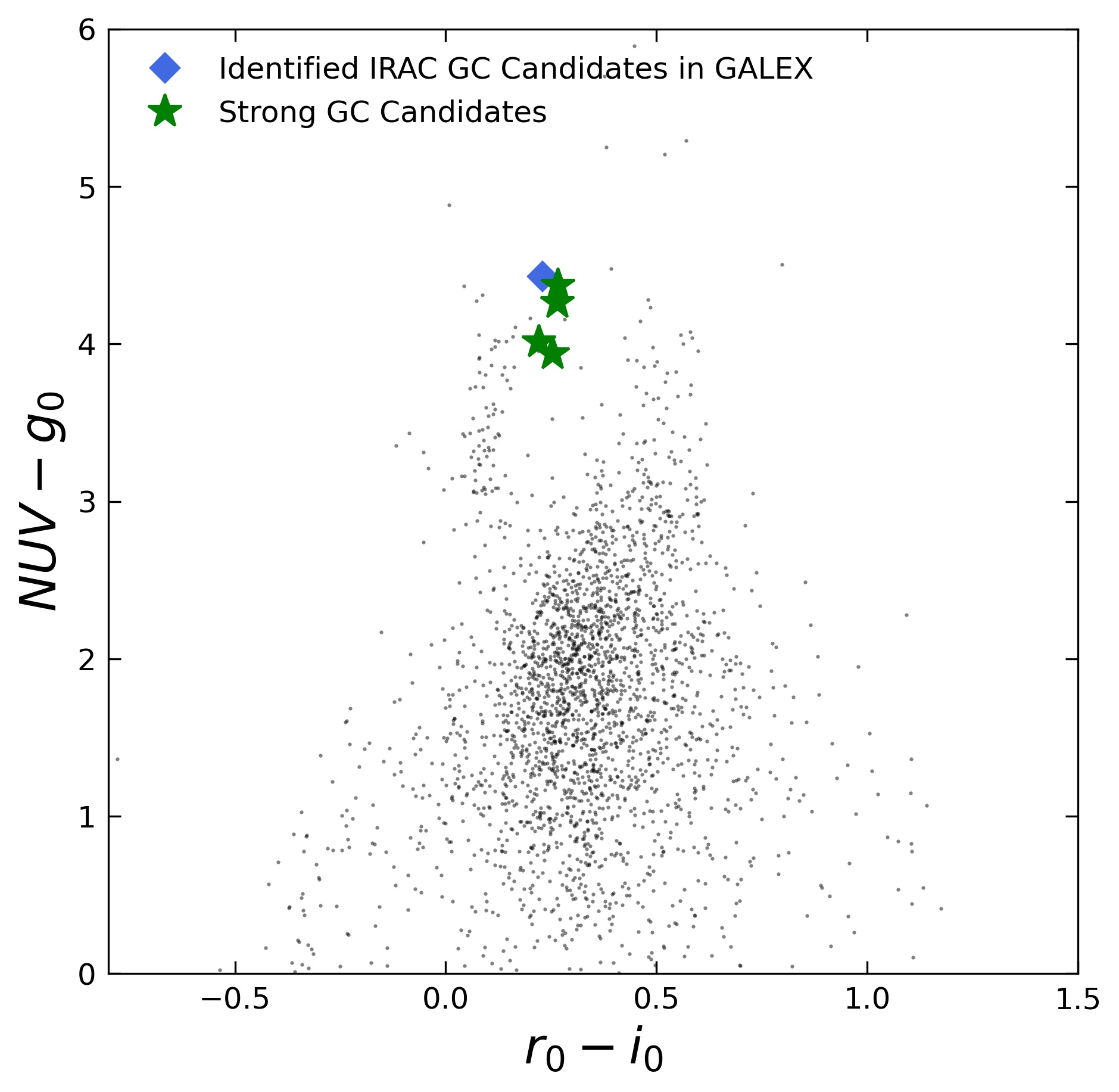}
    \caption{GALEX [NUV] band data combined with Subaru data of one IRAC GC candidate and four strong GC candidates. Black dots are objects crossed matched with Subaru data. Identified IRAC GC candidates and strong GC candidates have distinct [NUV]$-g$ colours from stars and galaxies. 
    }
    \label{fig:GALEX}
\end{figure}

GALEX has relatively bright magnitude limits and detects only five GC candidates (four of the strong GC candidates and one of the IRAC-selected GC candidates [ID 12]), so its import in this study is limited. We do find that the colour-colour diagram [NUV]$-g/r-i$ discriminates well between stars, background galaxies, and GCs, where globular clusters have dramatically different colours from both foreground stars and background galaxies as shown in Fig. \ref{fig:GALEX}. The resulting classification is very consistent with IRAC selection, but adds no decisive new insight into these five GC candidates. 

\section{HSC images for GC candidates}
\label{appendix:B}
We present HSC images for strong GC candidates, IRAC GC candidates, and galaxies in the IRAC candidates based on visual appearance. 
\begin{figure*}
    \begin{tabular}{lll}
    \includegraphics[width=.3\linewidth,valign=m]{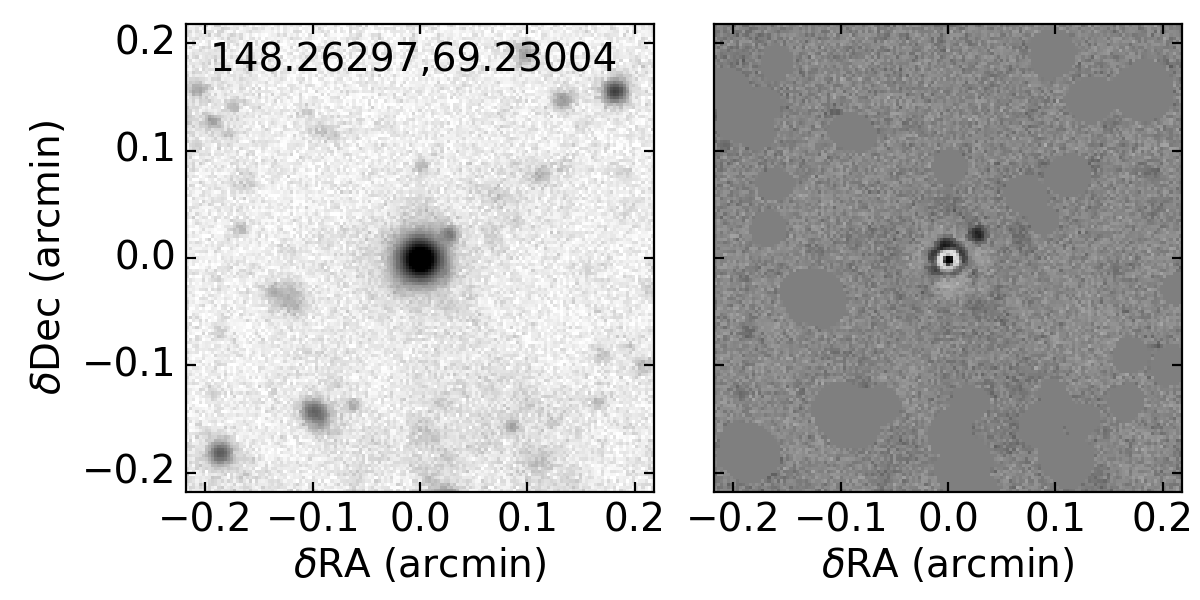} & \includegraphics[width=.3\linewidth,valign=m]{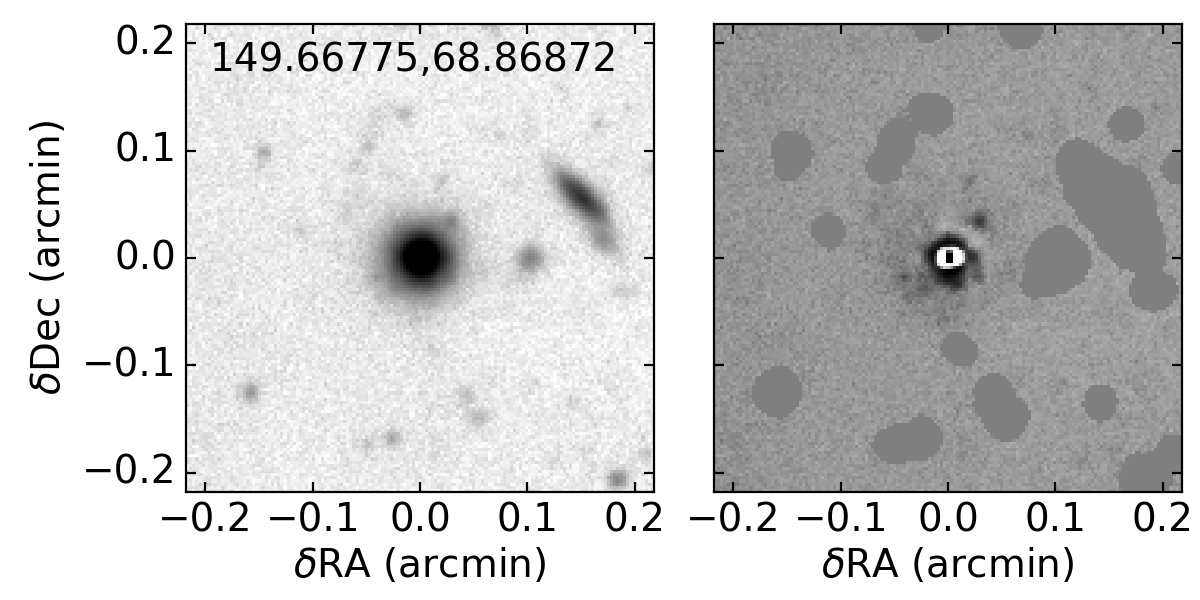} & \includegraphics[width=.3\linewidth,valign=m]{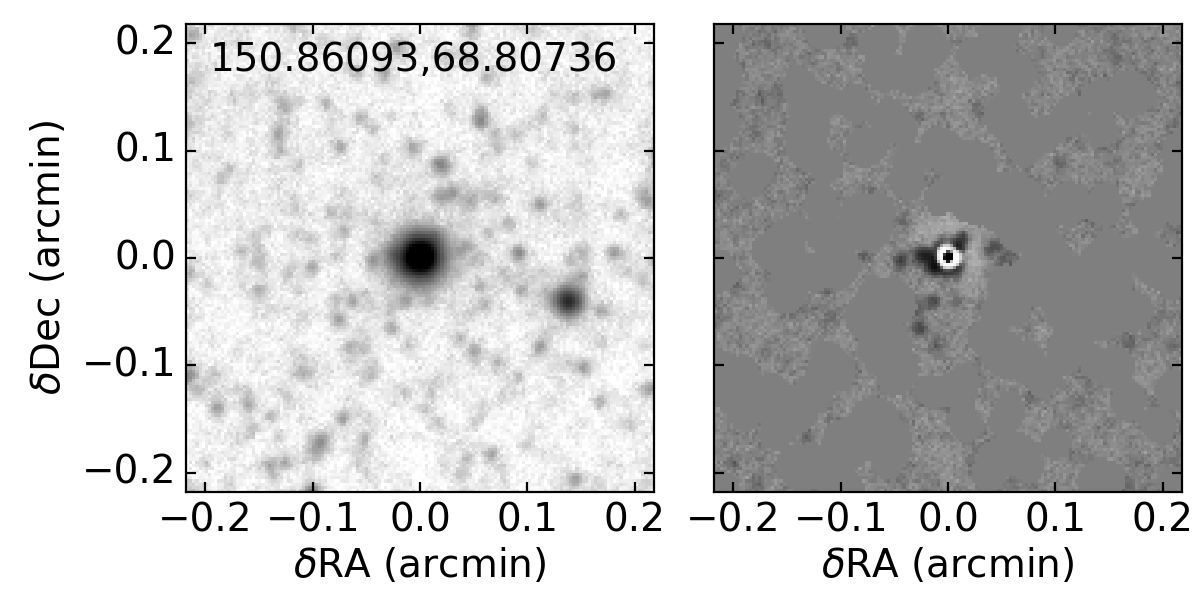}\\
    \includegraphics[width=.3\linewidth,valign=m]{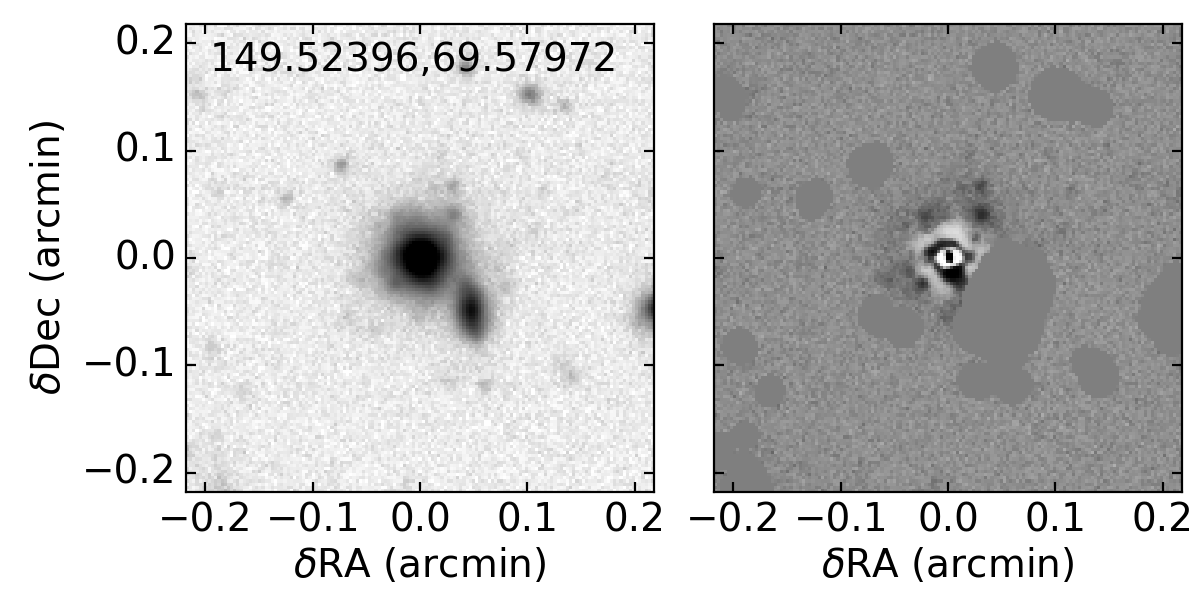} & \includegraphics[width=.3\linewidth,valign=m]{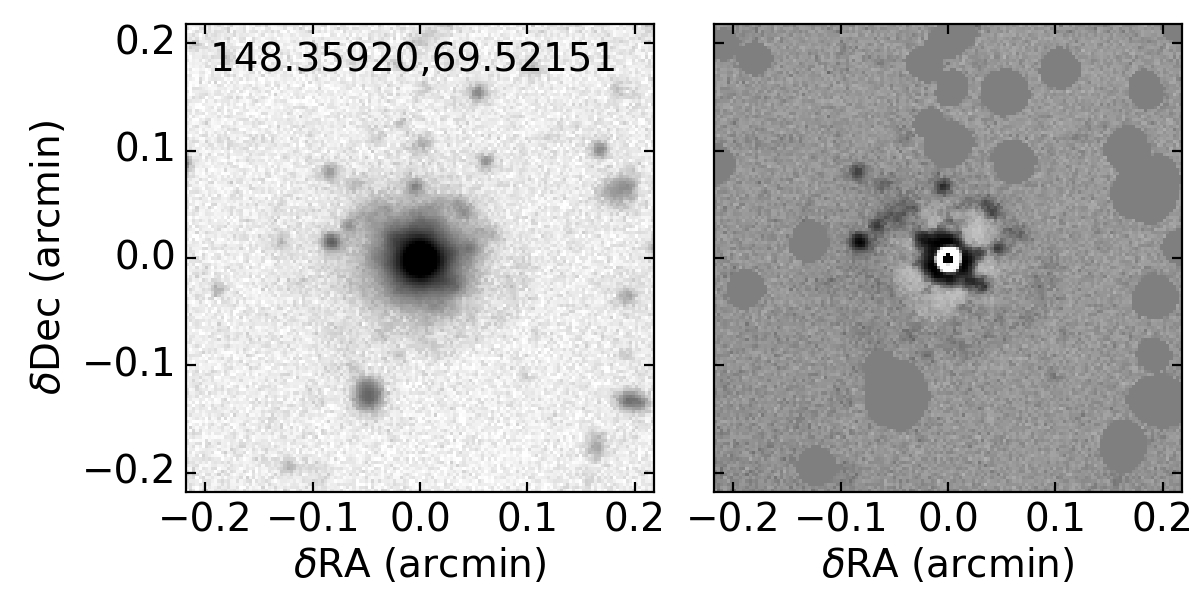} & \includegraphics[width=.3\linewidth,valign=m]{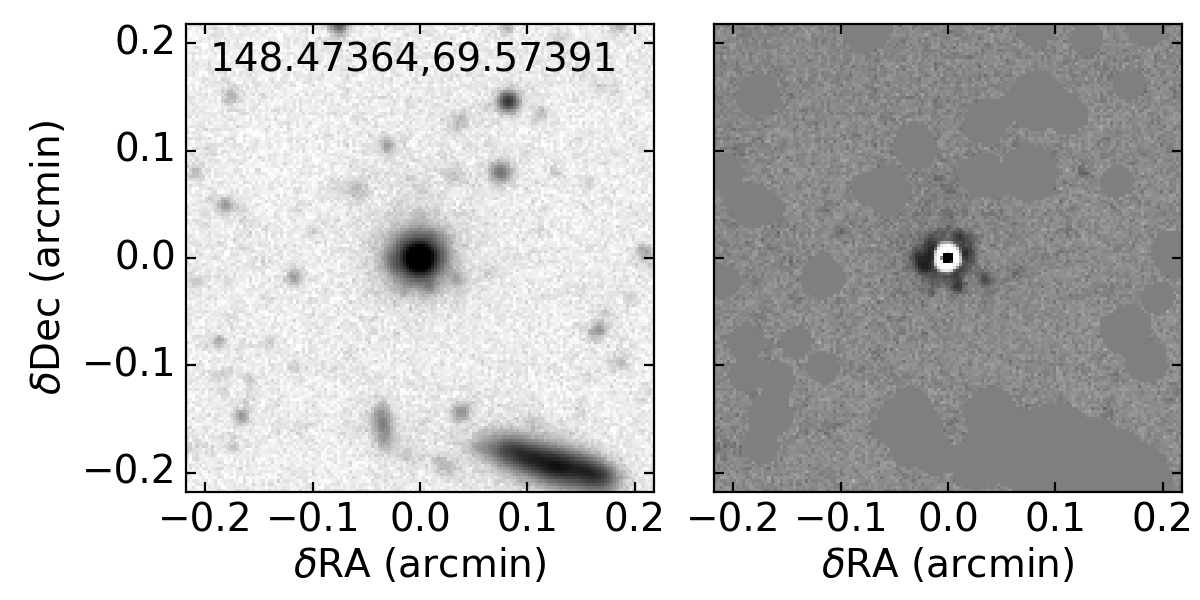}\\
    \includegraphics[width=.3\linewidth,valign=m]{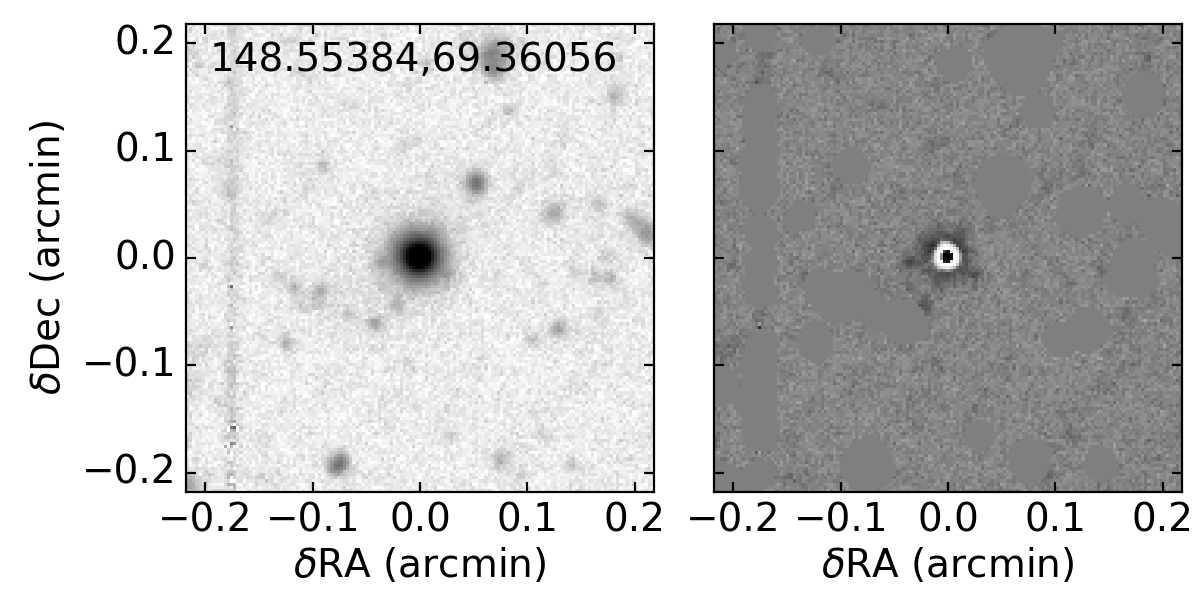} & \includegraphics[width=.3\linewidth,valign=m]{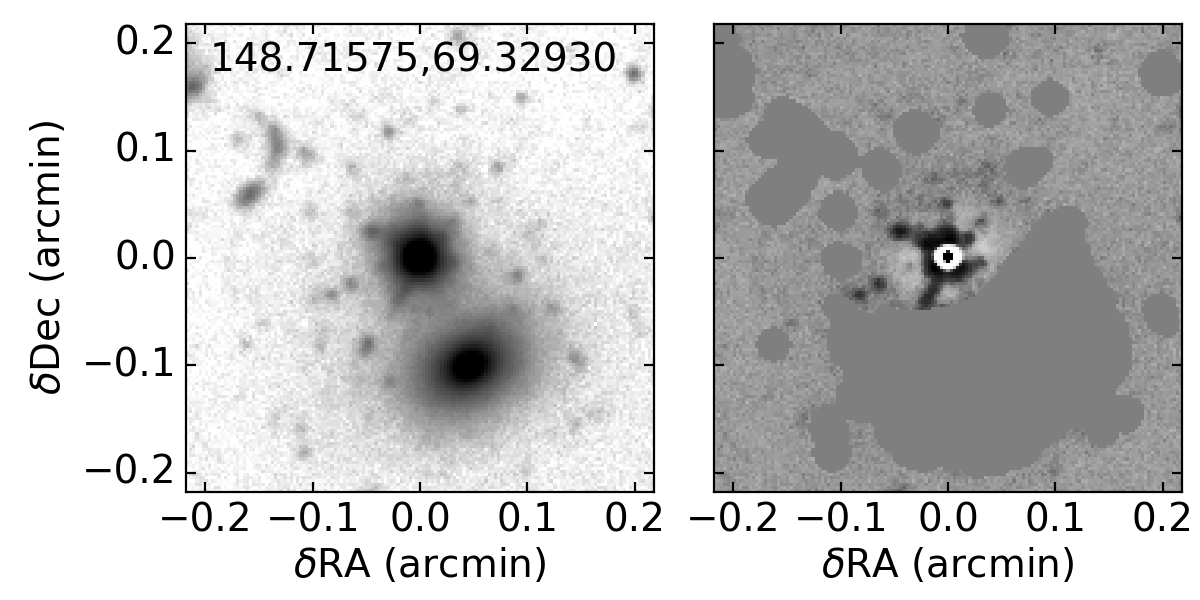} & \includegraphics[width=.3\linewidth,valign=m]{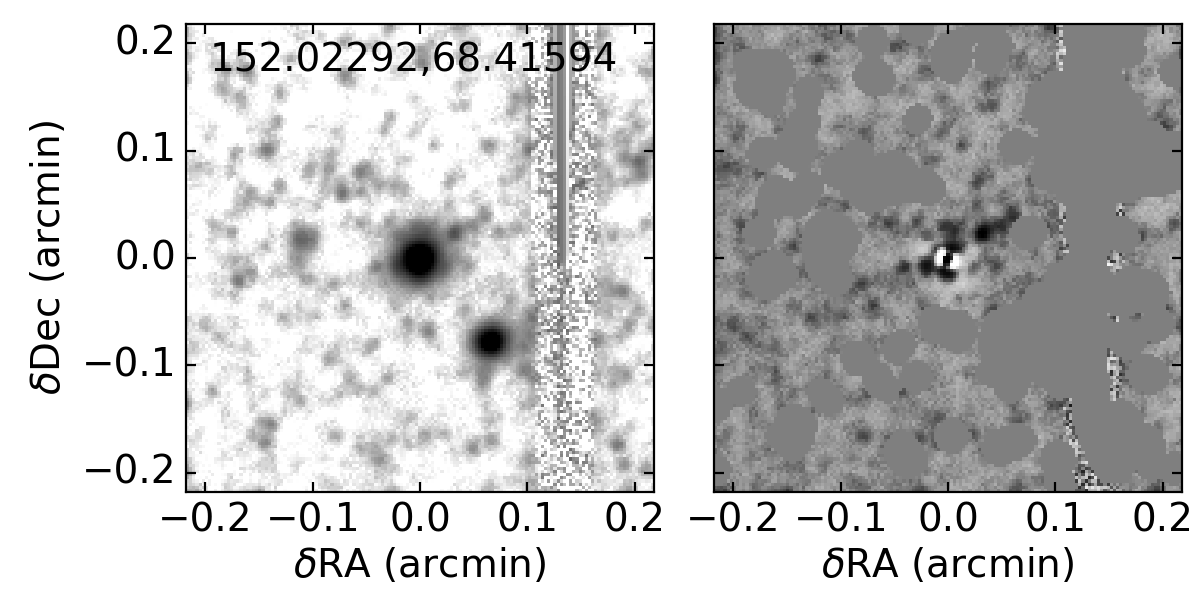}\\
    \end{tabular}
    \caption{Images of the 9 {\it strong} GC candidates that have three band coverage from Subaru. The left-hand panel for each candidate shows the image; the right hand panel shows the residuals after subtraction of a multi-gaussian expansion model that highlights non-axisymmetric features. Candidates are arranged as in Table \ref{tab:GCcatalog} from left to right and top to bottom, where candidate 6 is missing.  
}
    \label{fig:GC image}
\end{figure*}

\begin{figure*}
    \begin{tabular}{lll}
    \includegraphics[width=.3\linewidth,valign=m]{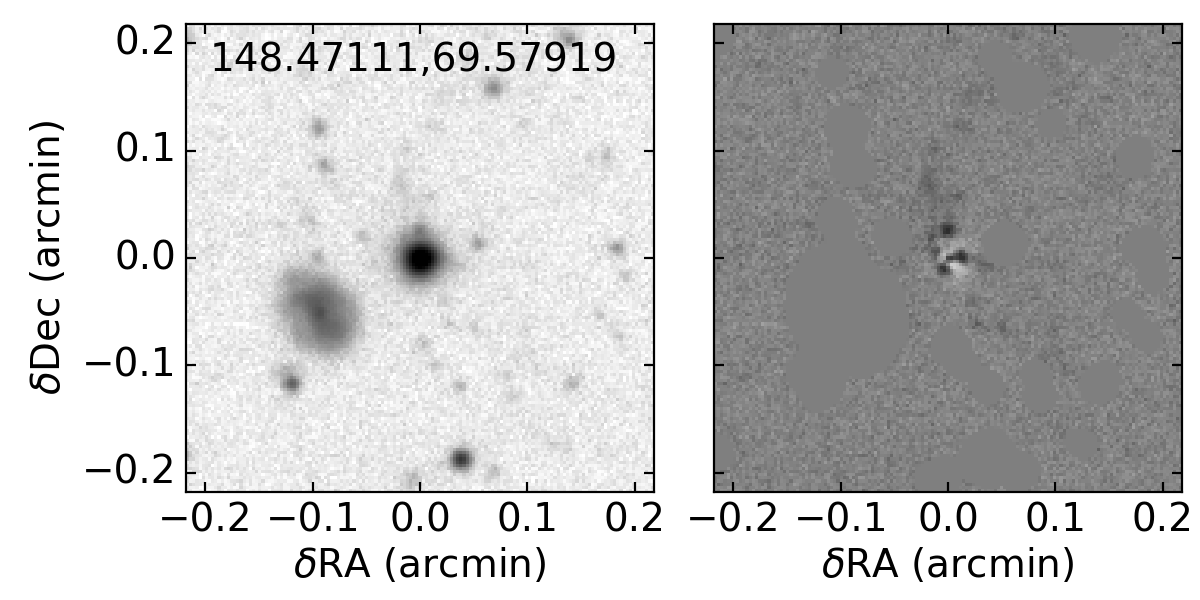} & \includegraphics[width=.3\linewidth,valign=m]{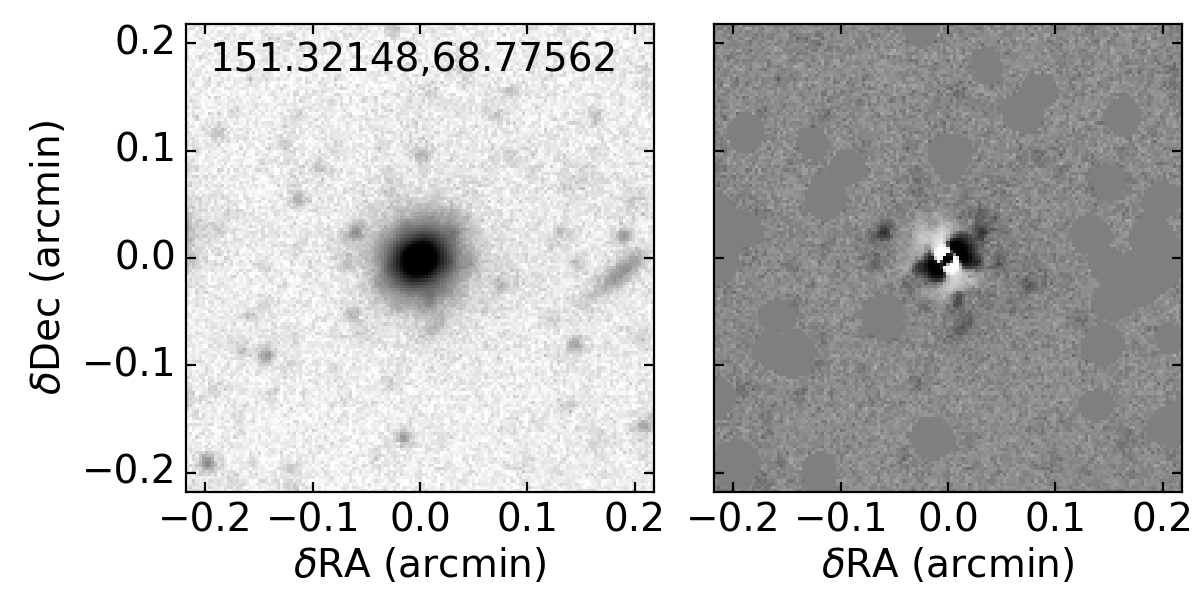} & \includegraphics[width=.3\linewidth,valign=m]{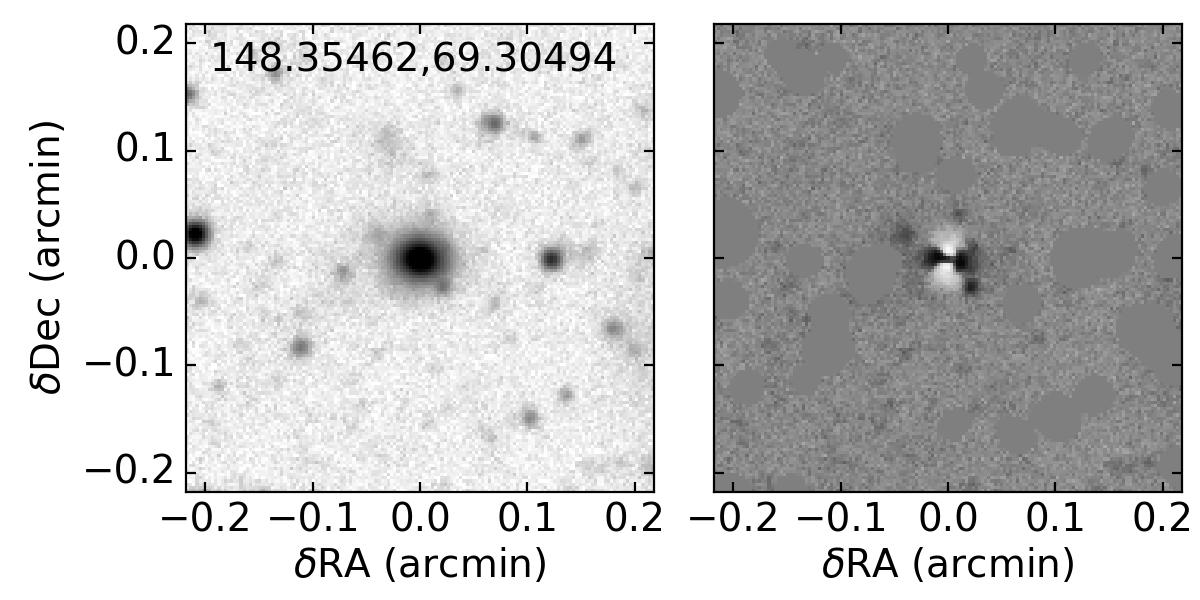} \\
     \includegraphics[width=.3\linewidth,valign=m]{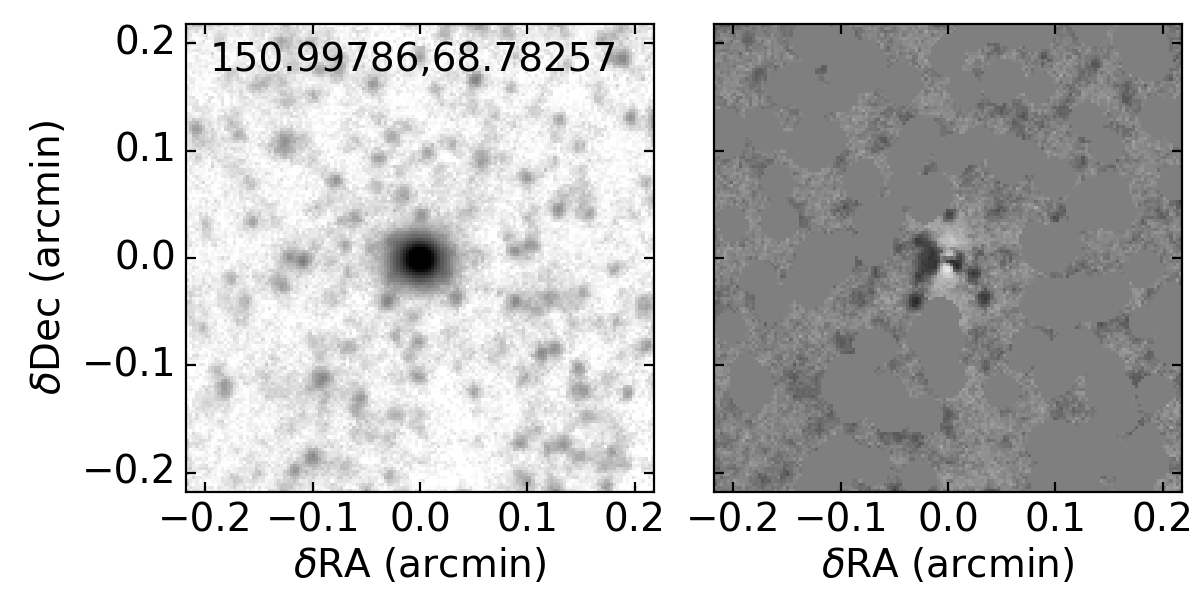} & \includegraphics[width=.3\linewidth,valign=m]{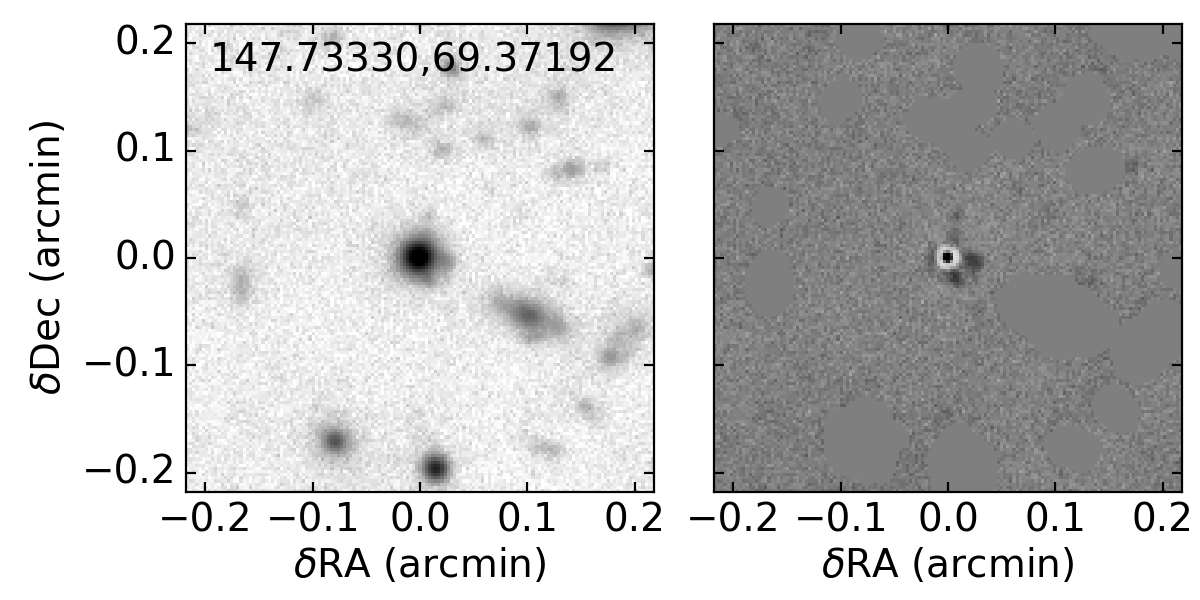} & \includegraphics[width=.3\linewidth,valign=m]{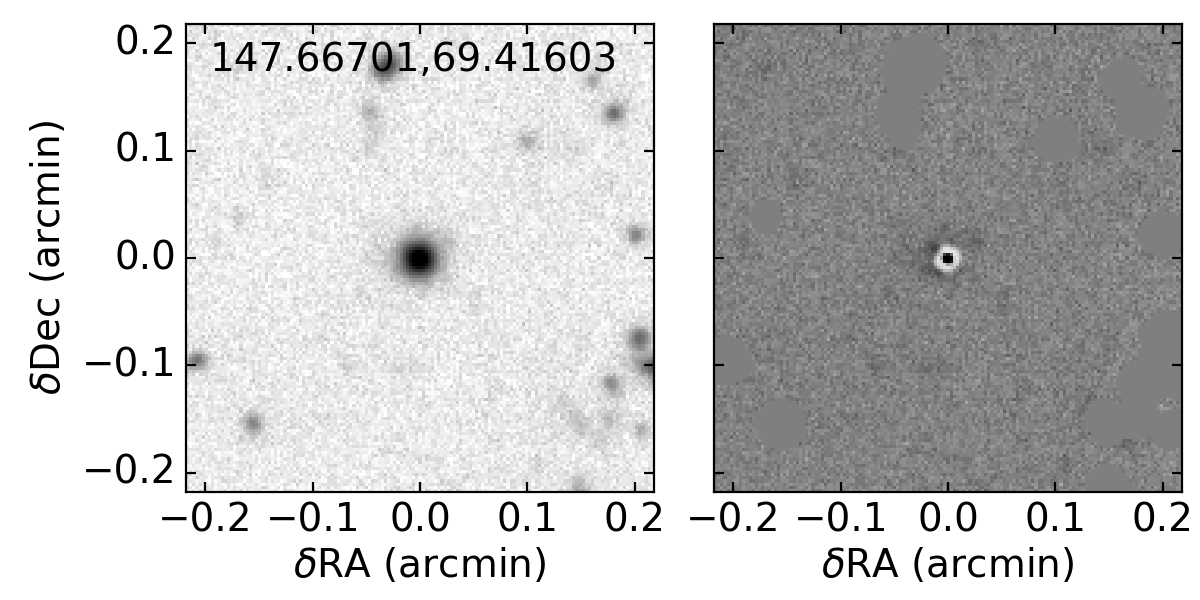} \\
     \includegraphics[width=.3\linewidth,valign=m]{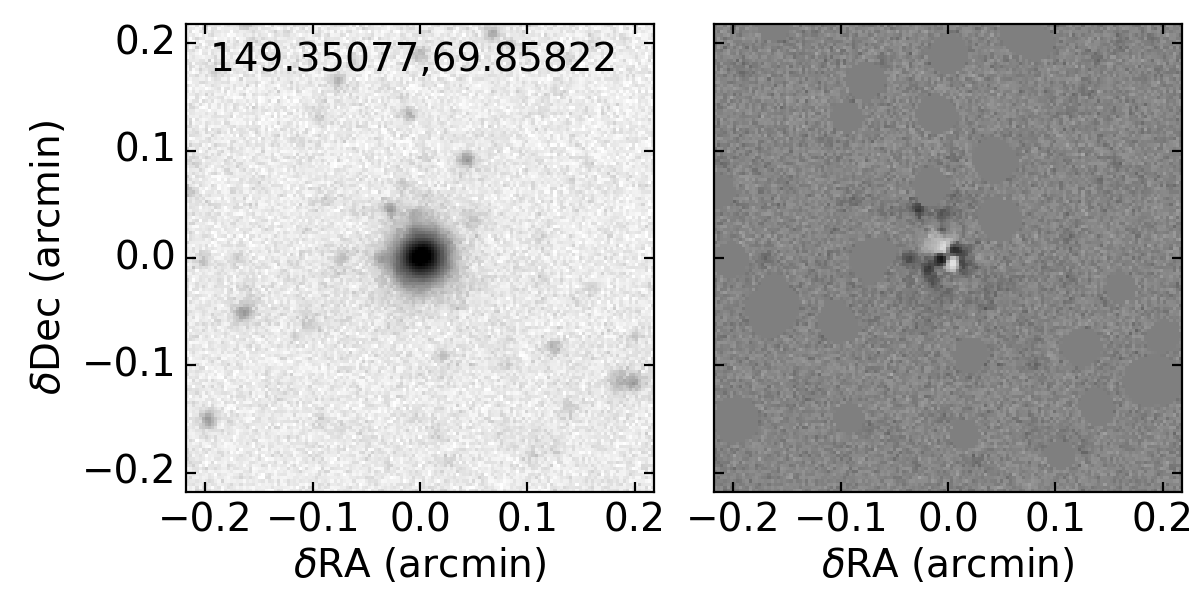} & \includegraphics[width=.3\linewidth,valign=m]{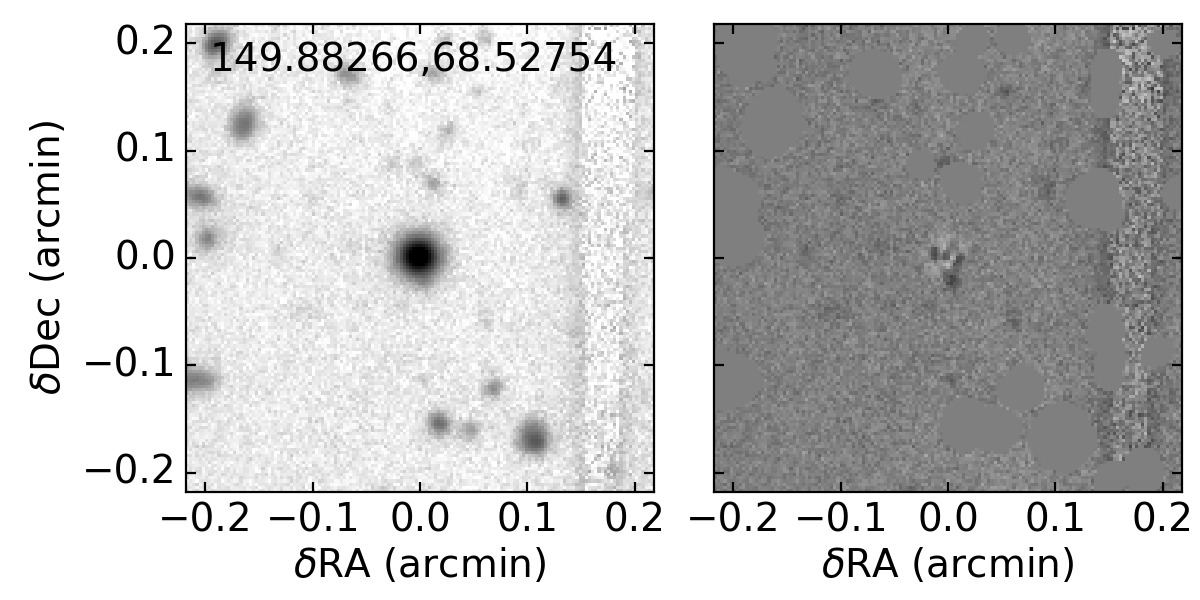} & \includegraphics[width=.3\linewidth,valign=m]{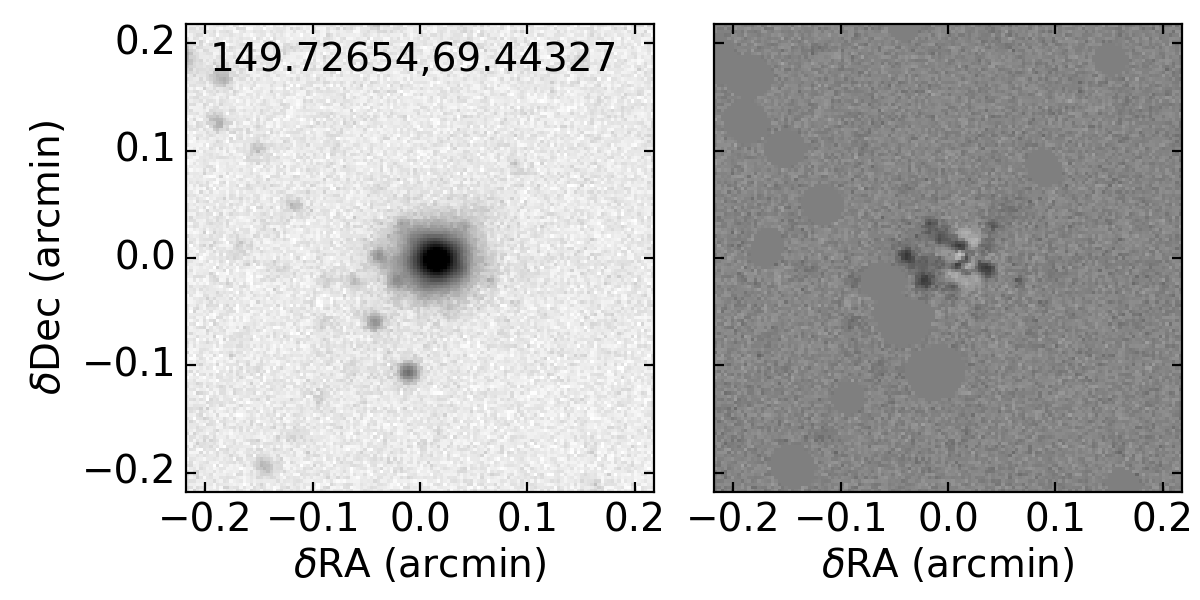} \\
    \includegraphics[width=.3\linewidth,valign=m]{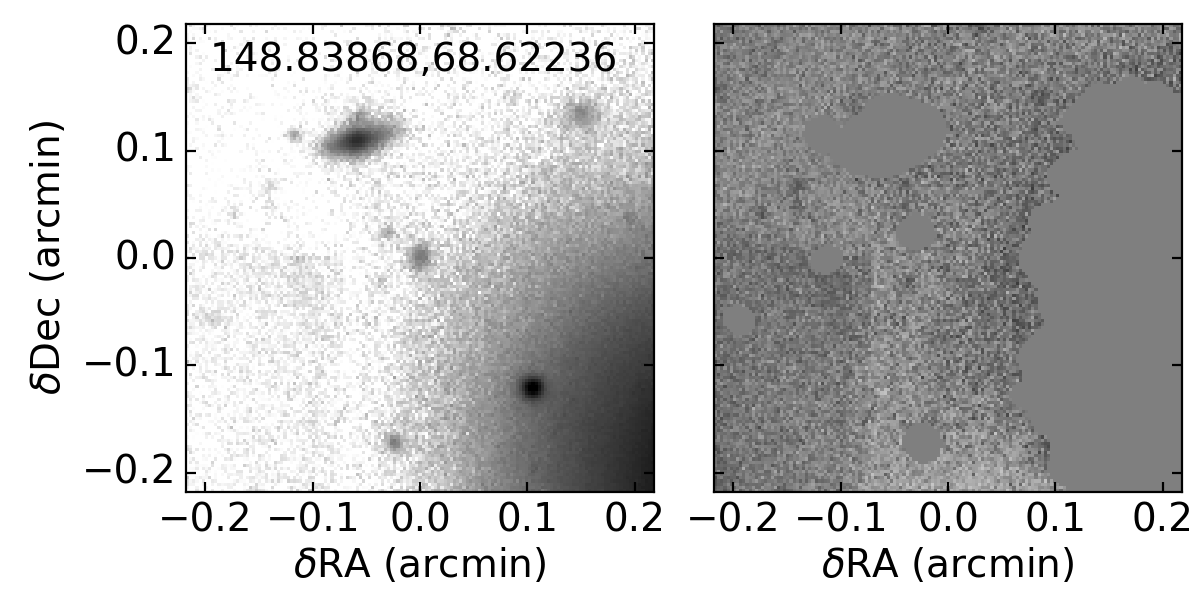} & \includegraphics[width=.3\linewidth,valign=m]{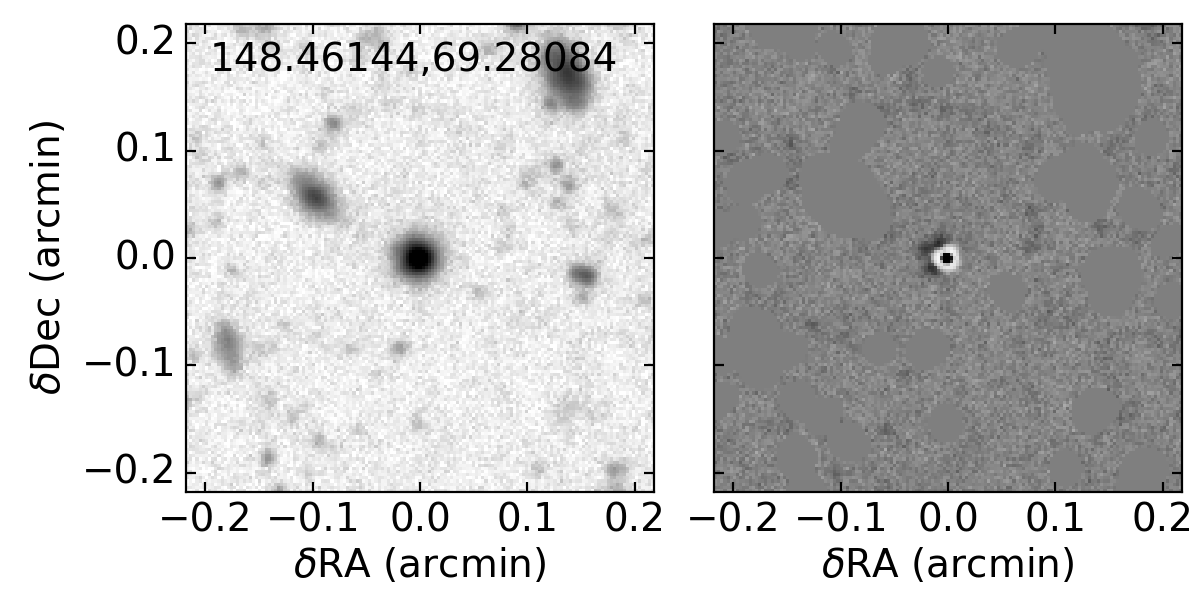} & \includegraphics[width=.3\linewidth,valign=m]{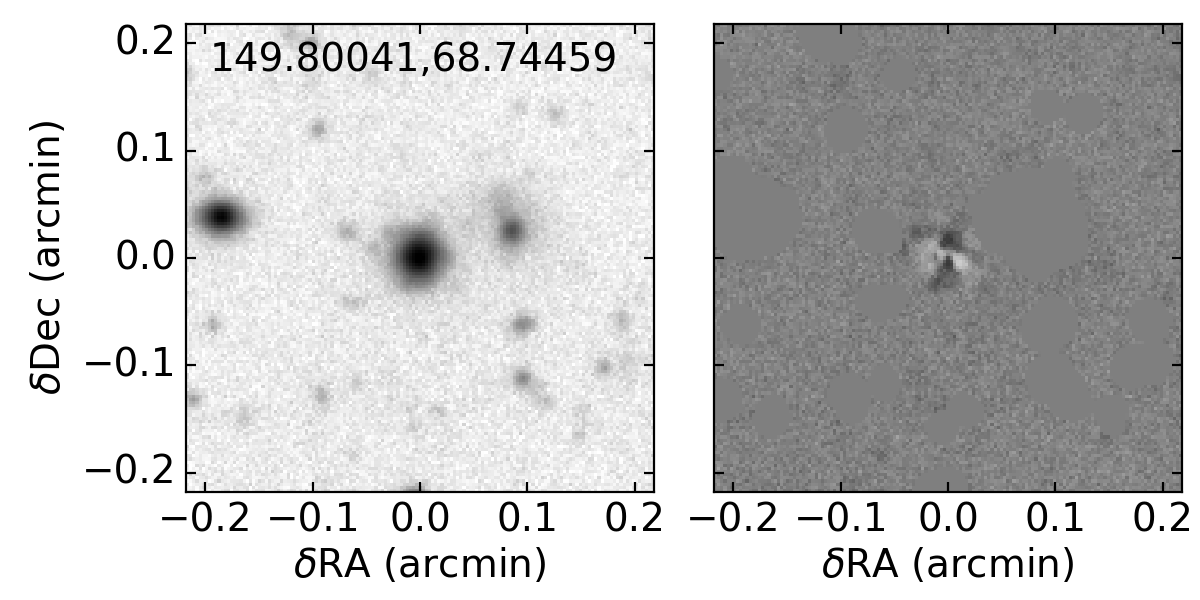} \\
    \includegraphics[width=.3\linewidth,valign=m]{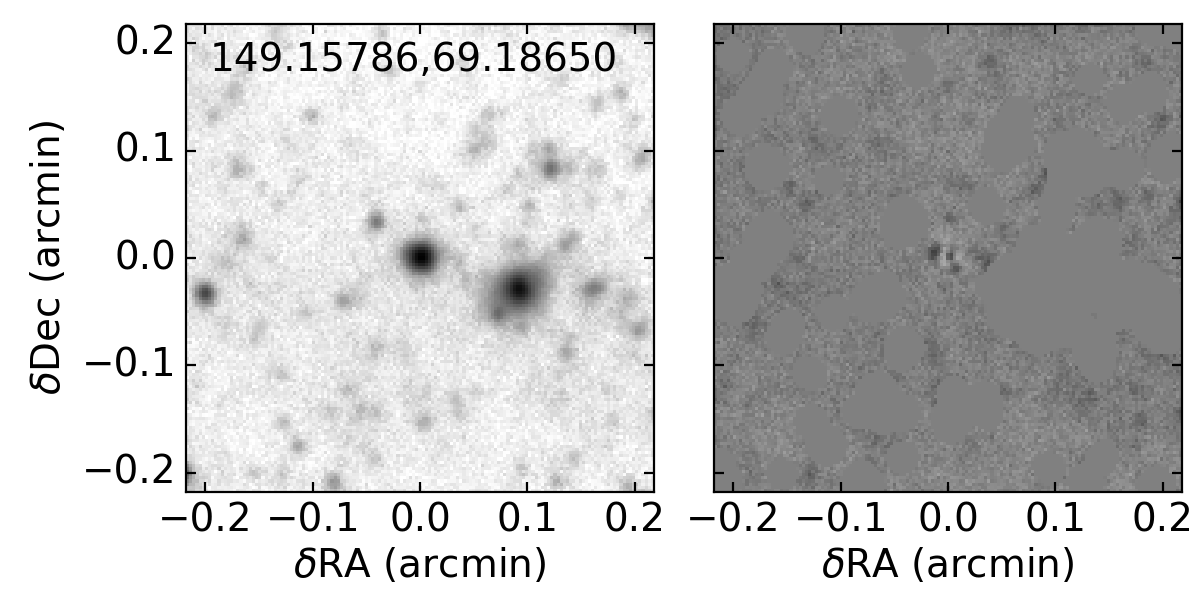} &  & \\
    \end{tabular}
    \caption{Images of the 13 IRAC GC candidates that have three band coverage from Subaru. The left-hand panel for each candidate shows the image; the right hand panel shows the residuals after subtraction of a multi-gaussian expansion model that highlights non-axisymmetric features. Candidates are arranged as in Table \ref{tab:GCcatalog} from left to right and top to bottom, starting with candidate 11 and ending with candidate 23.
}
    \label{fig:GC image IRAC}
\end{figure*}
\begin{figure*}
    \begin{tabular}{lll}
    \includegraphics[width=.3\linewidth,valign=m]{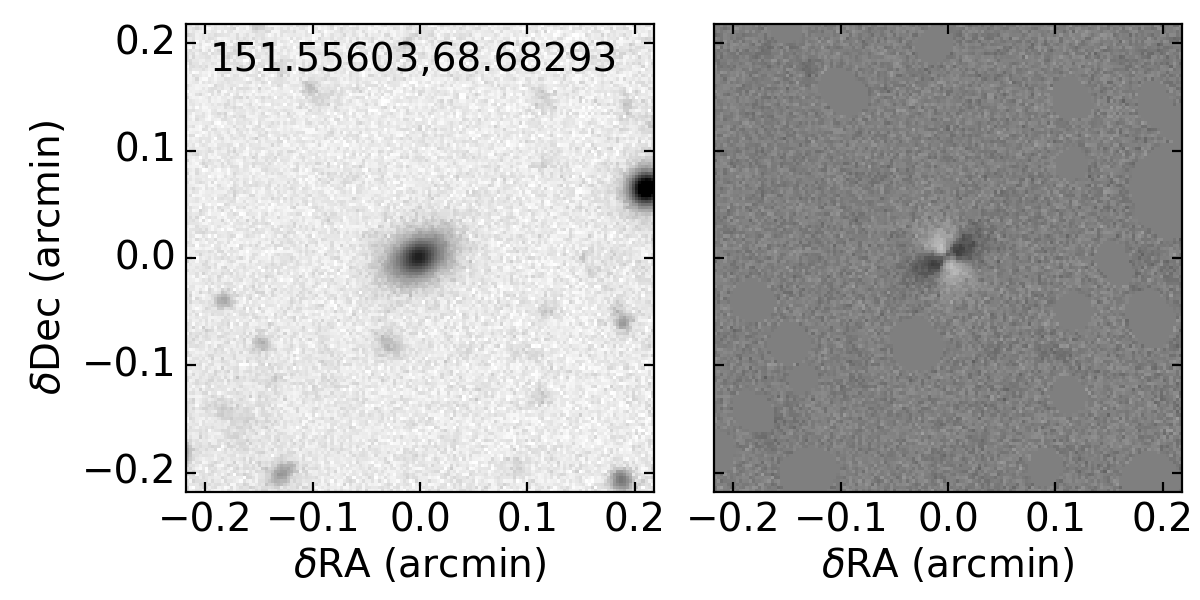} & \includegraphics[width=.3\linewidth,valign=m]{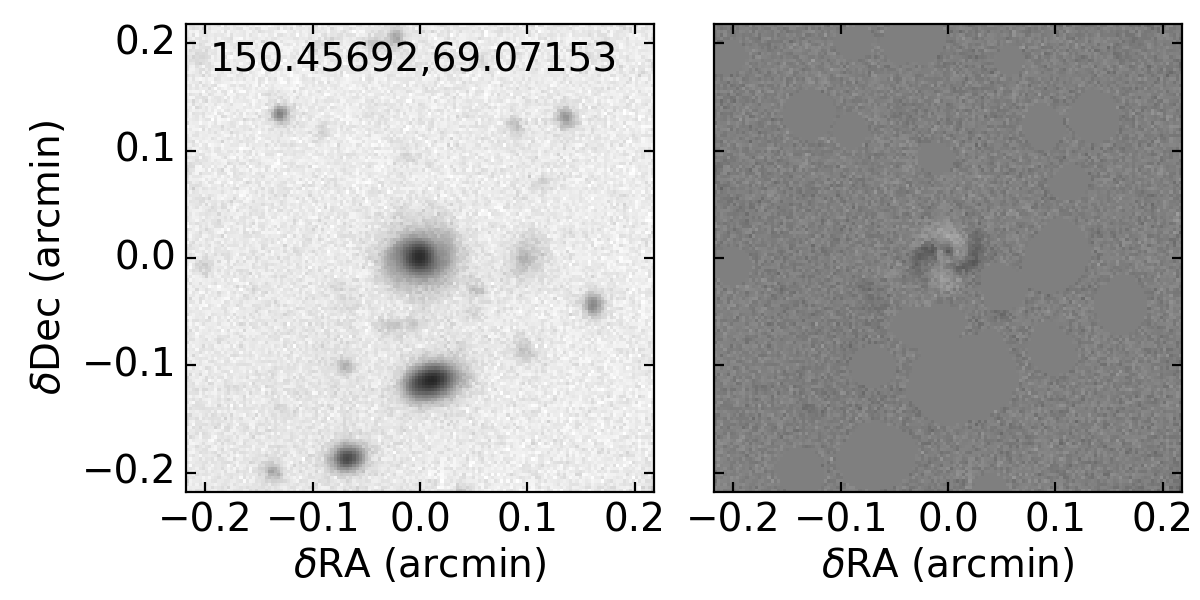} & \includegraphics[width=.3\linewidth,valign=m]{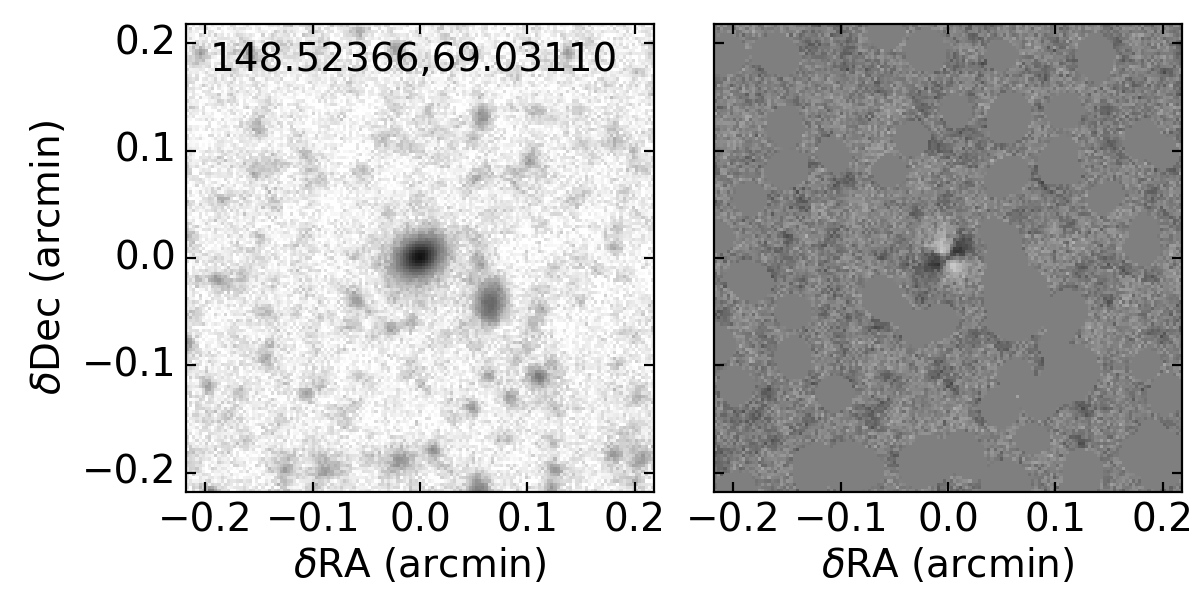}\\
     \includegraphics[width=.3\linewidth,valign=m]{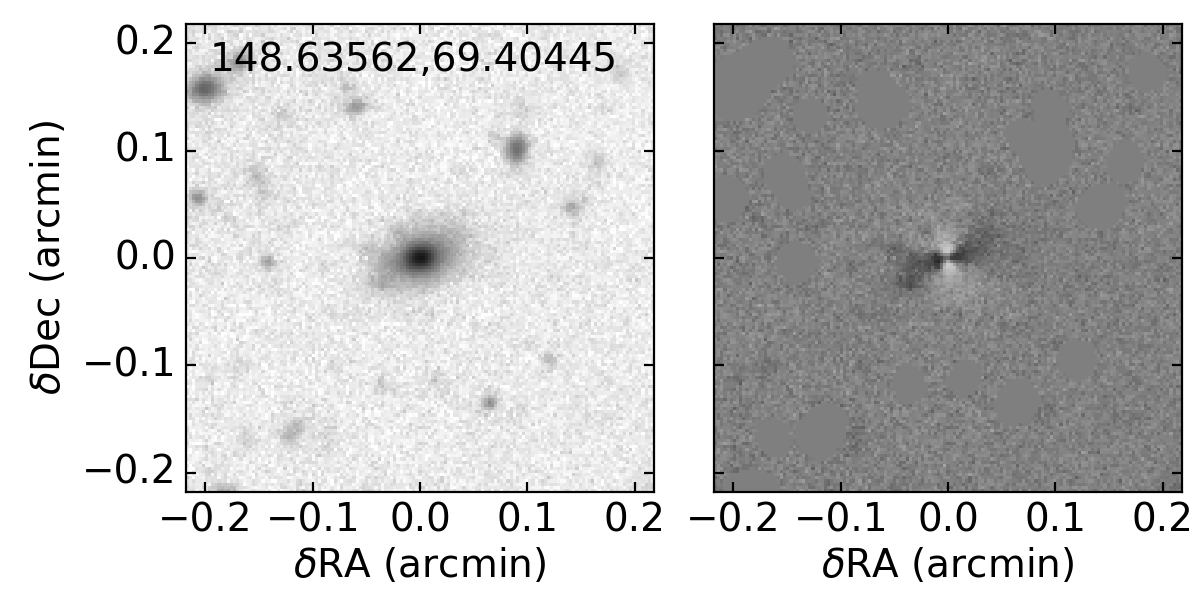} & \includegraphics[width=.3\linewidth,valign=m]{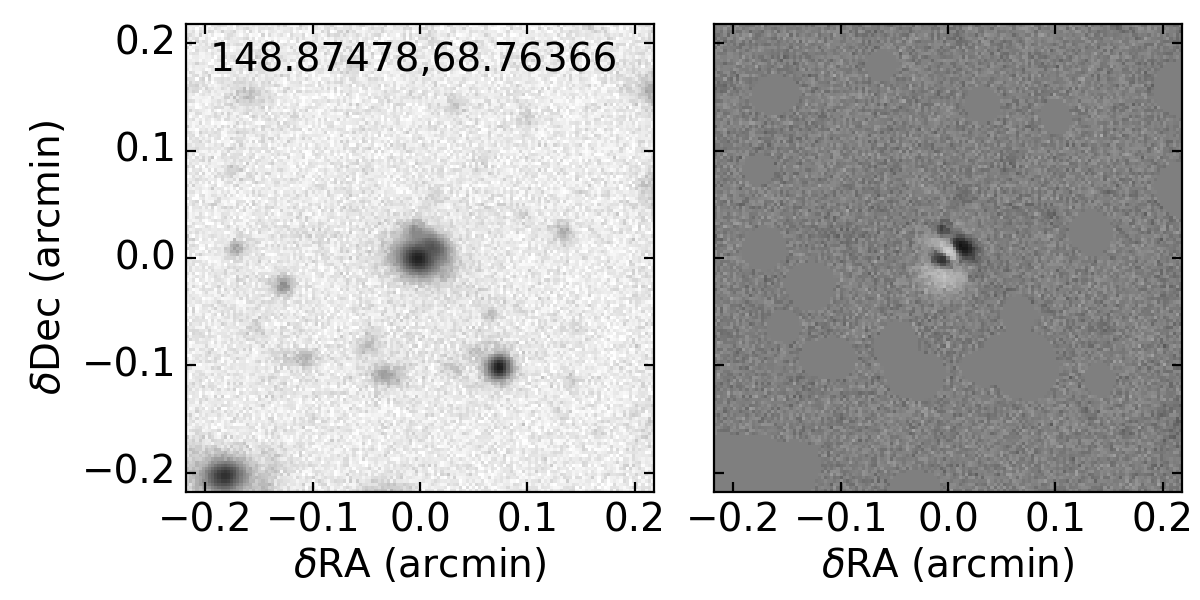} &  \includegraphics[width=.3\linewidth,valign=m]{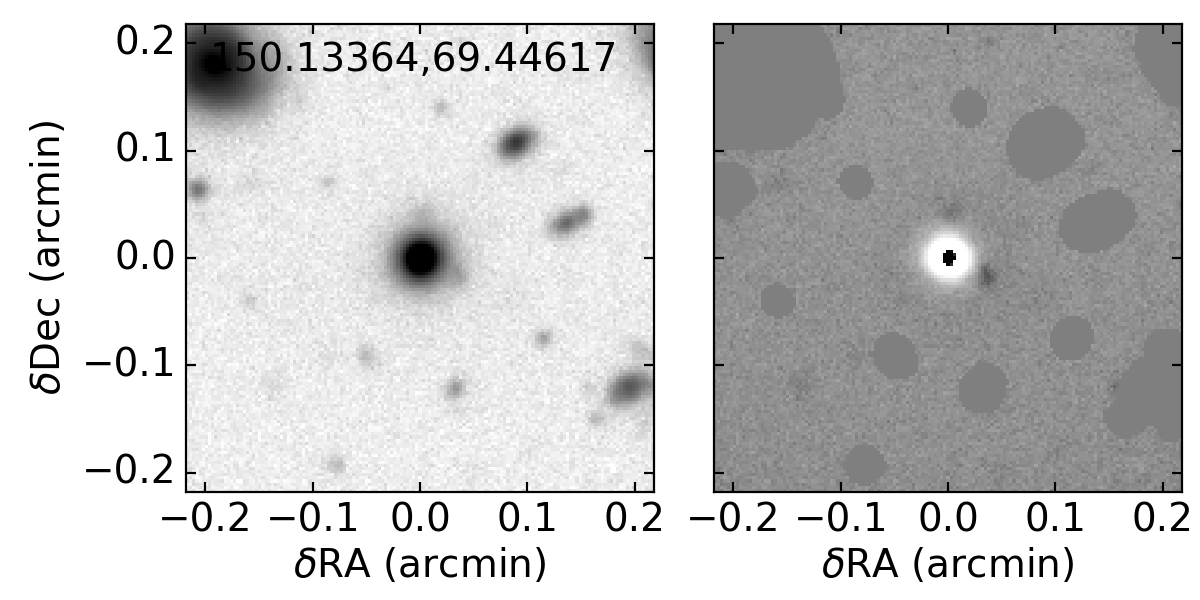}  \\
     \includegraphics[width=.3\linewidth,valign=m]{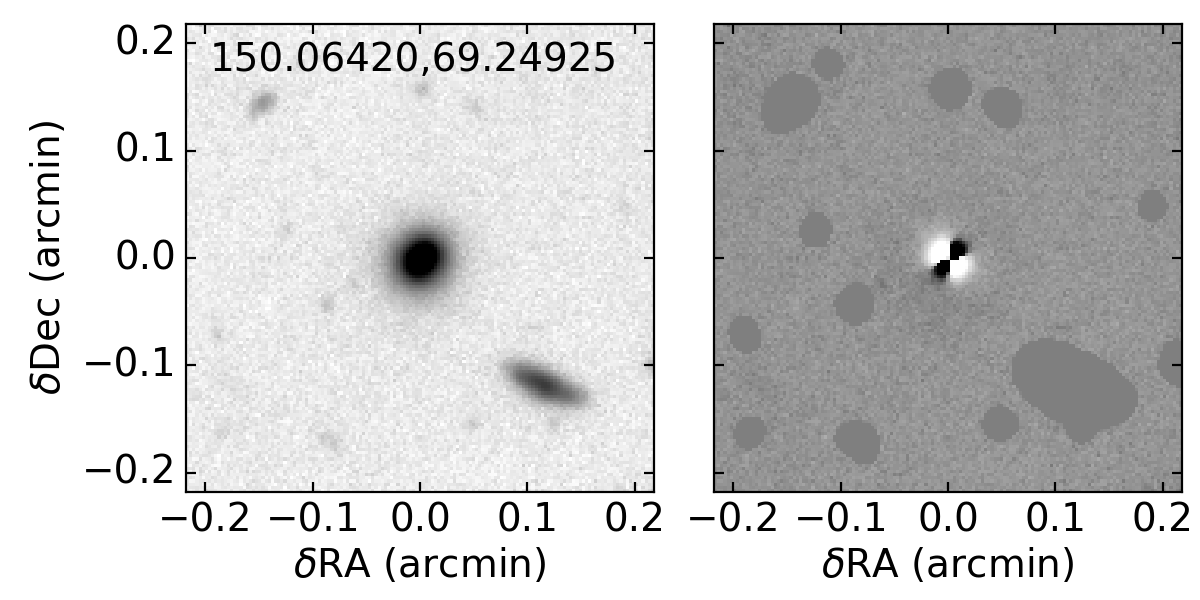} & &  \\
    \end{tabular}
    \caption{Images of the 7 objects that were selected via Subaru--IRAC colour and size selections to be GC candidates that appear to be galaxies on the basis of visual classification. The left-hand panel for each candidate shows the image; the right hand panel shows the residuals after subtraction of a multi-gaussian expansion model that highlights non-axisymmetric features. Candidates are arranged as in Table \ref{tab:GCgalaxy} from left to right and top to bottom.
}
    \label{fig:Galaxyimage IRAC}
\end{figure*}
\section{A best-effort compilation M81 GC candidates}
\label{appendix:A}
Here we present the catalogue for the M81 GC candidates in Table \ref{tab:bestM81}. For objects M81-1 to M81-17, V-band magnitudes were estimated following the process described in section \ref{sec:GCLF}.  For object M81-18 and M81-19, which are taken from the ZD15 catalogue, we identified them in the SDSS Data Release 16 (DR16) via CasJobs  \citep{sdssdr16}. We obtained dereddened model g, r, and i magnitudes for these two objects. We converted these magnitudes to V-band magnitude following \citet{Lupton} transformation equations for stars. The V-band magnitudes for objects from M81-20 to M81-121 were taken from \citet{Nantais11} for the same object with available V band magnitude, which are uncorrected for reddening. $R_{p}$ is the projected radius from M81 galaxy in kpc.

\begin{table*}
 \caption{Best-effort catalogue of M81 GC candidates}
 \label{tab:bestM81}
 \begin{tabular}{lccccccc}
  \hline
M81 ID  & This work &NH10 ID  &ZD15 ID & RA (J2000) & DEC (J2000)   & $R_{p}$&V  \\
&&&& (deg) & (deg) &(kpc)& (mag) \\
(1) &(2)& (3) & (4) & (5) &(6)&(7)&(8) \\
  \hline
M81-1 & 1 & 70349 & $\dots$ & 148.26297 & 69.23004 & 17.4 & 19.89 \\
M81-2 & 2 & $\dots$ & $\dots$ & 149.66775 & 68.86872 & 21.5 & 18.61 \\
M81-3 & 4 & BH91 HS01& $\dots$  & 148.71575 & 69.3293 & 17.0 & 18.9 \\
M81-4 & 3 & $\dots$ & $\dots$ & 148.55384 & 69.36056 & 20.0 & 20.02 \\
M81-5 & 16 & $\dots$ & $\dots$ & 148.83868 & 68.62236 & 27.9 & 20.1 \\
M81-6 & 17 & $\dots$ & $\dots$ & 148.46144 & 69.28084 & 16.6 & 20.49 \\
M81-7 & 11 & $\dots$ & $\dots$ & 148.35462 & 69.30494 & 19.2 & 19.93 \\
M81-8 & 14 & $\dots$ & $\dots$ & 149.88266 & 68.52754 & 40.6 & 20.41 \\
M81-9 & 18 & $\dots$ & $\dots$ & 149.80041 & 68.74459 & 28.8 & 20.52 \\
M81-10 & 19 & $\dots$ & $\dots$ & 149.15786 & 69.1865 & 9.7 & 21.5 \\
M81-11 & 12 & $\dots$ & $\dots$ & 147.7333 & 69.37192 & 32.2 & 20.79 \\
M81-12 & 13 & $\dots$ & $\dots$ & 147.66701 & 69.41603 & 35.0 & 20.92 \\
M81-13 & 15 & $\dots$ & $\dots$ & 149.72654 & 69.44327 & 30.2 & 19.87 \\
M81-14 & $\dots$ & 464 & $\dots$ & 148.73373 & 69.04195 & 3.8 & 19.41 \\
M81-15 & $\dots$ & 34 & $\dots$ & 148.52035 & 69.1551 & 10.0 & 19.67 \\
M81-16 & $\dots$ & $\dots$ & $\dots$ & 149.26089 & 68.59278 & 30.9 & 20.93 \\
M81-17 & $\dots$ & $\dots$ & $\dots$ & 148.83272 & 68.91008 & 9.8 & 20.93 \\
M81-18 & $\dots$ & $\dots$ & M81-C1 & 148.10805 & 68.80711 & 23.9 & 19.38 \\
M81-19 & $\dots$ & $\dots$ & M81-C2 & 148.57085 & 68.92275 & 11.5 & 17.62 \\
M81-20 & $\dots$ & 108 & $\dots$ & 148.60458 & 69.13438 & 7.7 & 20.17 \\
M81-21 & $\dots$ & 173 & $\dots$ & 148.66171 & 69.06951 & 5.1 & 19.16 \\
M81-22 & $\dots$ & 187 & $\dots$ & 148.66542 & 69.05735 & 5.0 & 20.07 \\
M81-23 & $\dots$ & 264 & $\dots$ & 148.69362 & 69.18165 & 8.5 & 20.57 \\
M81-24 & $\dots$ & 345 & $\dots$ & 148.70962 & 69.08567 & 4.2 & 20.9 \\
M81-25 & $\dots$ & 359 & $\dots$ & 148.71271 & 69.13058 & 5.7 & 20.1 \\
M81-26 & $\dots$ & 505 & $\dots$ & 148.74362 & 69.13569 & 5.5 & 20.75 \\
M81-27 & $\dots$ & 526 & $\dots$ & 148.74912 & 69.1576 & 6.6 & 20.48 \\
M81-28 & $\dots$ & 594 & $\dots$ & 148.76158 & 69.12483 & 4.7 & 19.91 \\
M81-29 & $\dots$ & 605 & $\dots$ & 148.76379 & 69.03992 & 3.2 & 20.78 \\
M81-30 & $\dots$ & 628 & $\dots$ & 148.76825 & 69.08773 & 3.0 & 19.65 \\
M81-31 & $\dots$ & 676 & $\dots$ & 148.78025 & 69.12617 & 4.5 & 20.73 \\
M81-32 & $\dots$ & 705 & $\dots$ & 148.78508 & 69.06976 & 2.3 & 20.31 \\
M81-33 & $\dots$ & 720 & $\dots$ & 148.78742 & 69.09765 & 3.0 & 19.27 \\
M81-34 & $\dots$ & 722 & $\dots$ & 148.78767 & 69.07455 & 2.3 & 19.15 \\
M81-35 & $\dots$ & 743 & $\dots$ & 148.79079 & 69.06879 & 2.2 & 17.53 \\
M81-36 & $\dots$ & 839 & $\dots$ & 148.8095 & 69.03509 & 2.6 & 20.19 \\
M81-37 & $\dots$ & 861 & $\dots$ & 148.81317 & 69.00717 & 4.0 & 20.12 \\
M81-38 & $\dots$ & 863 & $\dots$ & 148.8135 & 69.09005 & 2.3 & 20.0 \\
M81-39 & $\dots$ & 876 & $\dots$ & 148.81504 & 69.09661 & 2.6 & 19.86 \\
M81-40 & $\dots$ & 966 & $\dots$ & 148.83004 & 69.09725 & 2.4 & 19.74 \\
M81-41 & $\dots$ & 993 & $\dots$ & 148.834 & 69.09378 & 2.2 & 20.24 \\
M81-42 & $\dots$ & 1029 & $\dots$ & 148.84133 & 69.11045 & 3.0 & 16.92 \\
M81-43 & $\dots$ & 1089 & $\dots$ & 148.85496 & 69.12074 & 3.6 & 19.11 \\
M81-44 & $\dots$ & 1104 & $\dots$ & 148.85708 & 69.02776 & 2.5 & 20.78 \\
M81-45 & $\dots$ & 1154 & $\dots$ & 148.87137 & 69.00861 & 3.6 & 20.79 \\
M81-46 & $\dots$ & 1162 & $\dots$ & 148.87387 & 69.08658 & 1.4 & 18.95 \\
M81-47 & $\dots$ & 1172 & $\dots$ & 148.87542 & 69.03321 & 2.0 & 20.59 \\
M81-48 & $\dots$ & 1257 & $\dots$ & 148.89338 & 69.11173 & 2.9 & 20.79 \\
M81-49 & $\dots$ & 1265 & $\dots$ & 148.8955 & 68.97078 & 5.9 & 19.08 \\
M81-50 & $\dots$ & 1300 & $\dots$ & 148.90475 & 69.10987 & 2.8 & 19.02 \\
M81-51 & $\dots$ & 1301 & $\dots$ & 148.90533 & 69.03544 & 1.9 & 18.44 \\
M81-52 & $\dots$ & 1308 & $\dots$ & 148.90737 & 68.98825 & 4.9 & 20.18 \\
M81-53 & $\dots$ & 1309 & $\dots$ & 148.9075 & 69.05777 & 0.6 & 18.72 \\
M81-54 & $\dots$ & 1327 & $\dots$ & 148.91046 & 69.11528 & 3.2 & 19.88 \\
M81-55 & $\dots$ & 1341 & $\dots$ & 148.91396 & 69.09239 & 1.8 & 19.77 \\
M81-56 & $\dots$ & 1350 & $\dots$ & 148.91671 & 69.04154 & 1.6 & 21.0 \\
M81-57 & $\dots$ & 1352 & $\dots$ & 148.91675 & 69.06948 & 0.7 & 17.78 \\
M81-58 & $\dots$ & 1363 & $\dots$ & 148.91871 & 69.09019 & 1.7 & 19.81 \\
M81-59 & $\dots$ & 1393 & $\dots$ & 148.92471 & 68.91687 & 9.4 & 19.83 \\
M81-60 & $\dots$ & 1413 & $\dots$ & 148.93083 & 69.06434 & 1.0 & 19.36 \\
M81-61 & $\dots$ & 1428 & $\dots$ & 148.93429 & 69.07337 & 1.2 & 20.14 \\
M81-62 & $\dots$ & 1456 & $\dots$ & 148.94183 & 69.02381 & 2.9 & 20.13 \\

   \hline
  \end{tabular}%

\end{table*}

\begin{table*}
 \contcaption{.}
 \label{tab:continue}
 \begin{tabular}{lccccccc}
  \hline
M81 ID  & This work &NH10 ID  &ZD15 ID & RA (J2000) & DEC (J2000)   & $R_{p}$&V  \\
&&&& (deg) & (deg) &(kpc)& (mag) \\
(1) &(2)& (3) & (4) & (5) &(6)&(7)&(8) \\
  \hline
M81-63 & $\dots$ & 1490 & $\dots$ & 148.94867 & 69.10701 & 3.0 & 19.0 \\
M81-64 & $\dots$ & 1495 & $\dots$ & 148.95 & 69.1244 & 4.0 & 20.72 \\
M81-65 & $\dots$ & 1496 & $\dots$ & 148.95 & 69.06445 & 1.4 & 19.17 \\
M81-66 & $\dots$ & 1506 & $\dots$ & 148.95208 & 69.10333 & 2.8 & 20.02 \\
M81-67 & $\dots$ & 1512 & $\dots$ & 148.95317 & 69.08952 & 2.1 & 20.06 \\
M81-68 & $\dots$ & 1524 & $\dots$ & 148.95492 & 69.02093 & 3.2 & 19.0 \\
M81-69 & $\dots$ & 1537 & $\dots$ & 148.95942 & 68.973 & 6.0 & 19.6 \\
M81-70 & $\dots$ & 1563 & $\dots$ & 148.96383 & 69.05657 & 1.8 & 18.92 \\
M81-71 & $\dots$ & 1571 & $\dots$ & 148.96608 & 69.12764 & 4.3 & 18.79 \\
M81-72 & $\dots$ & 1627 & $\dots$ & 148.97712 & 69.04795 & 2.3 & 18.54 \\
M81-73 & $\dots$ & 1635 & $\dots$ & 148.97908 & 69.0156 & 3.7 & 18.6 \\
M81-74 & $\dots$ & 1643 & $\dots$ & 148.98025 & 69.06042 & 2.1 & 19.86 \\
M81-75 & $\dots$ & 1652 & $\dots$ & 148.98412 & 69.04125 & 2.6 & 20.94 \\
M81-76 & $\dots$ & 1816 & $\dots$ & 149.01275 & 69.1221 & 4.5 & 19.79 \\
M81-77 & $\dots$ & 1859 & $\dots$ & 149.02087 & 69.15591 & 6.4 & 19.9 \\
M81-78 & $\dots$ & 1946 & $\dots$ & 149.03621 & 69.04015 & 3.7 & 20.28 \\
M81-79 & $\dots$ & 1951 & $\dots$ & 149.03675 & 69.00655 & 5.0 & 20.85 \\
M81-80 & $\dots$ & 2081 & $\dots$ & 149.05892 & 69.08481 & 4.0 & 20.78 \\
M81-81 & $\dots$ & 2087 & $\dots$ & 149.05967 & 69.02499 & 4.6 & 20.23 \\
M81-82 & $\dots$ & 2163 & $\dots$ & 149.07296 & 68.95335 & 8.2 & 19.83 \\
M81-83 & $\dots$ & 2170 & $\dots$ & 149.07404 & 68.98852 & 6.4 & 19.9 \\
M81-84 & $\dots$ & 2171 & $\dots$ & 149.07408 & 69.05128 & 4.3 & 20.4 \\
M81-85 & $\dots$ & 2196 & $\dots$ & 149.07896 & 68.9987 & 6.0 & 19.95 \\
M81-86 & $\dots$ & 2219 & $\dots$ & 149.08537 & 69.04696 & 4.6 & $\dots$ \\
M81-87 & $\dots$ & 2230 & $\dots$ & 149.08787 & 69.03375 & 4.9 & 18.88 \\
M81-88 & $\dots$ & 2327 & $\dots$ & 149.11446 & 69.01942 & 5.8 & 17.63 \\
M81-89 & $\dots$ & 2355 & $\dots$ & 149.13221 & 69.06524 & 5.5 & 20.46 \\
M81-90 & $\dots$ & 2381 & $\dots$ & 149.15371 & 69.02954 & 6.4 & 19.75 \\
M81-91 & $\dots$ & 2490 & $\dots$ & 149.25767 & 68.94803 & 11.1 & 19.55 \\
M81-92 & $\dots$ & PBH30244 & $\dots$ & 149.47854 & 68.81678 & 20.5 & $\dots$ \\
M81-93 & $\dots$ & PBH40083 & $\dots$ & 149.16029 & 69.38064 & 20.7 & $\dots$ \\
M81-94 & $\dots$ & PBH40165 & $\dots$ & 148.766 & 69.2605 & 12.6 & $\dots$ \\
M81-95 & $\dots$ & PBH40181 & $\dots$ & 148.93371 & 69.23658 & 10.8 & $\dots$ \\
M81-96 & $\dots$ & PBH50037 & $\dots$ & 148.7765 & 68.94056 & 8.2 & $\dots$ \\
M81-97 & $\dots$ & PBH50225 & $\dots$ & 149.16912 & 68.99794 & 7.6 & $\dots$ \\
M81-98 & $\dots$ & PBH50233 & $\dots$ & 148.98262 & 69.00094 & 4.6 & 19.26 \\
M81-99 & $\dots$ & PBH50286 & $\dots$ & 148.74529 & 69.01617 & 4.5 & $\dots$ \\
M81-100 & $\dots$ & PBH50394 & $\dots$ & 148.56875 & 69.04289 & 7.3 & $\dots$ \\
M81-101 & $\dots$ & PBH50401 & $\dots$ & 149.13229 & 69.044 & 5.6 & $\dots$ \\
M81-102 & $\dots$ & PBH50696 & $\dots$ & 148.76021 & 69.09392 & 3.4 & 18.16 \\
M81-103 & $\dots$ & PBH50785 & $\dots$ & 148.6495 & 69.11197 & 6.1 & 18.96 \\
M81-104 & $\dots$ & PBH50826 & $\dots$ & 148.96746 & 69.11956 & 3.8 & 19.85 \\
M81-105 & $\dots$ & PBH50886 & $\dots$ & 148.87912 & 69.12739 & 3.9 & 18.16 \\
M81-106 & $\dots$ & PBH50960 & $\dots$ & 148.96692 & 69.13853 & 4.9 & 18.56 \\
M81-107 & $\dots$ & PBH51027 & $\dots$ & 148.58342 & 69.15308 & 8.8 & $\dots$ \\
M81-108 & $\dots$ & PBH60045 & $\dots$ & 148.98692 & 68.87039 & 12.4 & $\dots$ \\
M81-109 & $\dots$ & PBH80172 & $\dots$ & 148.46538 & 68.95122 & 11.9 & $\dots$ \\
M81-110 & $\dots$ & PBH90103 & $\dots$ & 148.416 & 68.80002 & 19.8 & $\dots$ \\
M81-111 & $\dots$ & BH91 HS02 & $\dots$ & 149.10725 & 69.12439 & 6.2 & $\dots$ \\
M81-112 & $\dots$ & BH91 HS06 & $\dots$ & 148.78367 & 68.83236 & 14.8 & $\dots$ \\
M81-113 & $\dots$ & BH91 R012 & $\dots$ & 148.86237 & 68.74822 & 19.9 & $\dots$ \\
M81-114 & $\dots$ & S04 & $\dots$ & 148.99113 & 69.03974 & 2.8 & 19.31 \\
M81-115 & $\dots$ & S08 & $\dots$ & 148.8429 & 69.08866 & 1.8 & 18.13 \\
M81-116 & $\dots$ & S09 & $\dots$ & 148.83995 & 69.09222 & 2.0 & 18.95 \\
M81-117 & $\dots$ & S10 & $\dots$ & 148.87615 & 69.1017 & 2.3 & 20.75 \\
M81-118 & $\dots$ & S11 & $\dots$ & 148.89933 & 69.10709 & 2.6 & $\dots$ \\
M81-119 & $\dots$ & S12 & $\dots$ & 148.88767 & 69.11108 & 2.9 & 18.77 \\
M81-120 & $\dots$ & S13 & $\dots$ & 149.0232 & 69.11194 & 4.2 & 19.19 \\
M81-121 & $\dots$ & S16 & $\dots$ & 148.91747 & 69.12523 & 3.8 & $\dots$ \\
   \hline
  \end{tabular}%

\end{table*}

 
\bsp	
\label{lastpage}
\end{document}